\pdfoutput=1

\documentclass[11pt]{article}

\usepackage{acl}

\usepackage{times}
\usepackage{hhline}
\usepackage{latexsym}
\usepackage{rotating}
\usepackage{tikz,lipsum}
\usepackage{cuted,tcolorbox,lipsum}
\usepackage{float}
\usepackage{amsthm}
\usepackage{soul}
\usepackage{amssymb}
\usepackage{pifont}
\usepackage{todonotes}
\usepackage{arydshln}

\usepackage{listings}
\newtheorem{definition}{Definition}
\usepackage{graphicx}

\lstdefinestyle{pythonstyle}{
    language=Python,
    basicstyle=\ttfamily\small,
    keywordstyle=\color{blue},
    commentstyle=\color{gray},
    stringstyle=\color{black},
    frame=single,
    breaklines=true,
    captionpos=b,
    numbers=left,
    numberstyle=\tiny\color{gray},
    numbersep=5pt,
    showstringspaces=false
}

\usepackage[T1]{fontenc}

\usepackage[utf8]{inputenc}

\usepackage{microtype}

\usepackage{inconsolata}

\usepackage{graphicx}
\usepackage{multirow}
\usepackage{multicol}
\usepackage{color, xcolor} 
\usepackage{colortbl}

\definecolor{peach}{HTML}{f5e2d5}

\DeclareRobustCommand{\xmark}{\ding{55}}

\title{
    \begin{minipage}{0.08\textwidth}
        \centering
        \includegraphics[width=\textwidth]{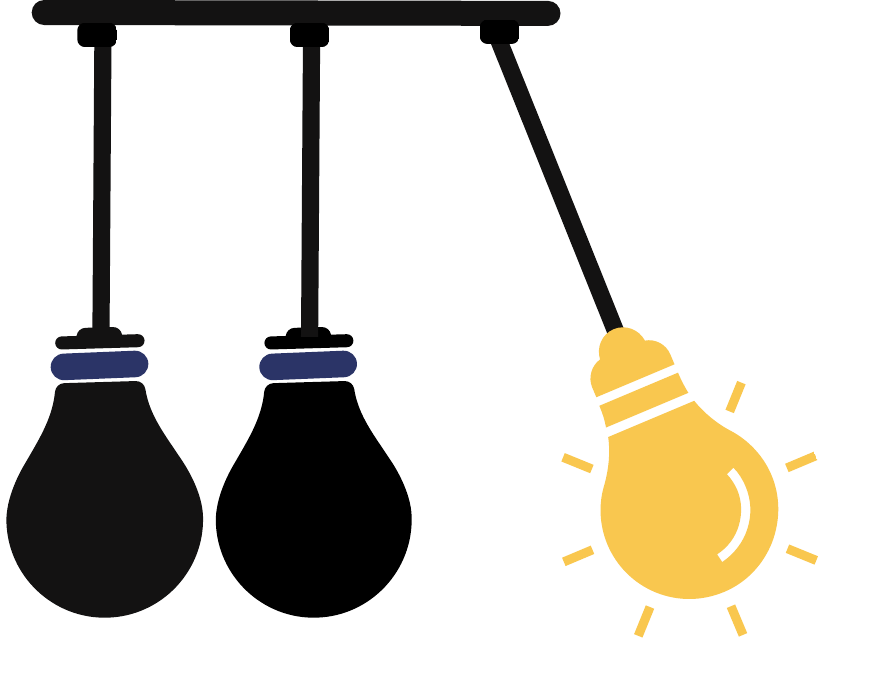}
    \end{minipage}
    \hspace{0.2cm}
    \begin{minipage}{0.8\textwidth}
        In-depth Research Impact Summarization\\  through Fine-Grained Temporal Citation Analysis
    \end{minipage}
}

\author{
 Hiba Arnaout$^1$,
 Noy Sternlicht$^2$,
 Tom Hope$^{2,3}$, 
 Iryna Gurevych$^1$\\
 \vspace{0.25em}\\
  $^1$UKP Lab, TU Darmstadt and Hessian Center for AI (hessian.AI)\\
  $^2$School of Computer Science and Engineering, Hebrew University of Jerusalem\\
  $^3$The Allen Institute for AI (AI2)
   \vspace{0.25em}\\
     \url{www.ukp.tu-darmstadt.de}\\
}

\begin{document}
\maketitle
\begin{abstract}

Understanding the impact of scientific publications is crucial for identifying breakthroughs and guiding future research. Traditional metrics based on citation counts often miss the nuanced ways a paper contributes to its field. In this work, we propose a new task: generating nuanced, expressive, and time-aware impact summaries that capture both praise (confirmation citations) and critique (correction citations) through the evolution of fine-grained citation intents. We introduce an evaluation framework tailored to this task, showing moderate to strong human correlation on subjective metrics such as insightfulness. Expert feedback from professors reveals a strong interest in these summaries and suggests future improvements. Data and code are made available. 
\footnote{\url{https://ukplab.github.io/acl2026-generating-impact-summaries}}

\end{abstract}

\section{Introduction}

\begin{figure}[!ht]
\centering
\includegraphics[width=\columnwidth]{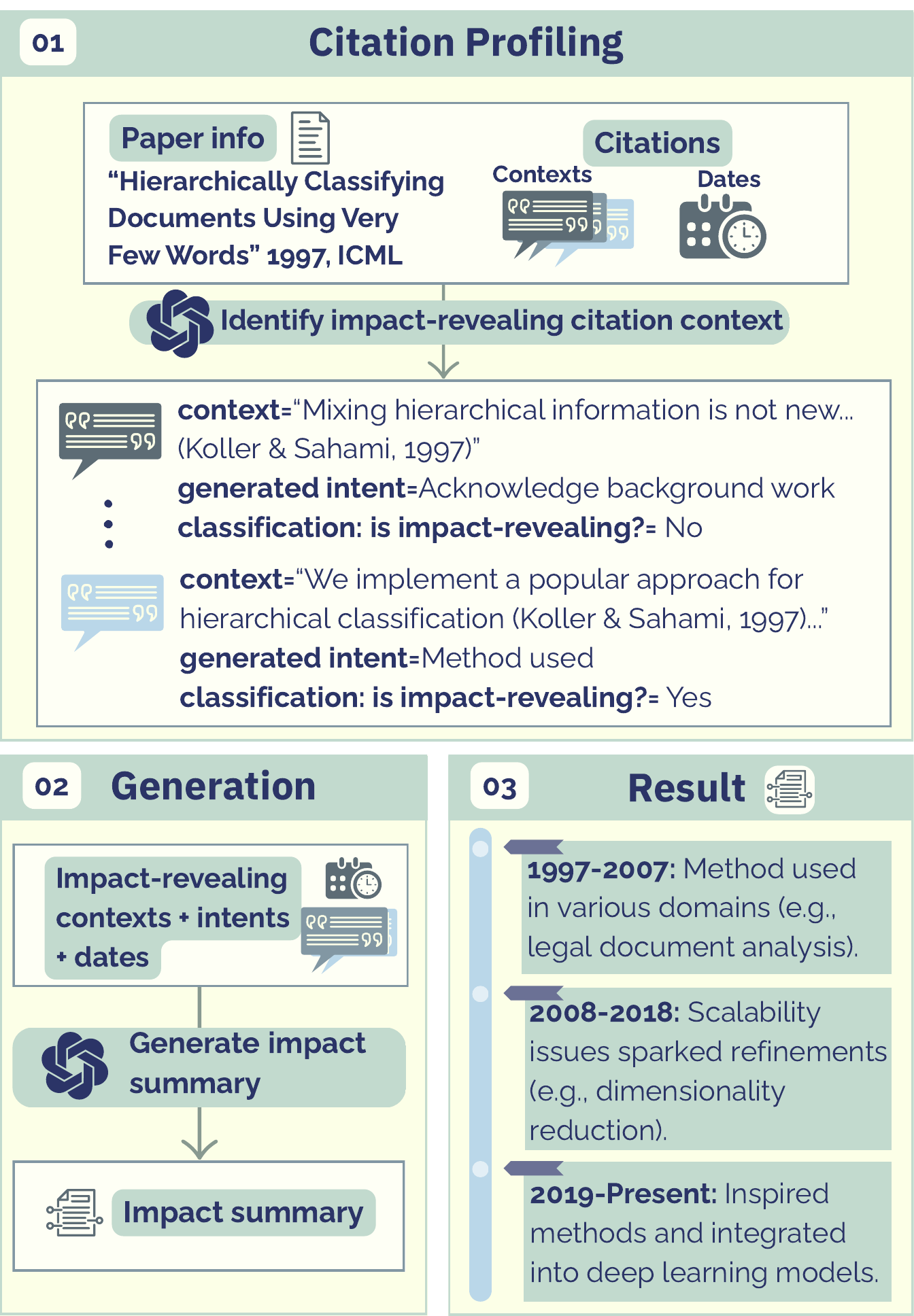}
  \caption{We propose a new task to summarize a paper’s evolving impact over time, by analyzing \textit{impact-revealing} citation contexts, reflecting both praise (confirming ideas) and critique (calls for correction). Our summary of ~\citet{koller1997hierarchically} reveals its  impact trajectory—adaptation, critique, and rediscovery—offering deeper insight than its $\sim$1.4k cite count.}
  \label{fig:motivation}
\end{figure}

Citation counts are a common proxy for measuring the impact of research papers, but they offer only a shallow view that fails to capture \textit{how} a paper has influenced subsequent work. A raw citation count does not reveal whether a paper was foundational, extended, critiqued, or merely mentioned in passing. To truly understand the impact of a research paper, we must go beyond simple counts and examine the \textit{context} in which it is cited, analyzing how its ideas have been discussed, applied, and evolved over time. Manually tracking how a paper is discussed across a very large number of publications and diverse domains is infeasible due to the scale and complexity of the task. To address this, we introduce a new task: \textbf{research impact summary generation}, which aims to generate concise, time-aware narratives that reflect a paper’s evolving scientific influence. These summaries can support practical use cases such as enhancing future research directions by helping researchers quickly assess a paper’s relevance and legacy, or assisting funding agencies, hiring committees, and research administrators in making more informed evaluations of scholarly contributions. Figure~\ref{fig:motivation} shows an example of an impact summary about a research paper~\cite{koller1997hierarchically}, published at The International Conference on Machine Learning in 1997. This paper’s impact has followed a dynamic trajectory, initially used for its method (1997–2007), later critiqued for scalability and accuracy limitations, then refined (2008–2018), and recently rediscovered as an inspiration for modern methods (2019–present). This example shows the need to look beyond the paper's $\sim$1.4k  citations to understand the deeper story of how ideas gain relevance, face scrutiny, and find renewed significance.

We break the task of generating impact summaries for research papers into two subtasks. First, given all the citation contexts of a paper, we identify the ``impact revealing'' citations that directly interact with the paper, and their specific intents, whether in confirmation of its ideas (e.g., ``method use'') or in correction (e.g., ``identifying limitations and proposing refinements''). Second, using \textit{only} the impact-revealing citation contexts, their dates and intents, we generate an impact summary that captures the paper's influence over time. To the best of our knowledge, this is the first attempt to express scientific impact through time-aware textual summaries derived from citation context analysis.  An overview is shown in Figure~\ref{fig:motivation}. To implement these subtasks, our method, in Section~\ref{sec:method}, uses in-context learning (ICL) with Large Language Models (LLMs) to generate fine-grained citation intents in the first stage. In the second stage, it filters for impact-revealing citation contexts based on the generated intents and feeds them to an LLM to generate semi-structured impact summaries under a pre-defined schema.

Existing work on scientific impact analysis predominantly focuses on citation and similar numerical counts as the main indicator of impact~\cite{gu2024forecasting}, ignoring the exact reasons a paper has been cited, which can often be ``unimpactful'', e.g., citing a paper to acknowledge background work on a topic. Work on citation context analysis~\cite{lauscher2022multicite} studies the role citations play, but is based on coarse categories and is not directly intended for impact analysis. The goal of our new task is to create a narrative about a paper's impact using fine-grained citation intent analysis. Moreover, existing work typically emphasizes praise (confirmatory citations) ~\cite{zhang2024pst,valenzuela2015identifying}, but scientific progress relies on both confirmation and correction \cite{catalini2015incidence}, as ideas advance through adoption as well as critique and refinement. In this work, we ensure that our generated impact summaries capture both aspects of impact (when applicable), as shown in Figure~\ref{fig:motivation}. To support this broader view of impact, we extend the existing PST-Bench data set~\cite{zhang2024pst}, which only focuses on ``motivation'' and ``inspiration'' intentions to consider an impact revealing citation, by introducing, in Section~\ref{pstbenchextended}, a new ground-truth data set of 4000 citation contexts labeled for whether they reveal impact. Our extended dataset also covers not only additional intents of confirmatory citations (e.g., method use), but also correction citations (e.g., method refinement). For detailed comparisons with related work, see Appendix~\ref{app:literaturetable} and Section~\ref{sec:related}.

Finally, in the absence of gold-standard summaries for this novel impact generation task, we propose an automated evaluation framework to measure the trustworthiness and informativeness of our impact summaries (Section~\ref{sec:metrics}) and demonstrate a moderate to strong correlation with human assessments (Section~\ref{sec:ablation}). To further validate our approach, we collect expert feedback from  professors on the usefulness of impact summaries about their own papers (Section~\ref{expertreview}), demonstrating the need for such summaries. For future research, we release the code and data used to develop this work.

\section{Method}
\label{sec:method}

\subsection{Definitions}
\label{sec:task}
We propose the task of generating time-aware impact summaries for research papers by identifying and analyzing impact-revealing citation contexts and their intents over time.

\begin{definition}  
\normalfont  \textbf{Citation context} refers to the specific part of a paper \( p' \) where another paper \( p \) is mentioned, including the surrounding text that explains how \( p \) is relevant to \( p' \).  
\end{definition}

\begin{definition}
\normalfont \textbf{Fine-grained citation intent} refers to the specific, nuanced reason (the ``why'') behind citing paper \( p \) in paper \( p' \), as inferred from the citation context. Unlike predefined or categorical intent schemes, fine-grained citation intent is expressed in free-form text and does not rely on a fixed set of labels or a predefined taxonomy.
\end{definition}
A few examples are in Table~\ref{iclexamples2}.

\begin{definition}
\normalfont An \textbf{Impact-revealing citation intent} is a citation intent where the cited paper \( p \) has a direct influence on the citing paper \( p' \) through: (i) \textit{Confirmation}, by adopting or building upon \( p \); or (ii) \textit{Critique}, by correcting or refining \( p \). Citation intents that do not fall under these categories are considered non-impact-revealing.  
\end{definition}  

Figure~\ref{fig:motivation} shows the intent ``acknowledge background work'', for the context that simply refers to prior work in a general manner, without actively building upon it. Hence, this is not considered impact-revealing. In contrast, the intent ``method used'', describing the implementation of a popular approach is indeed impact-revealing.

\begin{definition}  \label{def:scientificimpact}
\normalfont  A \textbf{scientific impact summary} of a research paper describes the impact paper \( p \) has had on subsequent research. In other words, how it has been directly used for both confirmation and critique. The summary is time-aware, taking into account the date of these citations, to track the evolution of its impact over time.
\end{definition}

A sample impact summary is in Figure~\ref{fig:motivation}, complete examples are in Appendix~\ref{app:impactexamples}.

\subsection{Identifying impact-revealing intents}
\label{iclmethod}
We use ICL with LLMs to generate fine-grained intents and classify them as either ``impact-revealing'' or ``other''.  To facilitate this, we manually craft a prompt that incorporates a set of examples covering various citation intents, and ensuring the model can learn from them to generate fine-grained intents and classify them correctly (prompt and examples are in Appendix~\ref{app:intentprompt} and Table~\ref{iclexamples2}, respectively). This prompt is designed to capture both the confirmatory and correction citations, allowing for a nuanced understanding of how the cited paper influences the citing paper. It aims to generate concise but expressive labels that clearly explain the reason behind the citation without being overly general or overly specific, e.g., ``identifying knowledge gap in literature'' (more examples in Appendix~\ref{trainingexamples}). To examine the quality of our method, we conduct experiments in Sections~\ref{sec:iclforintentgeneration} and ~\ref{sec:intentexternalmethods}.

\begin{table*}
  \centering
  \resizebox{0.9\linewidth}{!}{
  \begin{tabular}{|p{8.5cm}|p{3.5cm}|c|}
    \hline
    \textbf{\cellcolor{gray!15}Citation context}            & \textbf{\cellcolor{gray!15}Intent}   & \textbf{\cellcolor{gray!15}Class} \\
    \hline
   ..we apply a minimization process [1]. & use of minimization methodology & impact-revealing\\
\hline
Chiu and Nichols (2016) introduced convolutional neural networks for NER & background about NER methods & other\\
\hline
Quirk and Poon (2017) and Peng et al. (2017) build two distantly supervised datasets without human annotation, which may make the evaluation less reliable.. we present .. a large-scale
human-annotated document-level RE dataset.. & criticizing existing datasets and proposing a better one & impact-revealing\\
 \hline
  \end{tabular}}
  \caption{\label{iclexamples2}Training examples for fine-grained intent generation and classification. Full table in Appendix~\ref{trainingexamples}.}
\end{table*}

\subsection{Generating scientific impact summaries}

To generate scientific impact summaries, we begin by identifying impact-revealing citation contexts through the generated fine-grained  intents (as detailed in Section~\ref{iclmethod}). To ensure the generation of consistent and easily comparable summaries by LLMs, we design a tailored prompt (see Appendix~\ref{app:generateprompt}) that incorporates our definition of scientific impact (see Definition~\ref{def:scientificimpact}). This prompt includes the impact-revealing citation contexts of a given paper, augmented with corresponding fine-grained intents and citation timestamps (i.e., publication years). We evaluate the effectiveness of this approach through an ablation study (Section~\ref{sec:ablation}) and an expert review (Section~\ref{expertreview}).

\subsection{Evaluation metrics}
\label{sec:metrics}

Evaluating our summaries is challenging due to the lack of gold-standard datasets for this novel task. We explored sources like Wikipedia articles on influential papers~\footnote{E.g., Article for~\cite{vaswani2017attention}: \url{https://en.wikipedia.org/wiki/Attention_Is_All_You_Need}} and Test of Time Award pages\footnote{E.g., WSDM 2022 Test of Time Award Winner: \url{https://www.wsdm-conference.org/2022/timetable/event/wsdm-awards-program-test-of-time-presentation/}}, but these were insufficient, as they cover few papers and focus on content or general praise rather than usage-based insights. Web searches (e.g., ``\textit{what is the impact of paper X}'') mostly returned the original paper or explanatory blog posts. Given these limitations, we introduce a new automated, reference-free evaluation framework that measures trustworthiness and informativeness, and correlates moderately to strongly with human judgment (Appendix~\ref{app:humancorrelation}).

\noindent
\textbf{Trustworthiness.} Measuring summary correctness is challenging due to the large gap between input and output lengths, compressing thousands of citation contexts into a few sentences. Without a gold reference, determining what content should be included or excluded becomes difficult. Additionally, time-augmented impact descriptions must be accurate both in content and assigned time period.  Inspired by evaluation frameworks in the RAG setting~\cite{Ru2024RAGCheckerAF, Shahul2023RAGAsAE,Asai2024OpenScholarSS}, which aligns with our query-based summarization approach, we define:  \textbf{(1) Faithfulness:} Our time-aware faithfulness metric examines whether details in the impact summary, i.e., the impact description elaborating on the dominant intent of an impact period, is grounded in the paper's citation contexts (i.e., the LLM's input context). We first split the impact summaries into impact descriptions representing different time periods (as defined in the output schema in Appendix~\ref{app:generateprompt}). Next, we instruct an LLM to verify each against citations from the corresponding period.  Note that a summary impact description is verified against many citations at once and not in a simple pairwise entailment, since a single impact description can encompass information from multiple citations or discuss trends, which are only possible to verify by inspecting many sources (e.g., ``\textit{the paper is frequently used for...}''). The evaluator LLM is instructed to assess the impact description against the provided citations, determine its faithfulness, and justify the decision with a list of supporting citations. This procedure resembles the task of a human annotator evaluating the faithfulness of a machine-generated text~\cite{Kim2024FABLESEF}. \textbf{(2) Coverage:}  Coverage is defined as the ratio of impact-revealing intents mentioned in the summary. For example, if a paper is cited for its method use, for inspiring new research, and for exposing limitations and refinement, the summary should reflect all these intents, including relevant details. Inspired by the evaluation rubrics for coverage proposed in~\cite{Asai2024OpenScholarSS}, which defines it as the topics (themes) mentioned (i.e., the diversity of impact-revealing intents in our case), we develop a two-step evaluation. In the first step, we cluster the list of fine-grained intents under similar topics, e.g., ``\textit{method used in the legal domain}''. These labels might have slightly different granularity, which depends on how much details the members (i.e., intents) of that cluster offer. Next, the evaluator LLM takes both the impact summary and the list of cluster titles that this summary should ideally cover, and returns a list of topics that were actually covered. Finally, we divide the number of intents mentioned in the summary by the total number of impact-revealing intent clusters. Prompts for faithfulness and coverage are in Appendix~\ref{evalprompts}. \textbf{(3) Citation Year Compliance: } We implement this metric as a script that flags citations falling outside the target impact period. 

\noindent
\textbf{Informativeness. } We also assess the informativeness of impact summaries, as trustworthy ones may still be unhelpful if they lack meaningful insights. Grounded in our definitions (Section~\ref{sec:task}), the informativeness of an impact summary depends on its ability to clearly describe the paper's direct influence on other papers, track the types of influence (intents) over time, and provide details inferred from the citation contexts. We define the following metrics: \textbf{(1) Insightfulness} measures how well the impact summary articulates the paper’s direct influence on other works. \textbf{(2) Trend Awareness}  evaluates the extent to which the impact summary identifies and distinguishes between different periods of the paper’s influence over time. \textbf{(3) Specificity} assesses whether the impact summary includes concrete examples, such as techniques, influenced by the paper. To systematically evaluate these criteria, we leverage G-Eval~\cite{liu2023g}, a framework that uses LLMs with chain-of-thought (CoT) reasoning to assess the impact summaries. By integrating structured evaluation steps, G-Eval enables LLMs to reason step-by-step, improving the reliability and depth of assessments. We transform our criteria into prompts in Appendix~\ref{evalprompts}.

\subsection{A new dataset for identifying impact-revealing citation contexts}
\label{pstbenchextended}
To construct a new dataset for classifying citation contexts as ``impact-revealing'' or not, we start by building on the PST-Bench dataset~\cite{zhang2024pst}, which focuses on positive influence through ``inspiration'' and ``motivation''. We select 1k impact-revealing citations from PST-Bench. Next, we augment this by first manually crafting textual patterns for both confirmation and correction intents, and use GPT-4 to generate variations. After crawling 200k random citation contexts (from the Semantic Scholar Academic Graph (S2AG)\footnote{\url{https://www.semanticscholar.org/product/api}}) and checking if they match any of the textual patterns, we randomly sample 1k instances of impact-revealing contexts (a total of 2k impact-revealing with the original selection from PST-Bench, covering both confirmatory and correct citations). Additionally, we sample 2k non-impact-revealing citations (``other'') directly from the PST-Bench dataset (references that were \textit{not} annotated as influential). We also ensure none of these match any of our impact-revealing citation patterns. This results in a balanced 4k citation context dataset, which we release alongside this work. To validate the quality of our automatically collected impact-revealing examples, we manually reviewed a random sample of 100 instances and found that 90\% were correctly labeled. The remaining 10\% consisted of edge cases where surface patterns, such as the phrase ``motivated by'' (e.g., Voters are motivated by partisan social identities... (Greene 2004)), appeared to indicate impact but did not actually reflect the citation’s true intent. Details on the construction and textual patterns are in Appendix~\ref{gt}.

\section{Experiments}
\label{sec:experiments}

\subsection{In-context-learning for fine-grained intent generation}
\label{sec:iclforintentgeneration}
\noindent
\textbf{Zero-shot vs. ICL.} To evaluate LLMs' ability to generate fine-grained intents and classify them as ``impact-revealing'' or ``non-impact-revealing'', we manually select 10 citation contexts (Table~\ref{iclexamples2} and Appendix~\ref{trainingexamples}).  Our selection criteria was based on diversity, a set that contains impact-revealing (praise, critique) and non-impact-revealing (``other'') intents. The ``other'' class includes incidental mentions or contexts lacking sufficient information to understand the reason behind it. We use our prompt (Appendix~\ref{app:intentprompt}) with and without these examples to generate and classify intents, using GPT4o-mini, of 200 randomly sampled citation contexts. Two annotators (a postdoc and a PhD student) assess: (i) whether the classification as impact-revealing is correct, (ii) whether the generated intent accurately reflects the citation reason, and (iii) whether the intent is concise, as LLM results can be highly verbose. Results, including per-field metrics and qualitative examples, are in Appendix~\ref{iclresults}. We find that including  examples significantly improves performance across all metrics, with gains of 29\%, 55\%, and 39\% in precision, recall, and F1, respectively. Inter-annotator agreement shows a substantial Cohen's kappa score of 0.68 on impact-revealing intent classification.

\noindent
\textbf{Effect of different number of shots (K).} We find that performance improves significantly as the number of shots increases, plateauing around K = 50 with strong metrics (recall: 0.94, F1: 0.92); see Appendix~\ref{app:numberofshots} for details.

\noindent
\textbf{How does citation intent vary across fields?} Table~\ref{table:statsdiffsplits} presents an analysis of citation intents across 70k citation contexts from psychology, medicine, and computer science papers, with further breakdowns by recency and citation counts. Overall, psychology citations tend to lean towards impact-revealing, while computer science skews more toward non-impact-revealing (``other'') citations—except in more recent papers—and medicine shows a more balanced distribution. We observe that citations in psychology papers often exhibit a stronger subjective tone, especially in correction citations, such as: ``\textit{some of those assumptions have been \underline{controversial}}'' and ``\textit{\underline{researchers disagree} about whether the kinds of behaviors measured by particular implicit tests should be considered indicators of attitudes or something else}''. In contrast, computer science shows a notable shift in recent years toward more impact-revealing citations, likely driven by the novelty and immediate influence of AI research before transitioning into the \textit{legacy} phase, where citations become more background-oriented. This legacy effect is particularly evident in the highly cited subsets. 

\begin{table}
\centering
\begin{tabular}{|l|ccc|}
\hline
\rowcolor{gray!15}  \multirow{1}{*}{\textbf{Cited Papers}} &\multicolumn{3}{c|}{\textbf{Citation Intents (\%)}}\\
\rowcolor{gray!15} & PS & MD & CS\\
\hline
All & \multicolumn{1}{c|}{\colorbox{orange!40}{65}\colorbox{cyan!10}{35}} & \multicolumn{1}{c|}{\colorbox{orange!40}{53}\colorbox{cyan!10}{47}} & \multicolumn{1}{c|}{\colorbox{orange!40}{34}\colorbox{cyan!10}{66}}\\
\hline
Recent & \multicolumn{1}{c|}{\colorbox{orange!40}{65}\colorbox{cyan!10}{35}} & \multicolumn{1}{c|}{\colorbox{orange!40}{57}\colorbox{cyan!10}{43}} & \multicolumn{1}{c|}{\colorbox{orange!40}{58}\colorbox{cyan!10}{42}}\\
\cdashline{1-4}
Older & \multicolumn{1}{c|}{\colorbox{orange!40}{65}\colorbox{cyan!10}{35}} & \multicolumn{1}{c|}{\colorbox{orange!40}{52}\colorbox{cyan!10}{48}} & \multicolumn{1}{c|}{\colorbox{orange!40}{33}\colorbox{cyan!10}{67}}\\
\hline
Highly cited & \multicolumn{1}{c|}{\colorbox{orange!40}{66}\colorbox{cyan!10}{34}} &  \multicolumn{1}{c|}{\colorbox{orange!40}{36}\colorbox{cyan!10}{64}}  & \multicolumn{1}{c|}{\colorbox{orange!40}{38}\colorbox{cyan!10}{62}}\\
\cdashline{1-4}
Less cited &  \multicolumn{1}{c|}{\colorbox{orange!40}{64}\colorbox{cyan!10}{36}} & \multicolumn{1}{c|}{\colorbox{orange!40}{56}\colorbox{cyan!10}{44}} & \multicolumn{1}{c|}{\colorbox{orange!40}{32}\colorbox{cyan!10}{68}}\\
\hline
\end{tabular}
\caption{Recent = the last 5 years, Highly cited = top 20 \% by citation count in our dataset; orange for \colorbox{orange!40}{\texttt{impact-revealing}}, light blue for \colorbox{cyan!10}{\texttt{other}}; PS= psychology, MD = medicine, CS = computer science.}
\label{table:statsdiffsplits}
\end{table}

\begin{table*}
  \centering
  \resizebox{0.8\linewidth}{!}{
  \begin{tabular}{|l|c|c|c|c|}
    \hline
    \textbf{\cellcolor{gray!15}Intent classifier}            & \textbf{\cellcolor{gray!15}P}   & \textbf{\cellcolor{gray!15}R}   & \textbf{\cellcolor{gray!15}F1} & \textbf{\cellcolor{gray!15}Acc} \\
    \hline
    baseline=random  & 0.54 & 0.51 & 0.52 & 0.50\\
        baseline=always-impact-revealing & 0.53 & \textbf{1.0} & \textbf{0.69} & 0.53\\
                     \cdashline{1-5} 
            Structural Scaffolds~\cite{cohan-etal-2019-structural}          & 0.55 & 0.44 & 0.49 & 0.51   \\

                    Meaningful Citations~\cite{valenzuela2015identifying}            & \underline{0.72} & 0.46 & \underline{0.56} & \underline{0.62}   \\

            Multi-cite~\cite{lauscher2022multicite}            & 0.59 & 0.41 & 0.48 & 0.53  \\

                    \textbf{\cellcolor{gray!15}Ours}            & \textbf{\cellcolor{gray!15}0.74}  & \underline{\cellcolor{gray!15}0.65}  & \textbf{\cellcolor{gray!15}0.69}  & \textbf{\cellcolor{gray!15}0.69}    \\
         \hline 
 
  \end{tabular}}
  \caption{Results for intent classification (``impact-revealing'' or ``other''). Our classifier is the best at distinguishing between citations that help in understanding the impact of a paper, and those that serve a more general purpose, such as background references or standard acknowledgments.  \label{intent_classifier_external_methods}}
\end{table*}

\begin{table}[t]
  \centering
  \resizebox{\linewidth}{!}{
  \begin{tabular}{|l|c|c|}
    \hline
    \textbf{\cellcolor{gray!15}Intent classifier}            & \textbf{\cellcolor{gray!15}Confirmatory}   & \textbf{\cellcolor{gray!15}Correction} \\
    \hline
            Structural Scaffolds         & 0.83 & 0.50   \\

                    Meaningful Citations            & 0.41 & 0.65  \\

            Multi-cite            & 0.81 & 0.58 \\

                    \textbf{\cellcolor{gray!15}Ours}            & \textbf{\cellcolor{gray!15}0.88}  & \textbf{\cellcolor{gray!15}0.98} \\
         \hline 
 
  \end{tabular}}
  \caption{F1 scores for impact-revealing citations, broken down by confirmatory and correction categories.\label{tab:f1percitationtype}}
\end{table}

\begin{table*}
\centering
\resizebox{0.9\linewidth}{!}{
\begin{tabular}{|l c|c|c|c|c|c|c|c|c|}
\hline
\rowcolor{gray!15}  \multicolumn{2}{|c|}{\textbf{Prompt variant}} &\multicolumn{4}{c|}{\textbf{Trustworthiness}} &\multicolumn{3}{c|}{\textbf{Informativeness}} \\
\hline
\rowcolor{gray!15} \textbf{Citations} & \textbf{Intents} & \textbf{Faith.} & \textbf{Cov.} & \textbf{Cov.@3}  &\textbf{Cyc.}  & \textbf{Insi.} & \textbf{Trend.} & \textbf{Spec.}  \\ 
\hline
None & \xmark & 0.77 & 0.25 & 0.58 & n/a  &0.70& 0.94 & 0.75\\
All & \xmark & 0.83 & 0.32 & 0.74 & 0.55 &0.80&0.95&0.85\\
All & \checkmark & 0.84 & 0.32 & 0.73& 0.48 &0.80 &0.97&0.86\\
Impact-rev. & \xmark & 0.87  & 0.33& 0.73 & \textbf{0.59} & 0.80&0.96&0.87\\
\rowcolor{gray!15} Impact-rev. & \checkmark & \textbf{0.88} & \textbf{0.34} &\textbf{0.75} & 0.56& \textbf{0.83}&\textbf{0.98}&\textbf{0.88}\\
\hline
\end{tabular}}
\caption{Ablation results - Faithfulness: \textbf{Faith.}, Coverage: \textbf{Cov.}, Citation Year Compliance: \textbf{Cyc.}, Specificity: \textbf{Spec.}, Insightfulness: \textbf{Insi.}, Trend Awareness: \textbf{Trend.} These results show that adding impact-revealing contexts and their intents has an improvement on almost all metrics.}
\label{table:ablation_results}
\end{table*}

\subsection{Can existing intent classifiers detect impact-revealing citations?}
\label{sec:intentexternalmethods}

\noindent
\textbf{Comparison with existing methods. } We compare our classifier against the following popular intent classification methods: (i) Meaningful Citations~\cite{valenzuela2015identifying}: a supervised machine learning method to classify citation intents into meaningful or non-meaningful, by leveraging context and citation metadata;  (ii) Structural Scaffolds~\cite{cohan-etal-2019-structural}: a multi-task model which enhances the main task by integrating auxiliary tasks; (iii)  Multi-cite~\cite{lauscher2022multicite}: a multi-label classification method that predicts multiple intents using a fine-tuned   SciBERT~\cite{beltagy2019scibert}. We map their intent classes to either ``impact-revealing'' or ``other''. More details are in Appendix~\ref{externalsetup}. For our method, we use the ICL prompt, with K=50 and LLM=GPT4o-mini. For every test instance, we run our prompt 3 times (shuffling the order of the shots), and take a majority vote for the classification. Note that the results of the 3 runs have a full agreement of 72\%. By full agreement, we mean all three runs predicted the same class, indicating stability to in-context example order.

\setlength{\parskip}{0pt}

\noindent
\textbf{Results. } Numerical results are reported in Table~\ref{intent_classifier_external_methods}. Our method demonstrates the best performance, with the most significant improvement in recall, outperforming the second-best method by 19 percentage points. This is particularly crucial for our goal in generating impact summaries from impact-revealing citations because the highest recall in this context means that our method ensures that fewer meaningful citations are missed. The qualitative results are shown in the appendix~\ref{externalsetup}. For a deeper analysis on the performance of these classifiers on confirmatory vs. correction citations, we report the F1 scores on these splits in Table~\ref{tab:f1percitationtype}. The table shows that our strong performance is largely driven by the intent classifier’s effectiveness at identifying impact-revealing citations with correction-related intents (e.g., method refinement or highlighting research gaps). To ensure a fairer comparison with prior methods, we restrict the results in Table~\ref{tab:f1percitationtype} to citations from computer science papers, aligning with the domain used to train those models.

\subsection{Ablation study for impact summary generation}
\label{sec:ablation}

\noindent
\textbf{Data: papers, citations, intents.} We select 105 papers from the Semantic Scholar Academic Graph (S2AG)\footnote{\url{https://www.semanticscholar.org/product/api}}, from psychology, medicine and computer science (35 each), published between 1974 and 2022 at top venues in their respective fields. The citation count for each article ranges from $\sim$ 500 to $\sim$ 5000. We collect citations and their contexts using the citation lookup feature in the S2AG API. We first run our fine-grained intent generator (with K=50) and classifier for every citation context. This amounts to 70k citation contexts with their generated intents and their classes.

\noindent
\textbf{Prompt variants.}  We experiment with different inputs to assess their effect on the summary's quality, focusing on: (1) Citation contexts: including all, only impact-revealing, or none, to test what the LLM infers without citations and how varying context types affect the summary. Although full citation contexts can enrich summaries, they can also introduce noise from incidental references~\cite{Liu2023LostIT}. (2) Intents: whether to include the intents alongside citation contexts.  Table~\ref{table:ablation_results} lists all variants. In total, we generate 945 impact summaries\footnote{1 variant without citations + (4 citation-based variants * 2 orderings) * 105 papers}. Appendix~\ref{app:generateprompt} shows the prompt. For intent generation, we choose GPT-4o-mini due to its lower cost. For both impact generation and evaluation, the more advanced tasks, we use GPT-4o.

\noindent
\textbf{Results. } Table~\ref{table:ablation_results} shows the results for all metrics. Even though it shows a good understanding of how the paper was cited, likely due to the LLM's training data having access to a large corpus of scholarly articles, the baseline (no-knowledge) variant is the least faithful, and we observe that prompts with only impact-revealing citations generate more faithful summaries. This suggests that longer contexts might induce hallucinations in the generation process. Interestingly, providing the citation intents also gives a slight boost to the faithfulness of the  summaries, possibly because the intents encourage wording that aligns with the input paper's citations. 
 
Coverage is especially challenging, requiring a balance of completeness, relevance, and conciseness. Our best variant still outperforms the baseline by 9\%, showing that the LLM can capture key themes. Some inputs include over 100 distinct themes, explaining modest overall scores. Focusing only on the 3 most frequent themes (largest clusters), coverage rises significantly, as our best variant reaches 0.75 (+17\%), highlighting its strength in capturing core impact. To assess the coverage of \textit{meaningful low-frequency themes}, we analyze long-tail impact across 10 papers; Appendix~\ref{app:longtailcoverage} shows up to 50\% coverage when intents are included. Variants using only impact-revealing citations show higher citation year compliance; adding citation intents lowers it, likely due to greater focus on intent over timing. See Appendix~\ref{app:ablationexamples} for examples and Section~\ref{sec:limit} for discussion.
All variants show high trend awareness, with a slight edge for the impact-revealing + intents variant, showing LLMs can recognize distinct impact periods. This variant also improves insightfulness and specificity by 13\% over the no-knowledge baseline. Qualitative examples are in Appendix~\ref{app:ablationexamples}. G-Eval, via DeepEval\footnote{\url{https://github.com/confident-ai/deepeval}}, also offers reasons for each score (Appendix~\ref{app:gevalexamples}). We visualize sample summaries in Appendix~\ref{app:impactexamples}. To examine whether results change depending on the paper's field, we report results per field in Appendix~\ref{app:resultsperfield}. Finally, Appendix~\ref{app:erroranalysis} show insights into how intent misclassification can impact the quality of the generated summaries.

\noindent 
\textbf{Beyond GPTs.} We also generate impact summaries using Qwen and Gemini, to ensure model-agnosticism. Quantitative and qualitative results are provided in Appendix~\ref{app:otherllms}. Although GPT-4o served as our primary model, both Qwen and Gemini demonstrated strong performances, with Gemini in particular standing out in citation year compliance (0.93) and faithfulness (0.96). 

\noindent \textbf{LLM-Human correlation} We compute the human-LLM correlation, which shows moderate to strong relationships for the Spearman and Kendall-Tau metrics. Details are provided in Appendix~\ref{app:humancorrelation}.

\subsection{Practical applications}
\label{practical}
We showcase the usability of our work in two practical applications. First, by analyzing 10 diverse research topics, such as LLMs for code generation, open information extraction, commonsense knowledge mining, and comparing papers with similar citation counts, it shows that papers can have very different types of impact. For example, in open information extraction, some papers are cited to highlight limitations or motivate new work~\cite{cui2018neural}, while others are cited mainly for their methods~\cite{gashteovski2017minie,han2019opennre}. More details and examples are in Appendix~\ref{app:equalimpact}. Second, we explore generating author-level impact summaries by aggregating LLM-generated paper-level summaries for an author’s top-cited papers. Examples for two senior NLP researchers are shown in Appendix~\ref{app:authorlevel}. This method could be extended to institutions and venues, which we leave for future work.

\begin{figure}
  \includegraphics[width=0.9\columnwidth]{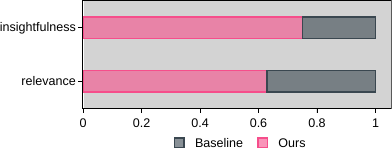}
  \caption{Pairwise comparison. \textit{Relevance}: Which summary better reflects the paper's actual impact? \textit{Insightfulness}: Which summary offers more valuable or novel information about the paper's usage? Professors consider our summaries more relevant and insightful.} 
  \label{fig:pairwise}
\end{figure}

Comprehensive details of the models and their configurations for each task/component are provided in Appendix~\ref{modelusage} to support reproducibility and attribution.

\section{Human Evaluation}
\label{expertreview}

\noindent
\textbf{Setup. } To gain deeper insights into the quality of our impact summaries and collect expert feedback for potential future improvements, we conduct a user study with 9 university professors from diverse backgrounds~\footnote{Nationalities: 2 German, 2 British, 2 Chinese, 1 American, 1 Czech, 1 Albanian; genders: 5 male, 4 female; research areas: AI, NLP, knowledge graphs, databases and information systems, computational social sciences, psychology and brain sciences.}. Each expert reviews impact summaries of \textit{their own papers} through a two-part evaluation: (1) a \textbf{pairwise comparison}, where they choose the better summary based on \textit{relevance} and \textit{insightfulness}, comparing the no-knowledge variant against our best variant. To prevent bias, we alternate the positions of the impact summaries. (2) a \textbf{perceived usefulness assessment}, where experts rate their agreement with given statements about an impact summary (generated by our best variant), using a 1-5 Likert scale. The professors evaluated the impact summaries for 82 papers and offered open-ended feedback.

\begin{figure*}
\centering
\includegraphics[width=\textwidth]{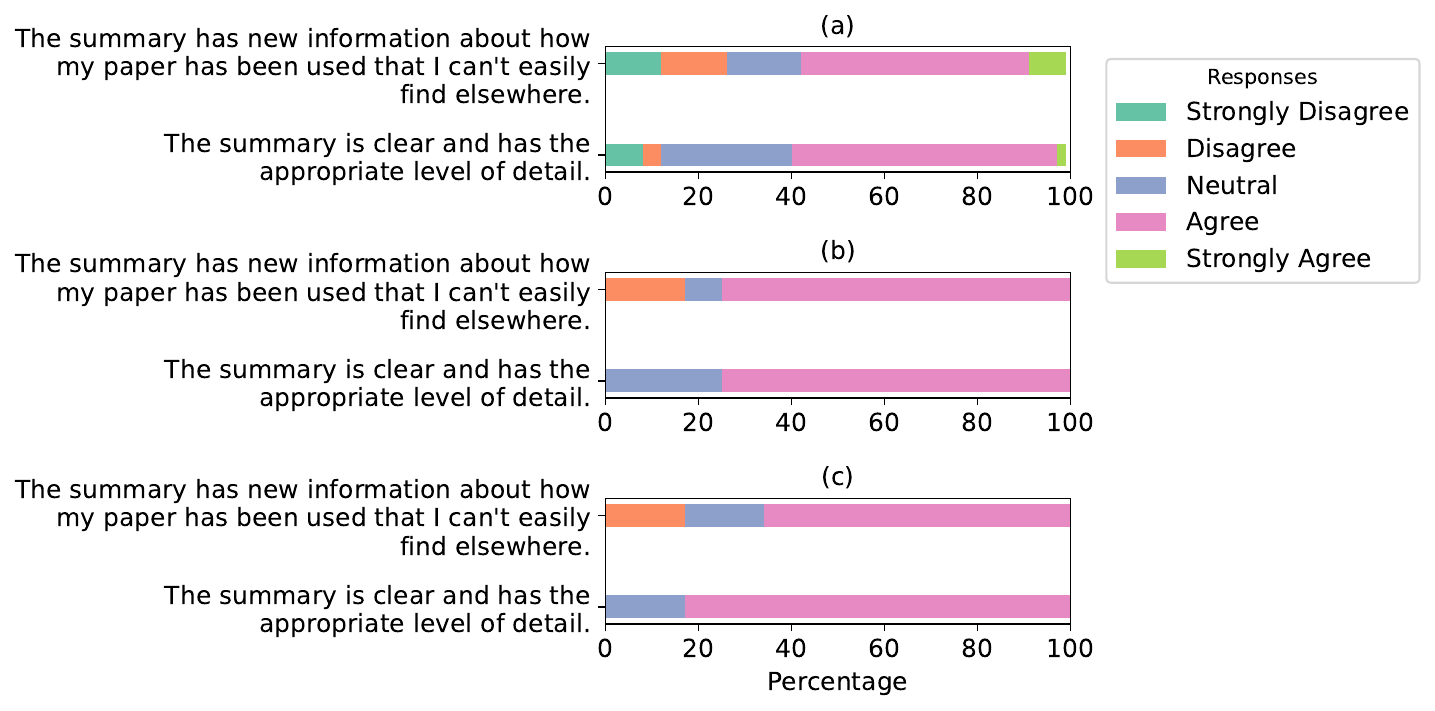}
  \caption{We show the overall results in (a), and results for two subsets of papers, namely papers with the most impact-revealing citations in (b), and papers with the highest number of citations in (c). Both subsets show impact summaries that are rated higher, likely because higher counts of impact-revealing citations and citations in general allow for richer contexts, which provide the model with clearer signals about the paper's impact, leading to more accurate and insightful summaries.}
  \label{fig:perceived}
\end{figure*}

\noindent
\textbf{Results. } Figure~\ref{fig:pairwise} shows results of the pairwise comparison. Impact summaries generated by our variant\footnote{Impact-revealing citations with intents.} demonstrate greater relevance (63\%) and insightfulness (75\%) than the baseline\footnote{no-knowledge variant}.  Results of the perceived usefulness, in Figure~\ref{fig:perceived}, show that approximately 60\% of professors agreed that the summaries provided an appropriate level of detail (clarity) and offered novel insights into how their papers were used, i.e., information not readily available elsewhere. Notably, for papers in the top 10\% based on impact-revealing citations, agreement on clarity and informativeness rose to 75\%. Finally, we ask the evaluators for open feedback. A couple professors found certain summaries too generic, failing to highlight specific strengths and limitations. They also pointed out that summaries based on citations from lesser-known conferences may not fully reflect a paper's true influence, and the quality of summaries often correlates with the size of impact-revealing citation context (a point further validated in Figure~\ref{fig:perceived} (b)). Additionally, one professor noted that for some papers, it was difficult to assess impact because the work was collaborative, and they were only familiar with certain aspects of the paper's contributions, while their co-authors might be more familiar with other aspects.  Other professors suggested improvements in both content and structure: they recommended incorporating citation counts within the text, which we plan to implement in the future, and enhancing the structure of our semi-structured summaries by adding elements like bullet points to improve readability.

\section{Related Work}
\label{sec:related}

\noindent
\textbf{Scientific impact analysis. } Studying scientific impact is vital for guiding research and recognizing contributions. While prior work relies on citation counts and related metrics \cite{hutchins2016relative,bos2019interdisciplinary,siudem2020three,min2021citation,wahle2023we,gu2024forecasting}, impact cannot be reduced to a single number~\cite{zhu2015measuring}. Few studies use citation context to assess scientific impact, and those that do focus on classifying citation intent individually rather than aggregating them in summaries~\cite{jurgens2018measuring,valenzuela2015identifying}. Our approach uses fine-grained citation context analysis to capture and aggregate specific citation reasons across impact-revealing citations. Other work links impact to novelty of the paper's content~\cite{rudiger2021explanatory,shi2023surprising,arts2021natural}, but overlooks how papers are used. Most also equate impact with praise~\cite{zhang2024pst,valenzuela2015identifying,zhu2015measuring}, whereas we distinguish confirmatory from correction citations, offering a fuller view. As detailed in Appendix~\ref{app:literaturetable}, we are the first to generate time-aware, multifaceted impact summaries from fine-grained citation analysis.

\noindent
\textbf{Citation intent prediction.} Prior work on citation intent classification uses pre-defined categories (e.g., method use, background) with early methods relying on supervised learning and linguistic features~\cite{teufel2006automatic,jurgens2018measuring,tuarob2019automatic}. Later approaches applied multi-task learning~\cite{cohan-etal-2019-structural}, while~\cite{lauscher2022multicite} introduced a multi-label method using fine-tuned SciBERT~\cite{beltagy2019scibert}. These rely on coarse categories; in contrast, we use in-context learning to generate fine-grained, free-form intent descriptions (Table~\ref{external_methods_qualitative}), offering more precise insights, an approach recently shown to aid related work generation~\cite{csahinucc2024systematic}.

\noindent
\textbf{Multi-document summarization in the era of LLMs.} Recent work in query-focused multi-document summarization uses LLMs to generate context-aware summaries~\cite{roy2023review}, with approaches like graph-based QA over private corpora~\cite{edge2024local} and retrieval-augmented summarization~\cite{zhang2025beyond}. In the scientific domain, prior efforts focus on summarizing papers for related work~\cite{csahinucc2024systematic,lu2020multi,chen2021capturing,wu2021towards} or reviews~\cite{kasanishi2023scireviewgen}. In contrast, we generate free-text summaries of scientific impact using a novel time-aware, multi-document approach.

\noindent
\textbf{LLM Evaluators. } Retrieval-Augmented Generation (RAG) systems condition generation on external information retrieved from user queries~\cite{Yu2024EvaluationOR, Gao2023RetrievalAugmentedGF}, but evaluating them is challenging due to multiple stages like chunking and retrieval. Tools like RagChecker~\cite{Ru2024RAGCheckerAF} use entailment-based metrics to assess faithfulness, while RAGAs~\cite{Shahul2023RAGAsAE} offers reference-free evaluation of relevance and faithfulness. Inspired by these, we assess our impact summaries’ faithfulness and coverage by validating them against the citation context corpus. While inspired by RAG methods, our approach introduces a distinct task focused on impact and temporal dynamics. By targeting impact-revealing citations and modeling shifts over time, we support more nuanced, citation-centric summarization. The LLM-as-a-judge paradigm, used for tasks like QA and writing evaluation~\cite{Zheng2023JudgingLW,Shao2024AssistingIW,Asai2024OpenScholarSS}, enables scalable assessment of complex qualities. Given the novelty of scientific impact summarization, we adopt G-Eval~\cite{Liu2023GEvalNE}, a CoT-based framework, to assess informativeness.

\section{Conclusion}
This work introduces a novel approach for generating time-aware impact summaries of research papers by analyzing evolving fine-grained confirmatory and correction citation intents. Our approach goes beyond citation counts, offering a more nuanced understanding of a paper’s impact over time. Expert evaluation highlights the value and potential for refinement of our summaries. We release our data and code to support research in this new task.

\section{Limitations}
\label{sec:limit}
\noindent
\textbf{Language. } Our work focuses on English papers, as it is the dominant language in most research fields. Extending this approach to a cross-lingual setting is a promising future direction.

\noindent
\textbf{Human evaluation. } Constructing a large-scale user study is challenging because assessors need deep knowledge of the papers' impact. We focused on experts evaluating their own work, but this limited our pool to 9 experts, as highly cited papers are typically authored by busy professors. We chose these experienced authors with multiple impactful papers to maximize the number of impact summaries evaluated per person. 

\noindent
\textbf{Trustworthiness evaluation. } Without restrictions, our approach shows unsatisfying coverage (0.34) and citation year compliance (0.59), compared to other metrics such as faithfulness. Regarding the year compliance, citation phrases often include multiple references accompanied by publication years, which can confuse the model and lead to incorrect citation year associations. In the future, we plan to remove or distinguish unrelated, potentially distracting numbers from the citation context (e.g., using well-crafted heuristics or prompt-based methods). For coverage, the LLM may focus on certain prominent themes while underrepresenting others, especially with complex topics. We briefly reported the coverage numbers for the most frequent themes, which showed a great improvement. In the future, we plan to explore this further, for instance, by ranking the importance of the themes and factor this into the coverage computation at different ranks. However, for this study, we decided to test full coverage to make sure the model captures a wide range of themes, without focusing on any specific ones, to check its overall understanding of the paper's impact. Finally, although checking potentially hundreds of citation contexts/intents at once raises concerns, the strong LLM-human agreement reported in Appendix~\ref{app:humancorrelation} suggests such issues are unlikely.

\noindent
\textbf{Experimenting with more LLMs.} In this paper, we prioritize introducing and exploring the new task in depth, which is why we only experimented with the long-context model GPT-4o. We leave testing a wider range of models for future work.  Although using the same LLM for both generation and evaluation can introduce bias, it will also promote consistency in task interpretation, and since our evaluation is not preference-based but scoring-based, bias is unlikely to affect results.

\noindent
\textbf{Broader spectrum of scientific impact.} While our study operationalizes scientific impact primarily through confirmation and correction, we acknowledge that real-world impact spans a broader spectrum (e.g., parallel development). Our framework is designed to be extensible, and future work can enrich the training examples with additional categories to capture a more nuanced landscape of scholarly influence, though we note that such categories may also introduce additional challenges and interpretive ambiguity.

\noindent
\textbf{Data reliability and quality controls. } We rely on the S2AG resource from Semantic Scholar, whose scale, coverage, active curation, and provision of citation contexts and structured metadata make it a uniquely reliable foundation for studying citations. While the data is generally high quality, our pipeline adds safeguards such as removing duplicate contexts, discarding citations lacking sufficient information, and enforcing temporal consistency to ensure robustness. We acknowledge that very large-scale datasets may have completeness limitations, but such issues are common across bibliographic resources, and further citation-quality filtering is a natural future enhancement rather than a prerequisite for the validity of our current findings.

\section*{Ethics Statement}

The data used in this study is publicly accessible and provided under open licenses, ensuring long-term reproducibility and building upon our work. For the expert study, we do not disclose the names of participating professors or publish individual votes and verbatim feedback. Instead, we report aggregated votes and rephrased feedback for suggestions of future improvements. While this study explores a promising new task,  we caution that inaccuracies in the generated summaries could mislead readers about a paper’s actual impact. Therefore, we do not recommend using our method directly and without human verification for critical decisions, such as academic hiring or grant allocations.

\section*{Acknowledgments}
This work has also been co-funded by the LOEWE Distinguished Chair “Ubiquitous Knowledge Processing”, LOEWE initiative, Hesse, Germany (Grant Number: LOEWE/4a//519/05/00.002(0002)/81 and by the European Union (ERC, InterText, 101054961). Views and opinions expressed are however those of the author(s) only and do not necessarily reflect those of the European Union or the European Research Council. Neither the European Union nor the granting authority can be held responsible for them.

\bibliography{acl_latex}

@inproceedings{cohan-etal-2019-structural,
  author       = {Arman Cohan and
                  Waleed Ammar and
                  Madeleine van Zuylen and
                  Field Cady},
  editor       = {Jill Burstein and
                  Christy Doran and
                  Thamar Solorio},
  title        = {Structural Scaffolds for Citation Intent Classification in Scientific
                  Publications},
  booktitle    = {Proceedings of the 2019 Conference of the North American Chapter of
                  the Association for Computational Linguistics: Human Language Technologies,
                  {NAACL-HLT} 2019, Minneapolis, MN, USA, June 2-7, 2019, Volume 1 (Long
                  and Short Papers)},
  pages        = {3586--3596},
  publisher    = {Association for Computational Linguistics},
  year         = {2019},
  url          = {https://doi.org/10.18653/v1/n19-1361},
  doi          = {10.18653/V1/N19-1361},
  bibsource    = {dblp computer science bibliography, https://dblp.org}
}

@inproceedings{liu2023g,
  author       = {Yang Liu and
                  Dan Iter and
                  Yichong Xu and
                  Shuohang Wang and
                  Ruochen Xu and
                  Chenguang Zhu},
  editor       = {Houda Bouamor and
                  Juan Pino and
                  Kalika Bali},
  title        = {G-Eval: {NLG} Evaluation using Gpt-4 with Better Human Alignment},
  booktitle    = {Proceedings of the 2023 Conference on Empirical Methods in Natural
                  Language Processing, {EMNLP} 2023, Singapore, December 6-10, 2023},
  pages        = {2511--2522},
  publisher    = {Association for Computational Linguistics},
  year         = {2023},
  url          = {https://doi.org/10.18653/v1/2023.emnlp-main.153},
  doi          = {10.18653/V1/2023.EMNLP-MAIN.153},
  bibsource    = {dblp computer science bibliography, https://dblp.org}
}

@article{comanici2025gemini,
  title={Gemini 2.5: Pushing the frontier with advanced reasoning, multimodality, long context, and next generation agentic capabilities},
  author={Comanici, Gheorghe and Bieber, Eric and Schaekermann, Mike and Pasupat, Ice and Sachdeva, Noveen and Dhillon, Inderjit and Blistein, Marcel and Ram, Ori and Zhang, Dan and Rosen, Evan and others},
  journal={arXiv preprint arXiv:2507.06261},
  year={2025}
}

@article{yang2025qwen3,
  author       = {An Yang and
                  Anfeng Li and
                  Baosong Yang and
                  Beichen Zhang and
                  Binyuan Hui and
                  Bo Zheng and
                  Bowen Yu and
                  Chang Gao and
                  Chengen Huang and
                  Chenxu Lv and
                  Chujie Zheng and
                  Dayiheng Liu and
                  Fan Zhou and
                  Fei Huang and
                  Feng Hu and
                  Hao Ge and
                  Haoran Wei and
                  Huan Lin and
                  Jialong Tang and
                  Jian Yang and
                  Jianhong Tu and
                  Jianwei Zhang and
                  Jian Yang and
                  Jiaxi Yang and
                  Jingren Zhou and
                  Jingren Zhou and
                  Junyang Lin and
                  Kai Dang and
                  Keqin Bao and
                  Kexin Yang and
                  Le Yu and
                  Lianghao Deng and
                  Mei Li and
                  Mingfeng Xue and
                  Mingze Li and
                  Pei Zhang and
                  Peng Wang and
                  Qin Zhu and
                  Rui Men and
                  Ruize Gao and
                  Shixuan Liu and
                  Shuang Luo and
                  Tianhao Li and
                  Tianyi Tang and
                  Wenbiao Yin and
                  Xingzhang Ren and
                  Xinyu Wang and
                  Xinyu Zhang and
                  Xuancheng Ren and
                  Yang Fan and
                  Yang Su and
                  Yichang Zhang and
                  Yinger Zhang and
                  Yu Wan and
                  Yuqiong Liu and
                  Zekun Wang and
                  Zeyu Cui and
                  Zhenru Zhang and
                  Zhipeng Zhou and
                  Zihan Qiu},
  title        = {Qwen3 Technical Report},
  journal      = {CoRR},
  volume       = {abs/2505.09388},
  year         = {2025},
  url          = {https://doi.org/10.48550/arXiv.2505.09388},
  doi          = {10.48550/ARXIV.2505.09388},
  eprinttype    = {arXiv},
  eprint       = {2505.09388},
  bibsource    = {dblp computer science bibliography, https://dblp.org}
}

@article{jurgens2018measuring,
  author       = {David Jurgens and
                  Srijan Kumar and
                  Raine Hoover and
                  Daniel A. McFarland and
                  Dan Jurafsky},
  title        = {Measuring the Evolution of a Scientific Field through Citation Frames},
  journal      = {Trans. Assoc. Comput. Linguistics},
  volume       = {6},
  pages        = {391--406},
  year         = {2018},
  url          = {https://doi.org/10.1162/tacl\_a\_00028},
  doi          = {10.1162/TACL\_A\_00028},
  bibsource    = {dblp computer science bibliography, https://dblp.org}
}

@inproceedings{valenzuela2015identifying,
  author       = {Marco Valenzuela and
                  Vu Ha and
                  Oren Etzioni},
  editor       = {Cornelia Caragea and
                  C. Lee Giles and
                  Narayan L. Bhamidipati and
                  Doina Caragea and
                  Sujatha Das Gollapalli and
                  Saurabh Kataria and
                  Huan Liu and
                  Feng Xia},
  title        = {Identifying Meaningful Citations},
  booktitle    = {Scholarly Big Data: {AI} Perspectives, Challenges, and Ideas, Papers
                  from the 2015 {AAAI} Workshop, Austin, Texas, USA, January, 2015},
  series       = {{AAAI} Technical Report},
  volume       = {{WS-15-13}},
  publisher    = {{AAAI} Press},
  year         = {2015},
  url          = {http://aaai.org/ocs/index.php/WS/AAAIW15/paper/view/10185},
  bibsource    = {dblp computer science bibliography, https://dblp.org}
}

@inproceedings{koller1997hierarchically,
  author       = {Daphne Koller and
                  Mehran Sahami},
  editor       = {Douglas H. Fisher},
  title        = {Hierarchically Classifying Documents Using Very Few Words},
  booktitle    = {Proceedings of the Fourteenth International Conference on Machine
                  Learning {(ICML} 1997), Nashville, Tennessee, USA, July 8-12, 1997},
  pages        = {170--178},
  publisher    = {Morgan Kaufmann},
  year         = {1997},
  bibsource    = {dblp computer science bibliography, https://dblp.org}
}

@article{he2018temporal,
  author       = {Jiangen He and
                  Chaomei Chen},
  title        = {Temporal Representations of Citations for Understanding the Changing
                  Roles of Scientific Publications},
  journal      = {Frontiers Res. Metrics Anal.},
  volume       = {3},
  pages        = {27},
  year         = {2018},
  url          = {https://doi.org/10.3389/frma.2018.00027},
  doi          = {10.3389/FRMA.2018.00027},
  bibsource    = {dblp computer science bibliography, https://dblp.org}
}

@inproceedings{zhang2025beyond,
  author       = {Weijia Zhang and
                  Jia{-}Hong Huang and
                  Svitlana Vakulenko and
                  Yumo Xu and
                  Thilina Rajapakse and
                  Evangelos Kanoulas},
  editor       = {Apostolos Antonacopoulos and
                  Subhasis Chaudhuri and
                  Rama Chellappa and
                  Cheng{-}Lin Liu and
                  Saumik Bhattacharya and
                  Umapada Pal},
  title        = {Beyond Relevant Documents: {A} Knowledge-Intensive Approach for Query-Focused
                  Summarization Using Large Language Models},
  booktitle    = {Pattern Recognition - 27th International Conference, {ICPR} 2024,
                  Kolkata, India, December 1-5, 2024, Proceedings, Part {XIX}},
  series       = {Lecture Notes in Computer Science},
  volume       = {15319},
  pages        = {89--104},
  publisher    = {Springer},
  year         = {2024},
  url          = {https://doi.org/10.1007/978-3-031-78495-8\_6},
  doi          = {10.1007/978-3-031-78495-8\_6},
  bibsource    = {dblp computer science bibliography, https://dblp.org}
}

@article{roy2023review,
  author       = {Prasenjeet Roy and
                  Suman Kundu},
  title        = {Review on Query-focused Multi-document Summarization {(QMDS)} with
                  Comparative Analysis},
  journal      = {{ACM} Comput. Surv.},
  volume       = {56},
  number       = {1},
  pages        = {5:1--5:38},
  year         = {2024},
  url          = {https://doi.org/10.1145/3597299},
  doi          = {10.1145/3597299},
  bibsource    = {dblp computer science bibliography, https://dblp.org}
}

@inproceedings{beltagy2019scibert,
  author       = {Iz Beltagy and
                  Kyle Lo and
                  Arman Cohan},
  editor       = {Kentaro Inui and
                  Jing Jiang and
                  Vincent Ng and
                  Xiaojun Wan},
  title        = {SciBERT: {A} Pretrained Language Model for Scientific Text},
  booktitle    = {Proceedings of the 2019 Conference on Empirical Methods in Natural
                  Language Processing and the 9th International Joint Conference on
                  Natural Language Processing, {EMNLP-IJCNLP} 2019, Hong Kong, China,
                  November 3-7, 2019},
  pages        = {3613--3618},
  publisher    = {Association for Computational Linguistics},
  year         = {2019},
  url          = {https://doi.org/10.18653/v1/D19-1371},
  doi          = {10.18653/V1/D19-1371},
  bibsource    = {dblp computer science bibliography, https://dblp.org}
}

@article{zhang2024pst,
  author       = {Fanjin Zhang and
                  Kun Cao and
                  Yukuo Cen and
                  Jifan Yu and
                  Da Yin and
                  Jie Tang},
  title        = {PST-Bench: Tracing and Benchmarking the Source of Publications},
  journal      = {CoRR},
  volume       = {abs/2402.16009},
  year         = {2024},
  url          = {https://doi.org/10.48550/arXiv.2402.16009},
  doi          = {10.48550/ARXIV.2402.16009},
  eprinttype    = {arXiv},
  eprint       = {2402.16009},
  bibsource    = {dblp computer science bibliography, https://dblp.org}
}

@article{gu2024forecasting,
  author       = {Xuemei Gu and
                  Mario Krenn},
  title        = {Forecasting high-impact research topics via machine learning on evolving
                  knowledge graphs},
  journal      = {CoRR},
  volume       = {abs/2402.08640},
  year         = {2024},
  url          = {https://doi.org/10.48550/arXiv.2402.08640},
  doi          = {10.48550/ARXIV.2402.08640},
  eprinttype    = {arXiv},
  eprint       = {2402.08640},
  bibsource    = {dblp computer science bibliography, https://dblp.org}
}

@article{bos2019interdisciplinary,
  author       = {Arthur R. Bos and
                  Sandrine Nitza},
  title        = {Interdisciplinary Comparison of Scientific Impact of Publications
                  Using the Citation-Ratio},
  journal      = {Data Sci. J.},
  volume       = {18},
  pages        = {19},
  year         = {2019},
  url          = {https://doi.org/10.5334/dsj-2019-019},
  doi          = {10.5334/DSJ-2019-019},
  bibsource    = {dblp computer science bibliography, https://dblp.org}
}

@article{zhu2015measuring,
  author       = {Xiaodan Zhu and
                  Peter D. Turney and
                  Daniel Lemire and
                  Andr{\'{e}} Vellino},
  title        = {Measuring academic influence: Not all citations are equal},
  journal      = {J. Assoc. Inf. Sci. Technol.},
  volume       = {66},
  number       = {2},
  pages        = {408--427},
  year         = {2015},
  url          = {https://doi.org/10.1002/asi.23179},
  doi          = {10.1002/ASI.23179},
  bibsource    = {dblp computer science bibliography, https://dblp.org}
}

@inproceedings{wahle2023we,
  author       = {Jan Philip Wahle and
                  Terry Ruas and
                  Mohamed Abdalla and
                  Bela Gipp and
                  Saif M. Mohammad},
  editor       = {Houda Bouamor and
                  Juan Pino and
                  Kalika Bali},
  title        = {We are Who We Cite: Bridges of Influence Between Natural Language
                  Processing and Other Academic Fields},
  booktitle    = {Proceedings of the 2023 Conference on Empirical Methods in Natural
                  Language Processing, {EMNLP} 2023, Singapore, December 6-10, 2023},
  pages        = {12896--12913},
  publisher    = {Association for Computational Linguistics},
  year         = {2023},
  url          = {https://doi.org/10.18653/v1/2023.emnlp-main.797},
  doi          = {10.18653/V1/2023.EMNLP-MAIN.797},
  bibsource    = {dblp computer science bibliography, https://dblp.org}
}

@article{rudiger2021explanatory,
  author       = {Matthias Sebastian R{\"{u}}diger and
                  David Antons and
                  Torsten{-}Oliver Salge},
  title        = {The explanatory power of citations: a new approach to unpacking impact
                  in science},
  journal      = {Scientometrics},
  volume       = {126},
  number       = {12},
  pages        = {9779--9809},
  year         = {2021},
  url          = {https://doi.org/10.1007/s11192-021-04103-w},
  doi          = {10.1007/S11192-021-04103-W},
  bibsource    = {dblp computer science bibliography, https://dblp.org}
}

@article{arts2025beyond,
  title={Beyond citations: Measuring novel scientific ideas and their impact in publication text},
  author={Arts, Sam and Melluso, Nicola and Veugelers, Reinhilde},
  journal={Review of Economics and Statistics},
  pages={1--33},
  year={2025},
  publisher={MIT Press 255 Main Street, 9th Floor, Cambridge, Massachusetts 02142, USA~…}
}

@inproceedings{lauscher2022multicite,
  author       = {Anne Lauscher and
                  Brandon Ko and
                  Bailey Kuehl and
                  Sophie Johnson and
                  Arman Cohan and
                  David Jurgens and
                  Kyle Lo},
  editor       = {Marine Carpuat and
                  Marie{-}Catherine de Marneffe and
                  Iv{\'{a}}n Vladimir Meza Ru{\'{\i}}z},
  title        = {MultiCite: Modeling realistic citations requires moving beyond the
                  single-sentence single-label setting},
  booktitle    = {Proceedings of the 2022 Conference of the North American Chapter of
                  the Association for Computational Linguistics: Human Language Technologies,
                  {NAACL} 2022, Seattle, WA, United States, July 10-15, 2022},
  pages        = {1875--1889},
  publisher    = {Association for Computational Linguistics},
  year         = {2022},
  url          = {https://doi.org/10.18653/v1/2022.naacl-main.137},
  doi          = {10.18653/V1/2022.NAACL-MAIN.137},
  bibsource    = {dblp computer science bibliography, https://dblp.org}
}

@article{Liu2023LostIT,
  author       = {Nelson F. Liu and
                  Kevin Lin and
                  John Hewitt and
                  Ashwin Paranjape and
                  Michele Bevilacqua and
                  Fabio Petroni and
                  Percy Liang},
  title        = {Lost in the Middle: How Language Models Use Long Contexts},
  journal      = {Trans. Assoc. Comput. Linguistics},
  volume       = {12},
  pages        = {157--173},
  year         = {2024},
  url          = {https://doi.org/10.1162/tacl\_a\_00638},
  doi          = {10.1162/TACL\_A\_00638},
  bibsource    = {dblp computer science bibliography, https://dblp.org}
}

@inproceedings{lu2020multi,
  author       = {Yao Lu and
                  Yue Dong and
                  Laurent Charlin},
  editor       = {Bonnie Webber and
                  Trevor Cohn and
                  Yulan He and
                  Yang Liu},
  title        = {Multi-XScience: {A} Large-scale Dataset for Extreme Multi-document
                  Summarization of Scientific Articles},
  booktitle    = {Proceedings of the 2020 Conference on Empirical Methods in Natural
                  Language Processing, {EMNLP} 2020, Online, November 16-20, 2020},
  pages        = {8068--8074},
  publisher    = {Association for Computational Linguistics},
  year         = {2020},
  url          = {https://doi.org/10.18653/v1/2020.emnlp-main.648},
  doi          = {10.18653/V1/2020.EMNLP-MAIN.648},
  bibsource    = {dblp computer science bibliography, https://dblp.org}
}

@inproceedings{chen2021capturing,
  author       = {Xiuying Chen and
                  Hind Alamro and
                  Mingzhe Li and
                  Shen Gao and
                  Xiangliang Zhang and
                  Dongyan Zhao and
                  Rui Yan},
  editor       = {Chengqing Zong and
                  Fei Xia and
                  Wenjie Li and
                  Roberto Navigli},
  title        = {Capturing Relations between Scientific Papers: An Abstractive Model
                  for Related Work Section Generation},
  booktitle    = {Proceedings of the 59th Annual Meeting of the Association for Computational
                  Linguistics and the 11th International Joint Conference on Natural
                  Language Processing, {ACL/IJCNLP} 2021, (Volume 1: Long Papers), Virtual
                  Event, August 1-6, 2021},
  pages        = {6068--6077},
  publisher    = {Association for Computational Linguistics},
  year         = {2021},
  url          = {https://doi.org/10.18653/v1/2021.acl-long.473},
  doi          = {10.18653/V1/2021.ACL-LONG.473},
  bibsource    = {dblp computer science bibliography, https://dblp.org}
}

@inproceedings{csahinucc2024systematic,
  author       = {Furkan Sahinu{\c{c}} and
                  Ilia Kuznetsov and
                  Yufang Hou and
                  Iryna Gurevych},
  editor       = {Lun{-}Wei Ku and
                  Andre Martins and
                  Vivek Srikumar},
  title        = {Systematic Task Exploration with LLMs: {A} Study in Citation Text
                  Generation},
  booktitle    = {Proceedings of the 62nd Annual Meeting of the Association for Computational
                  Linguistics (Volume 1: Long Papers), {ACL} 2024, Bangkok, Thailand,
                  August 11-16, 2024},
  pages        = {4832--4855},
  publisher    = {Association for Computational Linguistics},
  year         = {2024},
  url          = {https://doi.org/10.18653/v1/2024.acl-long.265},
  doi          = {10.18653/V1/2024.ACL-LONG.265},
  bibsource    = {dblp computer science bibliography, https://dblp.org}
}

@article{tuarob2019automatic,
  author       = {Suppawong Tuarob and
                  Sung Woo Kang and
                  Poom Wettayakorn and
                  Chanatip Pornprasit and
                  Tanakitti Sachati and
                  Saeed{-}Ul Hassan and
                  Peter Haddawy},
  title        = {Automatic Classification of Algorithm Citation Functions in Scientific
                  Literature},
  journal      = {{IEEE} Trans. Knowl. Data Eng.},
  volume       = {32},
  number       = {10},
  pages        = {1881--1896},
  year         = {2020},
  url          = {https://doi.org/10.1109/TKDE.2019.2913376},
  doi          = {10.1109/TKDE.2019.2913376},
  bibsource    = {dblp computer science bibliography, https://dblp.org}
}

@inproceedings{teufel2006automatic,
  author       = {Simone Teufel and
                  Advaith Siddharthan and
                  Dan Tidhar},
  editor       = {Dan Jurafsky and
                  {\'{E}}ric Gaussier},
  title        = {Automatic classification of citation function},
  booktitle    = {{EMNLP} 2006, Proceedings of the 2006 Conference on Empirical Methods
                  in Natural Language Processing, 22-23 July 2006, Sydney, Australia},
  pages        = {103--110},
  publisher    = {{ACL}},
  year         = {2006},
  url          = {https://aclanthology.org/W06-1613/},
  bibsource    = {dblp computer science bibliography, https://dblp.org}
}

@article{catalini2015incidence,
  title={The incidence and role of negative citations in science},
  author={Catalini, Christian and Lacetera, Nicola and Oettl, Alexander},
  journal={Proceedings of the National Academy of Sciences},
  volume={112},
  number={45},
  pages={13823--13826},
  year={2015},
  publisher={National Academy of Sciences}
}

@article{hutchins2016relative,
  title={Relative citation ratio (RCR): a new metric that uses citation rates to measure influence at the article level},
  author={Hutchins, B Ian and Yuan, Xin and Anderson, James M and Santangelo, George M},
  journal={PLoS biology},
  volume={14},
  number={9},
  pages={e1002541},
  year={2016},
  publisher={Public Library of Science San Francisco, CA USA}
}

@article{arts2021natural,
  title={Natural language processing to identify the creation and impact of new technologies in patent text: Code, data, and new measures},
  author={Arts, Sam and Hou, Jianan and Gomez, Juan Carlos},
  journal={Research policy},
  volume={50},
  number={2},
  pages={104144},
  year={2021},
  publisher={Elsevier}
}

@article{shi2023surprising,
  title={Surprising combinations of research contents and contexts are related to impact and emerge with scientific outsiders from distant disciplines},
  author={Shi, Feng and Evans, James},
  journal={Nature Communications},
  volume={14},
  number={1},
  pages={1641},
  year={2023},
  publisher={Nature Publishing Group UK London}
}

@article{edge2024local,
  author       = {Darren Edge and
                  Ha Trinh and
                  Newman Cheng and
                  Joshua Bradley and
                  Alex Chao and
                  Apurva Mody and
                  Steven Truitt and
                  Jonathan Larson},
  title        = {From Local to Global: {A} Graph {RAG} Approach to Query-Focused Summarization},
  journal      = {CoRR},
  volume       = {abs/2404.16130},
  year         = {2024},
  url          = {https://doi.org/10.48550/arXiv.2404.16130},
  doi          = {10.48550/ARXIV.2404.16130},
  eprinttype    = {arXiv},
  eprint       = {2404.16130},
  bibsource    = {dblp computer science bibliography, https://dblp.org}
}

@article{wu2021towards,
  author       = {Jia{-}Yan Wu and
                  Alexander Te{-}Wei Shieh and
                  Shih{-}Ju Hsu and
                  Yun{-}Nung Chen},
  title        = {Towards Generating Citation Sentences for Multiple References with
                  Intent Control},
  journal      = {CoRR},
  volume       = {abs/2112.01332},
  year         = {2021},
  url          = {https://arxiv.org/abs/2112.01332},
  eprinttype    = {arXiv},
  eprint       = {2112.01332},
  bibsource    = {dblp computer science bibliography, https://dblp.org}
}

@article{siudem2020three,
  title={Three dimensions of scientific impact},
  author={Siudem, Grzegorz and {\.Z}oga{\l}a-Siudem, Barbara and Cena, Anna and Gagolewski, Marek},
  journal={Proceedings of the National Academy of Sciences},
  volume={117},
  number={25},
  pages={13896--13900},
  year={2020},
  publisher={National Academy of Sciences}
}

@article{min2021citation,
  author       = {Chao Min and
                  Qingyu Chen and
                  Erjia Yan and
                  Yi Bu and
                  Jianjun Sun},
  title        = {Citation cascade and the evolution of topic relevance},
  journal      = {J. Assoc. Inf. Sci. Technol.},
  volume       = {72},
  number       = {1},
  pages        = {110--127},
  year         = {2021},
  url          = {https://doi.org/10.1002/asi.24370},
  doi          = {10.1002/ASI.24370},
  bibsource    = {dblp computer science bibliography, https://dblp.org}
}

@inproceedings{kasanishi2023scireviewgen,
  author       = {Tetsu Kasanishi and
                  Masaru Isonuma and
                  Junichiro Mori and
                  Ichiro Sakata},
  editor       = {Anna Rogers and
                  Jordan L. Boyd{-}Graber and
                  Naoaki Okazaki},
  title        = {SciReviewGen: {A} Large-scale Dataset for Automatic Literature Review
                  Generation},
  booktitle    = {Findings of the Association for Computational Linguistics: {ACL} 2023,
                  Toronto, Canada, July 9-14, 2023},
  pages        = {6695--6715},
  publisher    = {Association for Computational Linguistics},
  year         = {2023},
  url          = {https://doi.org/10.18653/v1/2023.findings-acl.418},
  doi          = {10.18653/V1/2023.FINDINGS-ACL.418},
  bibsource    = {dblp computer science bibliography, https://dblp.org}
}

@inproceedings{Zheng2023JudgingLW,
  author       = {Lianmin Zheng and
                  Wei{-}Lin Chiang and
                  Ying Sheng and
                  Siyuan Zhuang and
                  Zhanghao Wu and
                  Yonghao Zhuang and
                  Zi Lin and
                  Zhuohan Li and
                  Dacheng Li and
                  Eric P. Xing and
                  Hao Zhang and
                  Joseph E. Gonzalez and
                  Ion Stoica},
  editor       = {Alice Oh and
                  Tristan Naumann and
                  Amir Globerson and
                  Kate Saenko and
                  Moritz Hardt and
                  Sergey Levine},
  title        = {Judging LLM-as-a-Judge with MT-Bench and Chatbot Arena},
  booktitle    = {Advances in Neural Information Processing Systems 36: Annual Conference
                  on Neural Information Processing Systems 2023, NeurIPS 2023, New Orleans,
                  LA, USA, December 10 - 16, 2023},
  year         = {2023},
  url          = {http://papers.nips.cc/paper\_files/paper/2023/hash/91f18a1287b398d378ef22505bf41832-Abstract-Datasets\_and\_Benchmarks.html},
  bibsource    = {dblp computer science bibliography, https://dblp.org}
}

@article{Asai2024OpenScholarSS,
  author       = {Akari Asai and
                  Jacqueline He and
                  Rulin Shao and
                  Weijia Shi and
                  Amanpreet Singh and
                  Joseph Chee Chang and
                  Kyle Lo and
                  Luca Soldaini and
                  Sergey Feldman and
                  Mike D'Arcy and
                  David Wadden and
                  Matt Latzke and
                  Minyang Tian and
                  Pan Ji and
                  Shengyan Liu and
                  Hao Tong and
                  Bohao Wu and
                  Yanyu Xiong and
                  Luke Zettlemoyer and
                  Graham Neubig and
                  Daniel S. Weld and
                  Doug Downey and
                  Wen{-}tau Yih and
                  Pang Wei Koh and
                  Hannaneh Hajishirzi},
  title        = {OpenScholar: Synthesizing Scientific Literature with Retrieval-augmented
                  LMs},
  journal      = {CoRR},
  volume       = {abs/2411.14199},
  year         = {2024},
  url          = {https://doi.org/10.48550/arXiv.2411.14199},
  doi          = {10.48550/ARXIV.2411.14199},
  eprinttype    = {arXiv},
  eprint       = {2411.14199},
  bibsource    = {dblp computer science bibliography, https://dblp.org}
}

@inproceedings{Shao2024AssistingIW,
  author       = {Yijia Shao and
                  Yucheng Jiang and
                  Theodore A. Kanell and
                  Peter Xu and
                  Omar Khattab and
                  Monica S. Lam},
  editor       = {Kevin Duh and
                  Helena G{\'{o}}mez{-}Adorno and
                  Steven Bethard},
  title        = {Assisting in Writing Wikipedia-like Articles From Scratch with Large
                  Language Models},
  booktitle    = {Proceedings of the 2024 Conference of the North American Chapter of
                  the Association for Computational Linguistics: Human Language Technologies
                  (Volume 1: Long Papers), {NAACL} 2024, Mexico City, Mexico, June 16-21,
                  2024},
  pages        = {6252--6278},
  publisher    = {Association for Computational Linguistics},
  year         = {2024},
  url          = {https://doi.org/10.18653/v1/2024.naacl-long.347},
  doi          = {10.18653/V1/2024.NAACL-LONG.347},
  bibsource    = {dblp computer science bibliography, https://dblp.org}
}

@inproceedings{Liu2023GEvalNE,
  author       = {Yang Liu and
                  Dan Iter and
                  Yichong Xu and
                  Shuohang Wang and
                  Ruochen Xu and
                  Chenguang Zhu},
  editor       = {Houda Bouamor and
                  Juan Pino and
                  Kalika Bali},
  title        = {G-Eval: {NLG} Evaluation using Gpt-4 with Better Human Alignment},
  booktitle    = {Proceedings of the 2023 Conference on Empirical Methods in Natural
                  Language Processing, {EMNLP} 2023, Singapore, December 6-10, 2023},
  pages        = {2511--2522},
  publisher    = {Association for Computational Linguistics},
  year         = {2023},
  url          = {https://doi.org/10.18653/v1/2023.emnlp-main.153},
  doi          = {10.18653/V1/2023.EMNLP-MAIN.153},
  bibsource    = {dblp computer science bibliography, https://dblp.org}
}

@article{Yu2024EvaluationOR,
  author       = {Hao Yu and
                  Aoran Gan and
                  Kai Zhang and
                  Shiwei Tong and
                  Qi Liu and
                  Zhaofeng Liu},
  title        = {Evaluation of Retrieval-Augmented Generation: {A} Survey},
  journal      = {CoRR},
  volume       = {abs/2405.07437},
  year         = {2024},
  url          = {https://doi.org/10.48550/arXiv.2405.07437},
  doi          = {10.48550/ARXIV.2405.07437},
  eprinttype    = {arXiv},
  eprint       = {2405.07437},
  bibsource    = {dblp computer science bibliography, https://dblp.org}
}

@article{Gao2023RetrievalAugmentedGF,
  author       = {Yunfan Gao and
                  Yun Xiong and
                  Xinyu Gao and
                  Kangxiang Jia and
                  Jinliu Pan and
                  Yuxi Bi and
                  Yi Dai and
                  Jiawei Sun and
                  Qianyu Guo and
                  Meng Wang and
                  Haofen Wang},
  title        = {Retrieval-Augmented Generation for Large Language Models: {A} Survey},
  journal      = {CoRR},
  volume       = {abs/2312.10997},
  year         = {2023},
  url          = {https://doi.org/10.48550/arXiv.2312.10997},
  doi          = {10.48550/ARXIV.2312.10997},
  eprinttype    = {arXiv},
  eprint       = {2312.10997},
  bibsource    = {dblp computer science bibliography, https://dblp.org}
}

@inproceedings{Ru2024RAGCheckerAF,
  author       = {Dongyu Ru and
                  Lin Qiu and
                  Xiangkun Hu and
                  Tianhang Zhang and
                  Peng Shi and
                  Shuaichen Chang and
                  Cheng Jiayang and
                  Cunxiang Wang and
                  Shichao Sun and
                  Huanyu Li and
                  Zizhao Zhang and
                  Binjie Wang and
                  Jiarong Jiang and
                  Tong He and
                  Zhiguo Wang and
                  Pengfei Liu and
                  Yue Zhang and
                  Zheng Zhang},
  editor       = {Amir Globersons and
                  Lester Mackey and
                  Danielle Belgrave and
                  Angela Fan and
                  Ulrich Paquet and
                  Jakub M. Tomczak and
                  Cheng Zhang},
  title        = {RAGChecker: {A} Fine-grained Framework for Diagnosing Retrieval-Augmented
                  Generation},
  booktitle    = {Advances in Neural Information Processing Systems 38: Annual Conference
                  on Neural Information Processing Systems 2024, NeurIPS 2024, Vancouver,
                  BC, Canada, December 10 - 15, 2024},
  year         = {2024},
  url          = {http://papers.nips.cc/paper\_files/paper/2024/hash/27245589131d17368cccdfa990cbf16e-Abstract-Datasets\_and\_Benchmarks\_Track.html},
  bibsource    = {dblp computer science bibliography, https://dblp.org}
}

@inproceedings{Shahul2023RAGAsAE,
  author       = {Shahul ES and
                  Jithin James and
                  Luis Espinosa Anke and
                  Steven Schockaert},
  editor       = {Nikolaos Aletras and
                  Orph{\'{e}}e De Clercq},
  title        = {RAGAs: Automated Evaluation of Retrieval Augmented Generation},
  booktitle    = {Proceedings of the 18th Conference of the European Chapter of the
                  Association for Computational Linguistics, {EACL} 2024 - System Demonstrations,
                  St. Julians, Malta, March 17-22, 2024},
  pages        = {150--158},
  publisher    = {Association for Computational Linguistics},
  year         = {2024},
  url          = {https://aclanthology.org/2024.eacl-demo.16},
  bibsource    = {dblp computer science bibliography, https://dblp.org}
}

@inproceedings{gashteovski2017minie,
  author       = {Kiril Gashteovski and
                  Rainer Gemulla and
                  Luciano Del Corro},
  editor       = {Martha Palmer and
                  Rebecca Hwa and
                  Sebastian Riedel},
  title        = {MinIE: Minimizing Facts in Open Information Extraction},
  booktitle    = {Proceedings of the 2017 Conference on Empirical Methods in Natural
                  Language Processing, {EMNLP} 2017, Copenhagen, Denmark, September
                  9-11, 2017},
  pages        = {2630--2640},
  publisher    = {Association for Computational Linguistics},
  year         = {2017},
  url          = {https://doi.org/10.18653/v1/d17-1278},
  doi          = {10.18653/V1/D17-1278},
  bibsource    = {dblp computer science bibliography, https://dblp.org}
}

@inproceedings{cui2018neural,
  author       = {Lei Cui and
                  Furu Wei and
                  Ming Zhou},
  editor       = {Iryna Gurevych and
                  Yusuke Miyao},
  title        = {Neural Open Information Extraction},
  booktitle    = {Proceedings of the 56th Annual Meeting of the Association for Computational
                  Linguistics, {ACL} 2018, Melbourne, Australia, July 15-20, 2018, Volume
                  2: Short Papers},
  pages        = {407--413},
  publisher    = {Association for Computational Linguistics},
  year         = {2018},
  url          = {https://aclanthology.org/P18-2065/},
  doi          = {10.18653/V1/P18-2065},
  bibsource    = {dblp computer science bibliography, https://dblp.org}
}

@inproceedings{han2019opennre,
  author       = {Xu Han and
                  Tianyu Gao and
                  Yuan Yao and
                  Deming Ye and
                  Zhiyuan Liu and
                  Maosong Sun},
  editor       = {Sebastian Pad{\'{o}} and
                  Ruihong Huang},
  title        = {OpenNRE: An Open and Extensible Toolkit for Neural Relation Extraction},
  booktitle    = {Proceedings of the 2019 Conference on Empirical Methods in Natural
                  Language Processing and the 9th International Joint Conference on
                  Natural Language Processing, {EMNLP-IJCNLP} 2019, Hong Kong, China,
                  November 3-7, 2019 - System Demonstrations},
  pages        = {169--174},
  publisher    = {Association for Computational Linguistics},
  year         = {2019},
  url          = {https://doi.org/10.18653/v1/D19-3029},
  doi          = {10.18653/V1/D19-3029},
  bibsource    = {dblp computer science bibliography, https://dblp.org}
}

@inproceedings{davison2019commonsense,
  author       = {Joe Davison and
                  Joshua Feldman and
                  Alexander M. Rush},
  editor       = {Kentaro Inui and
                  Jing Jiang and
                  Vincent Ng and
                  Xiaojun Wan},
  title        = {Commonsense Knowledge Mining from Pretrained Models},
  booktitle    = {Proceedings of the 2019 Conference on Empirical Methods in Natural
                  Language Processing and the 9th International Joint Conference on
                  Natural Language Processing, {EMNLP-IJCNLP} 2019, Hong Kong, China,
                  November 3-7, 2019},
  pages        = {1173--1178},
  publisher    = {Association for Computational Linguistics},
  year         = {2019},
  url          = {https://doi.org/10.18653/v1/D19-1109},
  doi          = {10.18653/V1/D19-1109},
  bibsource    = {dblp computer science bibliography, https://dblp.org}
}

@incollection{speer2013conceptnet,
  author       = {Robyn Speer and
                  Catherine Havasi},
  editor       = {Iryna Gurevych and
                  Jungi Kim},
  title        = {ConceptNet 5: {A} Large Semantic Network for Relational Knowledge},
  booktitle    = {The People's Web Meets NLP, Collaboratively Constructed Language Resources},
  series       = {Theory and Applications of Natural Language Processing},
  pages        = {161--176},
  publisher    = {Springer},
  year         = {2013},
  url          = {https://doi.org/10.1007/978-3-642-35085-6\_6},
  doi          = {10.1007/978-3-642-35085-6\_6},
  bibsource    = {dblp computer science bibliography, https://dblp.org}
}

@inproceedings{hwang2021comet,
  author       = {Jena D. Hwang and
                  Chandra Bhagavatula and
                  Ronan Le Bras and
                  Jeff Da and
                  Keisuke Sakaguchi and
                  Antoine Bosselut and
                  Yejin Choi},
  title        = {(Comet-) Atomic 2020: On Symbolic and Neural Commonsense Knowledge
                  Graphs},
  booktitle    = {Thirty-Fifth {AAAI} Conference on Artificial Intelligence, {AAAI}
                  2021, Thirty-Third Conference on Innovative Applications of Artificial
                  Intelligence, {IAAI} 2021, The Eleventh Symposium on Educational Advances
                  in Artificial Intelligence, {EAAI} 2021, Virtual Event, February 2-9,
                  2021},
  pages        = {6384--6392},
  publisher    = {{AAAI} Press},
  year         = {2021},
  url          = {https://doi.org/10.1609/aaai.v35i7.16792},
  doi          = {10.1609/AAAI.V35I7.16792},
  bibsource    = {dblp computer science bibliography, https://dblp.org}
}

@inproceedings{asai2023self,
  author       = {Akari Asai and
                  Zeqiu Wu and
                  Yizhong Wang and
                  Avirup Sil and
                  Hannaneh Hajishirzi},
  title        = {Self-RAG: Learning to Retrieve, Generate, and Critique through Self-Reflection},
  booktitle    = {The Twelfth International Conference on Learning Representations,
                  {ICLR} 2024, Vienna, Austria, May 7-11, 2024},
  publisher    = {OpenReview.net},
  year         = {2024},
  url          = {https://openreview.net/forum?id=hSyW5go0v8},
  bibsource    = {dblp computer science bibliography, https://dblp.org}
}

@article{ram2023context,
  author       = {Ori Ram and
                  Yoav Levine and
                  Itay Dalmedigos and
                  Dor Muhlgay and
                  Amnon Shashua and
                  Kevin Leyton{-}Brown and
                  Yoav Shoham},
  title        = {In-Context Retrieval-Augmented Language Models},
  journal      = {Trans. Assoc. Comput. Linguistics},
  volume       = {11},
  pages        = {1316--1331},
  year         = {2023},
  url          = {https://doi.org/10.1162/tacl\_a\_00605},
  doi          = {10.1162/TACL\_A\_00605},
  bibsource    = {dblp computer science bibliography, https://dblp.org}
}

@inproceedings{mallen2023not,
  author       = {Alex Mallen and
                  Akari Asai and
                  Victor Zhong and
                  Rajarshi Das and
                  Daniel Khashabi and
                  Hannaneh Hajishirzi},
  editor       = {Anna Rogers and
                  Jordan L. Boyd{-}Graber and
                  Naoaki Okazaki},
  title        = {When Not to Trust Language Models: Investigating Effectiveness of
                  Parametric and Non-Parametric Memories},
  booktitle    = {Proceedings of the 61st Annual Meeting of the Association for Computational
                  Linguistics (Volume 1: Long Papers), {ACL} 2023, Toronto, Canada,
                  July 9-14, 2023},
  pages        = {9802--9822},
  publisher    = {Association for Computational Linguistics},
  year         = {2023},
  url          = {https://doi.org/10.18653/v1/2023.acl-long.546},
  doi          = {10.18653/V1/2023.ACL-LONG.546},
  bibsource    = {dblp computer science bibliography, https://dblp.org}
}

@inproceedings{chanchateval,
  author       = {Chi{-}Min Chan and
                  Weize Chen and
                  Yusheng Su and
                  Jianxuan Yu and
                  Wei Xue and
                  Shanghang Zhang and
                  Jie Fu and
                  Zhiyuan Liu},
  title        = {ChatEval: Towards Better LLM-based Evaluators through Multi-Agent
                  Debate},
  booktitle    = {The Twelfth International Conference on Learning Representations,
                  {ICLR} 2024, Vienna, Austria, May 7-11, 2024},
  publisher    = {OpenReview.net},
  year         = {2024},
  url          = {https://openreview.net/forum?id=FQepisCUWu},
  bibsource    = {dblp computer science bibliography, https://dblp.org}
}

@inproceedings{fu2024gptscore,
  author       = {Jinlan Fu and
                  See{-}Kiong Ng and
                  Zhengbao Jiang and
                  Pengfei Liu},
  editor       = {Kevin Duh and
                  Helena G{\'{o}}mez{-}Adorno and
                  Steven Bethard},
  title        = {GPTScore: Evaluate as You Desire},
  booktitle    = {Proceedings of the 2024 Conference of the North American Chapter of
                  the Association for Computational Linguistics: Human Language Technologies
                  (Volume 1: Long Papers), {NAACL} 2024, Mexico City, Mexico, June 16-21,
                  2024},
  pages        = {6556--6576},
  publisher    = {Association for Computational Linguistics},
  year         = {2024},
  url          = {https://doi.org/10.18653/v1/2024.naacl-long.365},
  doi          = {10.18653/V1/2024.NAACL-LONG.365},
  bibsource    = {dblp computer science bibliography, https://dblp.org}
}

@inproceedings{wang2024large,
  author       = {Peiyi Wang and
                  Lei Li and
                  Liang Chen and
                  Zefan Cai and
                  Dawei Zhu and
                  Binghuai Lin and
                  Yunbo Cao and
                  Lingpeng Kong and
                  Qi Liu and
                  Tianyu Liu and
                  Zhifang Sui},
  editor       = {Lun{-}Wei Ku and
                  Andre Martins and
                  Vivek Srikumar},
  title        = {Large Language Models are not Fair Evaluators},
  booktitle    = {Proceedings of the 62nd Annual Meeting of the Association for Computational
                  Linguistics (Volume 1: Long Papers), {ACL} 2024, Bangkok, Thailand,
                  August 11-16, 2024},
  pages        = {9440--9450},
  publisher    = {Association for Computational Linguistics},
  year         = {2024},
  url          = {https://doi.org/10.18653/v1/2024.acl-long.511},
  doi          = {10.18653/V1/2024.ACL-LONG.511},
  bibsource    = {dblp computer science bibliography, https://dblp.org}
}

@article{gu2021domain,
  author       = {Yu Gu and
                  Robert Tinn and
                  Hao Cheng and
                  Michael Lucas and
                  Naoto Usuyama and
                  Xiaodong Liu and
                  Tristan Naumann and
                  Jianfeng Gao and
                  Hoifung Poon},
  title        = {Domain-Specific Language Model Pretraining for Biomedical Natural
                  Language Processing},
  journal      = {{ACM} Trans. Comput. Heal.},
  volume       = {3},
  number       = {1},
  pages        = {2:1--2:23},
  year         = {2022},
  url          = {https://doi.org/10.1145/3458754},
  doi          = {10.1145/3458754},
  bibsource    = {dblp computer science bibliography, https://dblp.org}
}

@article{singhal2023large,
  author       = {Karan Singhal and
                  Shekoofeh Azizi and
                  Tao Tu and
                  S. Sara Mahdavi and
                  Jason Wei and
                  Hyung Won Chung and
                  Nathan Scales and
                  Ajay Kumar Tanwani and
                  Heather Cole{-}Lewis and
                  Stephen Pfohl and
                  Perry Payne and
                  Martin Seneviratne and
                  Paul Gamble and
                  Chris Kelly and
                  Nathaneal Sch{\"{a}}rli and
                  Aakanksha Chowdhery and
                  Philip Andrew Mansfield and
                  Blaise Ag{\"{u}}era y Arcas and
                  Dale R. Webster and
                  Gregory S. Corrado and
                  Yossi Matias and
                  Katherine Chou and
                  Juraj Gottweis and
                  Nenad Tomasev and
                  Yun Liu and
                  Alvin Rajkomar and
                  Joelle K. Barral and
                  Christopher Semturs and
                  Alan Karthikesalingam and
                  Vivek Natarajan},
  title        = {Large Language Models Encode Clinical Knowledge},
  journal      = {CoRR},
  volume       = {abs/2212.13138},
  year         = {2022},
  url          = {https://doi.org/10.48550/arXiv.2212.13138},
  doi          = {10.48550/ARXIV.2212.13138},
  eprinttype    = {arXiv},
  eprint       = {2212.13138},
  bibsource    = {dblp computer science bibliography, https://dblp.org}
}

@article{thirunavukarasu2023large,
  title={Large language models in medicine},
  author={Thirunavukarasu, Arun James and Ting, Darren Shu Jeng and Elangovan, Kabilan and Gutierrez, Laura and Tan, Ting Fang and Ting, Daniel Shu Wei},
  journal={Nature medicine},
  volume={29},
  number={8},
  pages={1930--1940},
  year={2023},
  publisher={Nature Publishing Group US New York}
}

@article{bernal2009cultural,
  title={Cultural adaptation of treatments: A resource for considering culture in evidence-based practice.},
  author={Bernal, Guillermo and Jim{\'e}nez-Chafey, Mar{\'\i}a I and Domenech Rodr{\'\i}guez, Melanie M},
  journal={Professional Psychology: Research and Practice},
  volume={40},
  number={4},
  pages={361},
  year={2009},
  publisher={American Psychological Association}
}

@article{sue2001multidimensional,
  title={Multidimensional facets of cultural competence},
  author={Sue, Derald Wing},
  journal={The counseling psychologist},
  volume={29},
  number={6},
  pages={790--821},
  year={2001},
  publisher={Sage Publications Sage CA: Thousand Oaks, CA}
}

@article{sue1998search,
  title={In search of cultural competence in psychotherapy and counseling.},
  author={Sue, Stanley},
  journal={American psychologist},
  volume={53},
  number={4},
  pages={440},
  year={1998},
  publisher={American Psychological Association}
}

@inproceedings{sun2020treegen,
  author       = {Zeyu Sun and
                  Qihao Zhu and
                  Yingfei Xiong and
                  Yican Sun and
                  Lili Mou and
                  Lu Zhang},
  title        = {TreeGen: {A} Tree-Based Transformer Architecture for Code Generation},
  booktitle    = {The Thirty-Fourth {AAAI} Conference on Artificial Intelligence, {AAAI}
                  2020, The Thirty-Second Innovative Applications of Artificial Intelligence
                  Conference, {IAAI} 2020, The Tenth {AAAI} Symposium on Educational
                  Advances in Artificial Intelligence, {EAAI} 2020, New York, NY, USA,
                  February 7-12, 2020},
  pages        = {8984--8991},
  publisher    = {{AAAI} Press},
  year         = {2020},
  url          = {https://doi.org/10.1609/aaai.v34i05.6430},
  doi          = {10.1609/AAAI.V34I05.6430},
  bibsource    = {dblp computer science bibliography, https://dblp.org}
}

@inproceedings{le2022coderl,
  author       = {Hung Le and
                  Yue Wang and
                  Akhilesh Deepak Gotmare and
                  Silvio Savarese and
                  Steven Chu{-}Hong Hoi},
  editor       = {Sanmi Koyejo and
                  S. Mohamed and
                  A. Agarwal and
                  Danielle Belgrave and
                  K. Cho and
                  A. Oh},
  title        = {CodeRL: Mastering Code Generation through Pretrained Models and Deep
                  Reinforcement Learning},
  booktitle    = {Advances in Neural Information Processing Systems 35: Annual Conference
                  on Neural Information Processing Systems 2022, NeurIPS 2022, New Orleans,
                  LA, USA, November 28 - December 9, 2022},
  year         = {2022},
  url          = {http://papers.nips.cc/paper\_files/paper/2022/hash/8636419dea1aa9fbd25fc4248e702da4-Abstract-Conference.html},
  bibsource    = {dblp computer science bibliography, https://dblp.org}
}

@inproceedings{mastropaolo2021studying,
  author       = {Antonio Mastropaolo and
                  Simone Scalabrino and
                  Nathan Cooper and
                  David Nader{-}Palacio and
                  Denys Poshyvanyk and
                  Rocco Oliveto and
                  Gabriele Bavota},
  title        = {Studying the Usage of Text-To-Text Transfer Transformer to Support
                  Code-Related Tasks},
  booktitle    = {43rd {IEEE/ACM} International Conference on Software Engineering,
                  {ICSE} 2021, Madrid, Spain, 22-30 May 2021},
  pages        = {336--347},
  publisher    = {{IEEE}},
  year         = {2021},
  url          = {https://doi.org/10.1109/ICSE43902.2021.00041},
  doi          = {10.1109/ICSE43902.2021.00041},
  bibsource    = {dblp computer science bibliography, https://dblp.org}
}

@inproceedings{bian2020rumor,
  author       = {Tian Bian and
                  Xi Xiao and
                  Tingyang Xu and
                  Peilin Zhao and
                  Wenbing Huang and
                  Yu Rong and
                  Junzhou Huang},
  title        = {Rumor Detection on Social Media with Bi-Directional Graph Convolutional
                  Networks},
  booktitle    = {The Thirty-Fourth {AAAI} Conference on Artificial Intelligence, {AAAI}
                  2020, The Thirty-Second Innovative Applications of Artificial Intelligence
                  Conference, {IAAI} 2020, The Tenth {AAAI} Symposium on Educational
                  Advances in Artificial Intelligence, {EAAI} 2020, New York, NY, USA,
                  February 7-12, 2020},
  pages        = {549--556},
  publisher    = {{AAAI} Press},
  year         = {2020},
  url          = {https://doi.org/10.1609/aaai.v34i01.5393},
  doi          = {10.1609/AAAI.V34I01.5393},
  bibsource    = {dblp computer science bibliography, https://dblp.org}
}

@inproceedings{li2021align,
  author       = {Junnan Li and
                  Ramprasaath R. Selvaraju and
                  Akhilesh Gotmare and
                  Shafiq R. Joty and
                  Caiming Xiong and
                  Steven Chu{-}Hong Hoi},
  editor       = {Marc'Aurelio Ranzato and
                  Alina Beygelzimer and
                  Yann N. Dauphin and
                  Percy Liang and
                  Jennifer Wortman Vaughan},
  title        = {Align before Fuse: Vision and Language Representation Learning with
                  Momentum Distillation},
  booktitle    = {Advances in Neural Information Processing Systems 34: Annual Conference
                  on Neural Information Processing Systems 2021, NeurIPS 2021, December
                  6-14, 2021, virtual},
  pages        = {9694--9705},
  year         = {2021},
  url          = {https://proceedings.neurips.cc/paper/2021/hash/505259756244493872b7709a8a01b536-Abstract.html},
  bibsource    = {dblp computer science bibliography, https://dblp.org}
}

@article{pang2019antibiotic,
  title={Antibiotic resistance in Pseudomonas aeruginosa: mechanisms and alternative therapeutic strategies},
  author={Pang, Zheng and Raudonis, Renee and Glick, Bernard R and Lin, Tong-Jun and Cheng, Zhenyu},
  journal={Biotechnology advances},
  volume={37},
  number={1},
  pages={177--192},
  year={2019},
  publisher={Elsevier}
}

@article{aslam2018antibiotic,
  title={Antibiotic resistance: a rundown of a global crisis},
  author={Aslam, Bilal and Wang, Wei and Arshad, Muhammad Imran and Khurshid, Mohsin and Muzammil, Saima and Rasool, Muhammad Hidayat and Nisar, Muhammad Atif and Alvi, Ruman Farooq and Aslam, Muhammad Aamir and Qamar, Muhammad Usman and others},
  journal={Infection and drug resistance},
  pages={1645--1658},
  year={2018},
  publisher={Taylor \& Francis}
}

@inproceedings{vaswani2017attention,
  author       = {Ashish Vaswani and
                  Noam Shazeer and
                  Niki Parmar and
                  Jakob Uszkoreit and
                  Llion Jones and
                  Aidan N. Gomez and
                  Lukasz Kaiser and
                  Illia Polosukhin},
  editor       = {Isabelle Guyon and
                  Ulrike von Luxburg and
                  Samy Bengio and
                  Hanna M. Wallach and
                  Rob Fergus and
                  S. V. N. Vishwanathan and
                  Roman Garnett},
  title        = {Attention is All you Need},
  booktitle    = {Advances in Neural Information Processing Systems 30: Annual Conference
                  on Neural Information Processing Systems 2017, December 4-9, 2017,
                  Long Beach, CA, {USA}},
  pages        = {5998--6008},
  year         = {2017},
  url          = {https://proceedings.neurips.cc/paper/2017/hash/3f5ee243547dee91fbd053c1c4a845aa-Abstract.html},
  bibsource    = {dblp computer science bibliography, https://dblp.org}
}

@inproceedings{liu2016coupled,
  author       = {Ming{-}Yu Liu and
                  Oncel Tuzel},
  editor       = {Daniel D. Lee and
                  Masashi Sugiyama and
                  Ulrike von Luxburg and
                  Isabelle Guyon and
                  Roman Garnett},
  title        = {Coupled Generative Adversarial Networks},
  booktitle    = {Advances in Neural Information Processing Systems 29: Annual Conference
                  on Neural Information Processing Systems 2016, December 5-10, 2016,
                  Barcelona, Spain},
  pages        = {469--477},
  year         = {2016},
  url          = {https://proceedings.neurips.cc/paper/2016/hash/502e4a16930e414107ee22b6198c578f-Abstract.html},
  bibsource    = {dblp computer science bibliography, https://dblp.org}
}

@inproceedings{kim2021vilt,
  author       = {Wonjae Kim and
                  Bokyung Son and
                  Ildoo Kim},
  editor       = {Marina Meila and
                  Tong Zhang},
  title        = {ViLT: Vision-and-Language Transformer Without Convolution or Region
                  Supervision},
  booktitle    = {Proceedings of the 38th International Conference on Machine Learning,
                  {ICML} 2021, 18-24 July 2021, Virtual Event},
  series       = {Proceedings of Machine Learning Research},
  volume       = {139},
  pages        = {5583--5594},
  publisher    = {{PMLR}},
  year         = {2021},
  url          = {http://proceedings.mlr.press/v139/kim21k.html},
  bibsource    = {dblp computer science bibliography, https://dblp.org}
}

@article{monti2019fake,
  author       = {Federico Monti and
                  Fabrizio Frasca and
                  Davide Eynard and
                  Damon Mannion and
                  Michael M. Bronstein},
  title        = {Fake News Detection on Social Media using Geometric Deep Learning},
  journal      = {CoRR},
  volume       = {abs/1902.06673},
  year         = {2019},
  url          = {http://arxiv.org/abs/1902.06673},
  eprinttype    = {arXiv},
  eprint       = {1902.06673},
  bibsource    = {dblp computer science bibliography, https://dblp.org}
}

@inproceedings{ma2018rumor,
  author       = {Jing Ma and
                  Wei Gao and
                  Kam{-}Fai Wong},
  editor       = {Iryna Gurevych and
                  Yusuke Miyao},
  title        = {Rumor Detection on Twitter with Tree-structured Recursive Neural Networks},
  booktitle    = {Proceedings of the 56th Annual Meeting of the Association for Computational
                  Linguistics, {ACL} 2018, Melbourne, Australia, July 15-20, 2018, Volume
                  1: Long Papers},
  pages        = {1980--1989},
  publisher    = {Association for Computational Linguistics},
  year         = {2018},
  url          = {https://aclanthology.org/P18-1184/},
  doi          = {10.18653/V1/P18-1184},
  bibsource    = {dblp computer science bibliography, https://dblp.org}
}

@article{Kim2024FABLESEF,
  author       = {Yekyung Kim and
                  Yapei Chang and
                  Marzena Karpinska and
                  Aparna Garimella and
                  Varun Manjunatha and
                  Kyle Lo and
                  Tanya Goyal and
                  Mohit Iyyer},
  title        = {{FABLES:} Evaluating faithfulness and content selection in book-length
                  summarization},
  journal      = {CoRR},
  volume       = {abs/2404.01261},
  year         = {2024},
  url          = {https://doi.org/10.48550/arXiv.2404.01261},
  doi          = {10.48550/ARXIV.2404.01261},
  eprinttype    = {arXiv},
  eprint       = {2404.01261},
  bibsource    = {dblp computer science bibliography, https://dblp.org}
}

\clearpage
\newpage

\appendix
\section{Existing Work on Scientific Impact Analysis}
\label{app:literaturetable}
In Table~\ref{comparisontable}, we compare our work with existing research on scientific impact analysis. More on this in Section~\ref{sec:related}.

\begin{table*}
  \centering
\begin{tabular}{|m{2.2cm}|m{2.2cm}|m{1.1cm}|m{0.9cm}|m{2.2cm}|m{1.4cm}|m{2.2cm}|}
    \hline
    \textbf{\cellcolor{gray!15}Work}            & \textbf{\cellcolor{gray!15}Impact as}   & 
    \textbf{\cellcolor{gray!15}Facets}   & \textbf{\cellcolor{gray!15}Time}   & 
    \textbf{\cellcolor{gray!15}Method}   & 
    \textbf{\cellcolor{gray!15}Intents}   & \textbf{\cellcolor{gray!15}Fields}\\
    \hline
    \cite{valenzuela2015identifying} & Influential-citation counts & P. &  \xmark & Supervised Machine Learning & coarse & Computer science\\
    \hline
    \cite{zhu2015measuring} & Influential-citation counts & P. & \xmark & Supervised Machine Learning & \xmark & 10 fields\\
    \hline
 \cite{hutchins2016relative} & Relative citation ratio & n/a & \checkmark & Statistical measures & \xmark & Medicine\\
    \hline
    \cite{he2018temporal} & Quantified change rate of citation context & n/a & \checkmark & unsupervised temporal embedding analysis & \xmark & Biomedical\\
    \hline
\cite{jurgens2018measuring} & Citation counts & n/a & \checkmark & Supervised Machine Learning & coarse & Computer Science\\
    \hline
\cite{bos2019interdisciplinary} & Citation ratio  & n/a & \xmark & Statistical measures & \xmark & 13 fields\\
    \hline
\cite{siudem2020three} & Bibliometric indexes & n/a & \xmark & Statistical measures & \xmark & Computer science\\
    \hline
\cite{rudiger2021explanatory} & Word lists & n/a & \xmark & Text clustering & \xmark & Information systems\\
    \hline
\cite{arts2021natural} & Concept combinations & P. & \xmark & Text processing & \xmark & U.S. patents\\
    \hline
\cite{min2021citation} & Citation cascades & n/a & \checkmark & Network analysis & \xmark & Physics\\
    \hline
\cite{wahle2023we} & Citation Field Diversity Index & n/a & \checkmark & Statistical measures&  \xmark & 23 fields, focus on Computer science\\
    \hline
\cite{shi2023surprising} & Concept combinations & P. & \checkmark & Hypergraph model & \xmark & Medicine, Physics\\
    \hline
\cite{zhang2024pst} & Influential-citation counts& P. & \xmark & LLM & \xmark & Computer science\\
    \hline
\cite{gu2024forecasting} & Citation counts & n/a & \checkmark & FNN & \xmark  & Physics\\
    \hline
\cite{arts2025beyond} & Paper counts & P. & \checkmark &Statistical measures& \xmark & Nobel Prize-winning papers\\
    \hline
                    \cellcolor{gray!15}This work          & \cellcolor{gray!15}Textual Summary  & \cellcolor{gray!15}P., C. & \cellcolor{gray!15}\checkmark  &  \cellcolor{gray!15} LLM   &\cellcolor{gray!15} fine-grained   &  \cellcolor{gray!15}Computer science, Psychology, Medicine \\
         \hline 
 
  \end{tabular}
  \caption{Comparison with existing work on scientific impact analysis\label{comparisontable}. Our work is the first to express scientific impact through time-aware textual summaries derived from fine-grained citation context analysis, which covers both praise (confirmation) and critique (correction).}
\end{table*}

\section{Prompt for Identifying Impact-revealing Citation Intents}
\label{app:intentprompt}
The prompt used is shown in Figure~\ref{fig:prompt_intent_generation}.

\begin{figure*}
\begin{tcolorbox}[colback=white!10!white,colframe=black!90!black,title=Identifying impact-revealing intents]
  A \textbf{citation context} in a scientific paper refers to the specific part of a paper \( p' \) where another paper \( p \) is mentioned, including the surrounding sentences or paragraphs that explains how and why \( p \) is relevant to \( p' \).  
  
        A citation context with an \colorbox{orange!40}{\texttt{impact-revealing}} intent is a type of citation in scientific writing that highlights the significance or influence of a previously published work, often emphasizing its contribution or importance to the current research or the broader field, e.g., its role in inspiring, motivating, supporting, filling gaps, critically analyzing, or contributing methods, tools, data, extensions, or benchmarks for the current research.\\
        \colorbox{cyan!10}{\texttt{Other}} types of intents include a reference to prior work in a scientific paper that provides background or context without emphasizing the impact, significance, or influence of the cited work. It acknowledges the source in a routine or supporting role rather than showcasing its importance to the research.\\
        Given a citation context, describe, in a few words, the intention behind this citation phrase. Then, decide on the category of this intention. In particular, whether the intention behind this citation phrase is impact-revealing or not (i.e., incidental or that there isn't enough information to realize the real intention behind it). For the intention category, only return one of the following two labels \colorbox{orange!40}{\texttt{impact-revealing}} or \colorbox{cyan!10}{\texttt{other}}.\\
        Below are examples:\\
        \texttt{\$examples\$}

\end{tcolorbox}
\caption{Prompt for generating and classifying fine-grained intents.}
 \label{fig:prompt_intent_generation}
\end{figure*}

\section{Prompt for Generating Impact Summaries and Output Schema}
\label{app:generateprompt}

To ensure that the generated output contain all the components required to construct the impact summaries, (namely the impact periods, their dominating intent, details about the period, and citations used as evidence), we make use of OpenAI's structured output feature\footnote{\url{https://platform.openai.com/docs/guides/structured-outputs}}. Moreover, this facilitates a more systematic automated evaluation for our prompt variants. The prompt design is illustrated in Figure~\ref{fig:prompt_impact_generation}, and the corresponding structured output schema is presented in Figure~\ref{lst:schema}.

\begin{figure*}
\begin{tcolorbox}[colback=white!10!white,colframe=black!90!black,title=Generating an impact summary about a research paper]
  The \textbf{scientific impact summary of a research paper} describes the impact a given paper had on other papers, including both praise and critique. To understand the impact of a paper, one needs to understand how exactly it has been utilized and discussed by other papers. This is normally referred to as citation intents. One also needs to understand the evolution of the impact and citation intents over time.\\
  Given an input paper's title, its publication year, and its citation context, describe the impact of that paper.\\
  The citation context includes five components: $<$citation ID, citation title, citation year, citation context, citation intent$>$.\\

  Given the input paper with id \texttt{\$paperId\$} titled \texttt{\$title\$} published in \texttt{\$year\$}, and the following list of papers citing it:\\
  \texttt{\$citation\_context\$}

  Generate an impact summary about the input paper.
\end{tcolorbox}
\caption{Prompt for generating an impact summary about a research paper.}
 \label{fig:prompt_impact_generation}
\end{figure*}

\begin{figure*}
\begin{lstlisting}[style=pythonstyle]
# Schema for structured output
{"name": "impact_statement",
"schema": {"type": "object",
            "properties": {
            "input_paper_info": {
                "type": "object",
                "items": {
                "input_paper_id": "id of input paper",
                "input_paper_title": "title of input paper",
                "input_paper_year": "year of input paper"}},
            "impact_periods":{
              "type": "array",
              "items": {
                  "impact_period": "start year - end year",
                  "aspect_of_period": "the dominating citation intent(s) of that period",
                  "impact_description": "a paragraph to describe the impact of that period",
                  "evidence": "citing papers from that period to back up the described impact aspect"}}
            }
        }
}
\end{lstlisting}
\caption{Output schema for impact summary generation.}
 \label{lst:schema}
\end{figure*}

\section{Prompts for Evaluating Impact Summaries}
\label{evalprompts}
Trustworthiness metrics: The prompt we use for our faithfulness evaluation is shown in Figure~\ref{fig:faith_eval_prompt}, and for our two-step coverage evaluation in Figure~\ref{fig:prompt_coverage}. Informativeness metrics: The evaluation steps are shown in Figure~\ref{lst:qualityevalprompts}. 

\begin{figure*}
\begin{tcolorbox}
[colback=white!10!white,colframe=black!90!black,title=Faithfulness evaluation prompt]
**Task:** Verify the faithfulness of an impact description regarding the paper "\{\{PAPER\_NAME\}\}". It is faithful if it can be supported by one or more of the paper's citations from the provided list. Each citation is formatted as <title>:citation\_text, where the title indicates the title of the citing paper.\newline
\newline
**Impact description to Verify:**\newline
<impact-description>\newline
\{\{DESCRIPTION\}\}\newline
</impact-description>\newline
\newline
**Citation List:**\newline
<citations>\newline
\{\{SOURCES\}\}\newline
</citations>\newline
\newline
**Steps to Complete the Task:**\newline
\newline
1. Understand the impact description and its specified time period.\newline
2. Review each citation in the provided list. \newline
3. Determine if the impact description can be supported by any single citation or a combination of citations. \newline
4. If the impact description is supported, identify the relevant citations that support it. \newline
5. If the impact description cannot be supported or is contradicted by the citations, determine it as unfaithful. \newline
\newline
**Response Format:** \newline
\newline
- **Analysis:** <analysis> [Provide your analysis here] </analysis> \newline
- **Answer:** <answer> [yes/no] </answer> \newline
- **Proof:** <proof> [List the exact text of the citations that support the impact description, or "none" if unfaithful] </proof>\newline
\newline
**Additional Guidelines:**\newline
\newline
- The answer should be "yes" or "no" only.\newline
- In the proof section, include only the exact text of relevant citations without explanations.\newline
- List all necessary citations if multiple are needed to support the impact description.\newline
- If the impact description is unfaithful, state "none" in the proof section.\newline
- Avoid additional commentary outside the specified format.
\end{tcolorbox}
\caption{Prompt for verifying the time-aware faithfulness of  scientific impact summaries.}
\label{fig:faith_eval_prompt}
\end{figure*}

\begin{figure*}
\begin{tcolorbox}[colback=white!10!white,colframe=black!90!black,title=Prompt for coverage evaluation]
  Step 1: Given this list of phrases \texttt{\$listOfPhrases\$}, cluster highly similar phrases and give a label (expressive theme) for every cluster.
  \tcblower
  Step 2: Given this list of themes in the format of a python list: \texttt{\$listOfThemes\$}, determine how many of these themes were implicitly or explicitly mentioned in this summary \texttt{\$summary\$}, and list them.\\
\end{tcolorbox}
\caption{Prompt for measuring coverage of an impact summary.}
 \label{fig:prompt_coverage}
\end{figure*}

\begin{figure*}
\begin{lstlisting}[style=pythonstyle]
from deepeval.test_case import LLMTestCase, LLMTestCaseParams
from deepeval.metrics import GEval

insight_metric = GEval(
    name="Insightfulness",
    evaluation_steps=[
        "Determine whether the impact summary describes how the paper has been directly used by or influenced by other works.",
        "Assess how well the impact summary articulates the paper's influence with informative details.",
        "You should heavily penalize the impact summary for lack of insight."
            ]
)
trend_metric = GEval(
    name="Trend awareness",
    evaluation_steps=[
        "Determine whether the impact summary mentions how the impact of the paper has changed over time, ensuring each impact period is clearly identified with descriptive titles.",
        "You should heavily penalize the impact summary if the titles of consecutive impact periods are not diverse."
            ]
)
specif_metric = GEval(
    name="Specificity",
    evaluation_steps=[
        "Determine whether the impact summary mentions specific techniques, frameworks, or studies influenced by the paper, or if it remains broad and lacking supporting details.",
        "You should heavily penalize the impact summary if it only restates the title and abstract or provides vague, generic statements without concrete examples of the influence of the paper."
            ]
)
\end{lstlisting}
\caption{Code snippet for chain of thoughts (CoTs) evaluation steps for 3 metrics of the impact summary evaluation.}
 \label{lst:qualityevalprompts}
\end{figure*}

\section{Details about Zero-shot vs. ICL for Fine-grained Intent Generation}
\label{app:zsicldetails}

\subsection{Training examples}
\label{trainingexamples}
The list of manually annotated examples used in this experiment are in Table~\ref{iclexamples}.

\begin{table*}
  \centering
  \begin{tabular}{|p{7.5cm}|p{3.5cm}|c|}
    \hline
    \textbf{\cellcolor{gray!15}Citation context}            & \textbf{\cellcolor{gray!15}Intent}   & \textbf{\cellcolor{gray!15}Class} \\
    \hline
   In order to reduce the memory requirements, we apply a minimization process [1]. & use of minimization methodology & impact-revealing\\
\hline
       Motor adaptation is the process of re-shaping acquired motor skills through the reduction of errors (Hardwick and Celnik, 2014; Krakauer, 2009) & defines the term motor adaptation & other \\
       \hline
           Moreover, none of the above studies explored whether temporally primary PLEs are associated with an increased risk of subsequent insomnia [1,2].. In this study, we explored the changes in prevalence of insomnia and PLEs before and during the pandemic.       & identifying and addressing knowledge gap in literature& impact-revealing  \\
           \hline
Chiu and Nichols (2016) introduced convolutional neural networks for NER & background about NER methods & other\\
\hline
We employ the single-link method to compute the similarity be-tween two clusters, which has been applied widely in prior research (Bagga and Baldwin (1998); Mann and Yarowsky (2003)) & use of cluster similarity methods & impact-revealing\\
\hline
Quirk and Poon (2017) and Peng et al. (2017) build two distantly supervised datasets without human annotation, which may make the evaluation less reliable. In this paper, we present DocRED, a large-scale
human-annotated document-level RE dataset.. & criticizing existing datasets and proposing a better one & impact-revealing\\
\hline
In adults with severe malaria, increased Ang-2 plasma levels were associated with a decrease in NO bioavailability, higher lactate plasma concentrations, and patient mortality [101]. & reporting on existing studies about malaria & other\\
\hline
 , similarity measures [3]) to select an answer. & not enough information & other\\
 \hline
 Inspired by the reference [4], we proposed a novel algorithm called Soft-DDQN and applied it to the robot PAP skill learning problem & drawing inspiration from prior work to propose a new algorithm & impact-revealing\\
 \hline
 For instance, while Yoon et al. (2018) show that expanding the network capacity thorough width is helpful, they have not studied the impact of increasing capacity when the depth increases, nor why increasing the width is helpful. Overall, in this work, we are interested in understanding the impacts of network structure (e.g., width and
depth).. & highlighting gaps in existing research on network capacity & impact-revealing\\
 \hline
  \end{tabular}
  \caption{\label{iclexamples}Training examples (manually created from real contexts) for fine-grained intent generation and classification.}
\end{table*}

\subsection{More on setup and results}
\label{iclresults}

We use GPT-4o mini with temperature 0. As for the source of papers and citation context, we use the Semantic Scholar Academic Graph (S2AG)\footnote{\url{https://www.semanticscholar.org/product/api}}.

\begin{figure*}
  \includegraphics[width=\textwidth]{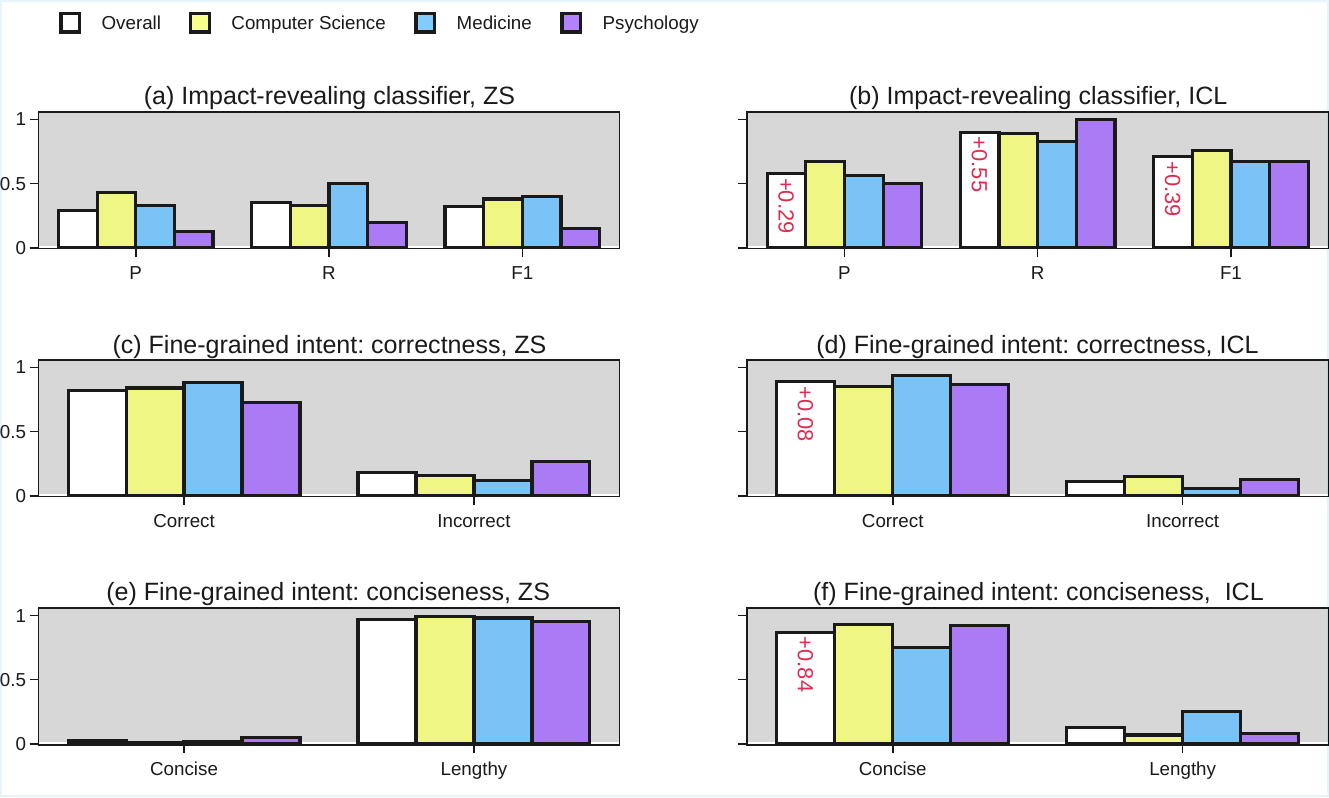}
  \caption{Results for fine-grained intent generations (ZS vs. ICL with 10 examples). Adding examples show significant improvements on metrics like precision, recall, F1.}
  \label{fig:fine_grained_intent_results}
\end{figure*}

\begin{figure*}[t]
\centering
  \includegraphics[width=0.8\textwidth]{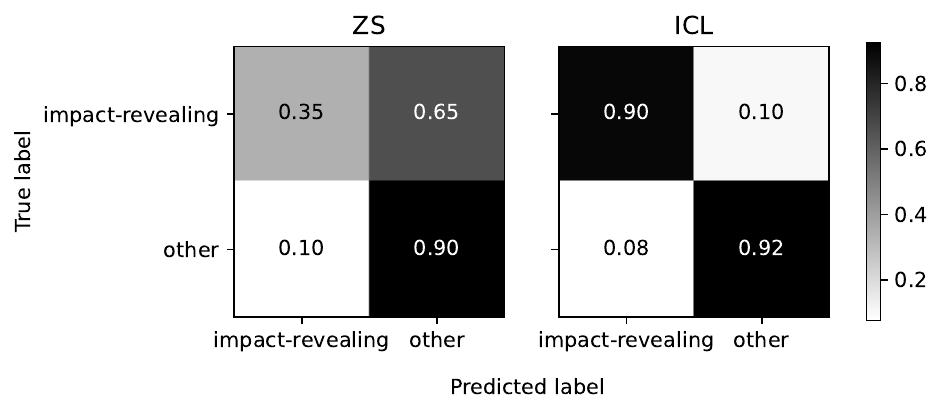}
  \caption{Confusion matrix for the impact-revealing classification task. More insights into misclassified instances in Section~\ref{iclresults}}
  \label{fig:cm_fine_grained_intent_generation}
\end{figure*}

Figure~\ref{fig:fine_grained_intent_results} shows the quantitative results per paper field, and Table~\ref{tab:examples_intent_generations} shows qualitative examples. We do not observe any large variations indicating that our method is not limited to a specific field. However, the highest recall (a perfect score of 1.0) was achieved in the psychology field. This may suggest that citation contexts in psychology papers tend to express intentions more explicitly, particularly those that reveal impact, e.g. ``\textit{Some of those assumptions have been controversial, as researchers disagree about whether the kinds of behaviors measured by particular implicit tests should be considered indicators of attitudes or something else}''. This seems to be in line with our observations made during error analysis. To understand where our ICL prompt failed, we plot the confusion matrices in Figure~\ref{fig:cm_fine_grained_intent_generation} and inspect the 10\% impact-revealing citation phrases that were misclassified as ``other''. We notice that the classifier considers \textit{minor} ``resource use'' intent as non-impact revealing, e.g., ``\textit{We retrieved 400 results for each keyword, resulting in a total of 145,682 stories downloaded in the JavaScript Object Notation (JSON) format [48].}''  On the same note, it appears that when ``inspiration'' is only hinted at in the citation context, the classifier is not able to pick up the signal, e.g., ``\textit{The search problem here corresponds closely to a binary dynamic constraint satisfaction problem [Mittal \& Falkenhainer , 1990]}''. Inter-annotator agreement on the annotated results is in Table~\ref{tab:agreement}. They range from fair to substantial. In this evaluation, annotators also indicate their confidence level in their decision on the impact-revealing classification as either ``low'', ``medium'', or ``high''. If there is a disagreement, the annotation with the higher confidence level takes precedence. If the confidence levels are tied, the tie is resolved randomly.

\begin{table*}
  \centering
  \begin{tabular}{|l|c|}
    \hline
    \textbf{\cellcolor{gray!15}Task}             & \textbf{\cellcolor{gray!15}Cohen's kappa}  \\
    \hline
    Impact-revealing classifier      & 0.68  \\
         \hline 
             Fine-grained intent: correctness         & 0.20 \\
\hline 
          Fine-grained intent: conciseness        & 0.34 \\
\hline
 
  \end{tabular}
  \caption{\label{tab:agreement}
   Inter-annotator agreement on fine-grained intent generation results (Section~\ref{sec:iclforintentgeneration}).
  }
\end{table*}

\begin{table*}
  \centering
  \begin{tabular}{|p{5cm}|p{4cm}|p{5cm}|}
    \hline
    \textbf{\cellcolor{gray!15}Citation context}           & \textbf{\cellcolor{gray!15}Baseline} & \textbf{\cellcolor{gray!15}Ours} \\
    \hline
Policy optimization is performed using an implementation of TRPO from rllab [62] with a step size of 0. & The intention behind this citation phrase is to acknowledge the use of a specific tool or method (TRPO from rllab) in the research without emphasizing its impact or significance \colorbox{cyan!10}{\texttt{other}} & reporting method used for policy optimization \colorbox{orange!40}{\texttt{impact-revealing}}\\
\hline
Alternative methods that instead rely on constraining model parameters, so-called regularization approaches [18,37], have in turn been shown to perform poorly on medical data. & This citation phrase acknowledges prior work related to regularization approaches and highlights that these methods perform poorly on medical data. \colorbox{cyan!10}{\texttt{other}} & highlighting the limitations of regularization approaches on medical data \colorbox{orange!40}{\texttt{impact-revealing}}\\
\hline
Additionally, suffering can occur due to illness, to the way in which illness is experienced, or, even, in absence of illness (Cassell, 1982; Frank, 2001). & The citation phrase acknowledges prior work to provide context and background for the concept of suffering in relation to illness and its experience. It doesn't emphasize the impact or significance of the cited works. \colorbox{cyan!10}{\texttt{other}} & providing context on the nature of suffering \colorbox{cyan!10}{\texttt{other}}\\
\hline
  \end{tabular}
  \caption{\label{tab:examples_intent_generations}
  Sample generated intents, baseline= ZS, Ours=ICL (10 examples).}
\end{table*}

\section{Effect of Different Number of Shots (K)} 
\label{app:numberofshots}
\begin{figure}[t]
  \includegraphics[width=\columnwidth]{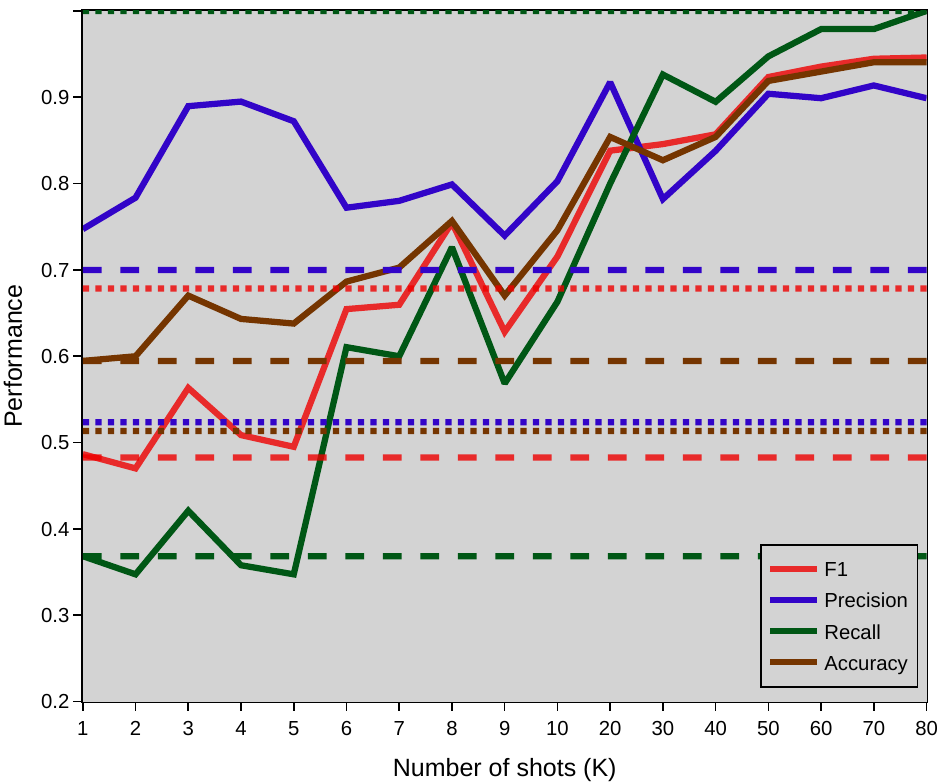}
  \caption{Effect of different number of shots (K) on detecting impact-revealing citations. Zero-shot:  horizontal dashed lines; \textit{always impact-revealing}: dotted lines. This shows that adding a reasonable number of examples (around 50) is sufficient to yield a 60\% improvement in recall.} 
  \label{fig:numofshots}
\end{figure}

We conduct an experiment to test our ICL prompt with different Ks. We manually polished the 200 instances from the previous experiment and randomly split the set into train 40\%, dev 30\%, test 30\%. For the number of shots, we test from K=1 to 80. For every dev and test instance, we run the prompt 5 times and report the average, using GPT4o-mini, shuffling the order of the shots. Results in Figure~\ref{fig:numofshots} show that increasing the number of shots has an effect on all metrics, outperforming both the zero-shot, and the always-impact-revealing baselines. Performance improves sharply up to K = 50, after which gains are minimal. At K = 50, metrics plateau with  precision at 0.90, recall at 0.94, F1 at 0.92, and accuracy at 0.92.

\subsection{A new dataset for identifying impact-revealing citations}
\label{gt}
Since our task on classifying citation context into ``impact-revealing'' (for both confirmation/praise and correction/critique) or ``other'' is new, there is a lack of groundtruth data. To construct such a resource, we build upon on an existing human annotated dataset, namely PST-Bench~\cite{zhang2024pst}. An instance in this dataset consists of a pair of research papers annotated for whether one of them influenced (impacted) the other. Influence here is defined only through a positive lens, and is further restricted to the following two intents: ``inspiration'' and ``motivation''.  To create our dataset, first, we randomly select 1k influential (impact-revealing)  papers, as labeled by the PST-Bench dataset annotators, and look up their aggregated citation contexts\footnote{The cited paper might be mentioned several times in the citing paper.}. Since the meaning of influential, according to PST-Bench, only focuses on two intents ``inspiration'' and ``motivation'', both limited to praise, we augment the dataset with other types for both confirmatory and correction citation intents. To do so, we manually craft a handful of textual patterns such as ``has led to open questions/challenges/unresolved issues'' to capture intents indicating research limitations, then ask GPT4-o to provide a longer list with variations of these patterns. We end up with a total of 33 patterns (examples shown in Figure~\ref{lst:impact_phrases}). Following this step, we crawl $\sim$ 200k aggregated citation contexts and search them using the list of impact-revealing phrases, then randomly sample 1k instances from the matching cases. To augment the data with non-impact-revealing (``other'') examples, we sample 2k non-influential instances from the PST-Bench dataset. These are citations that are listed under the category ``other citations'', as in non-influential ones. To ensure  the deemed non-influential examples are not in fact impact-revealing (ones which belong to intents other than ``motivation'' and ``inspiration''), we verify that there is no match with all the impact-revealing phrases. Our newly constructed dataset, which we release with this work, consists of 4k citation contexts augmented with either ``impact-revealing'' or ``other'' intent class (balanced). After manually inspecting a random sample of 100 instances, we found that 90\% were correctly identified as impact-revealing; the remaining 10\% included rare cases where phrases like motivated by'' appeared to signal impact (e.g., Voters are motivated by partisan social identities... (Greene 2004)'') but did not reflect the actual citation intent.

\begin{figure*}
\begin{lstlisting}[style=pythonstyle]
# Python list containing impact-revealing phrases (both praise and critique)
impact_revealing_phrases = [
    r"\b(builds? (upon|on)|extended by|extends the work of)\b",
    r"\b(serve|serves|served) as (a foundation|a basis|the groundwork) for\b",
    r"\b(has|have|had) (paved the way|opened avenues|led to advancements|sparked further research)\b",
    r"\b(is|was|has been) (criticized|challenged|questioned) for\b",
    r"\b(suffer(s|ed)? from|is|was|has been) (limited|constrained|hindered|flawed) by\b",
    r"\b(has|have|had) (left|created|highlighted|led to) (gaps|challenges|open questions|unresolved issues)\b",
   ...
]
\end{lstlisting}
\caption{Samples impact-revealing textual patterns (33 in total). These were used to extend the PST-Bench dataset to include more diverse impact-revealing citations (especially critique/correction). }
 \label{lst:impact_phrases}
\end{figure*}

\subsection{Details about comparisons with intent classifiers}
\label{externalsetup}
Mappings of the baselines' intents to ours are shown in Table~\ref{tab:schemamap}. Two of the external methods for intent classification, namely the work of ~\cite{cohan-etal-2019-structural} and ~\cite{lauscher2022multicite} allow for multi-label classification. This means: a citation context can receive both ``impact-revealing'' and ``other'' labels (e.g., with intents ``Result'' and ``Background''). However, in these cases, a ranking of the predictions can be utilized. For ~\cite{lauscher2022multicite}, we use the class with the highest prediction probability. For ~\cite{cohan-etal-2019-structural}, we count the frequency of every intent class describing the concatenated citation context, and most frequent class represented in the context. We break ties with a random pick.

\begin{table*}
  \centering
  \begin{tabular}{|l|l|}
    \hline
    \textbf{\cellcolor{gray!15}Intent Classifier}           & \textbf{\cellcolor{gray!15}Classes} \\
    \hline
\multirow{3}{*}{Structural Scaffolds~\cite{cohan-etal-2019-structural}} & Background  \colorbox{cyan!10}{\texttt{other}}\\
& Method \colorbox{orange!40}{\texttt{impact-revealing}}\\
& Result \colorbox{orange!40}{\texttt{impact-revealing}}\\
\hline
\multirow{2}{*}{Meaningful Citations~\cite{valenzuela2015identifying} } & non-meaningful  \colorbox{cyan!10}{\texttt{other}}\\
& meaningful \colorbox{orange!40}{\texttt{impact-revealing}}\\
\hline
\multirow{6}{*}{Multi-cite~\cite{lauscher2022multicite}}  & Background  \colorbox{cyan!10}{\texttt{other}}\\
& Motivation \colorbox{orange!40}{\texttt{impact-revealing}}\\
& Future Work \colorbox{cyan!10}{\texttt{other}}\\
& Similar/Difference \colorbox{orange!40}{\texttt{impact-revealing}}\\ 
& Uses \colorbox{orange!40}{\texttt{impact-revealing}}\\
& Extension \colorbox{orange!40}{\texttt{impact-revealing}}\\
         \hline 
  \end{tabular}
  \caption{\label{tab:schemamap}
  Mapping existing coarse intent classes of existing intent classifiers to our ``impact-revealing'' or ``other'' classes. }
\end{table*}

Sample results are shown in Table~\ref{external_methods_qualitative}.

\begin{table*}
  \centering
  \begin{tabular}{|p{3.2cm}|p{3.2cm}|p{3.2cm}|p{3.2cm}|}
    \hline
    \multicolumn{4}{|p{13cm}|}{Citation context: ``\textit{Inspired by the theory of hierarchical abstract machines (Parr and Russell 1998), we cast the task of profile reviser as a hierarchical Markov Decision Process (MDP).}''}\\
    \cdashline{1-4}
    \multicolumn{4}{|c|}{Intent predicted by:}\\
    \cdashline{1-4}
\textbf{\cellcolor{gray!15}Structural\newline Scaffolds}   & \textbf{\cellcolor{gray!15}Meaningful\newline Citations}   & \textbf{\cellcolor{gray!15}Multi-cite} & \textbf{\cellcolor{gray!15}Ours} \\     
        \cdashline{1-4}
            \colorbox{orange!40}{\texttt{impact-revealing}} (method) & \colorbox{cyan!10}{\texttt{other}}\newline(non-meaningful) & \colorbox{orange!40}{\texttt{impact-revealing}} (uses)  & \colorbox{orange!40}{\texttt{impact-revealing}}  (drawing on theoretical foundations for task formulation)  \\
         
         \hline 
                     \multicolumn{4}{|p{13cm}|}{Citation context: ``\textit{..on social network datasets, it is quite intuitive trying to extract information from text data to do ideology-detection, only a few paid attention to links [9, 13]... Even though some realized the importance of links [9, 13], they failed to provide an embedding.}''}\\
        \cdashline{1-4}
    \multicolumn{4}{|c|}{Intent predicted by:}\\
    \cdashline{1-4}
\textbf{\cellcolor{gray!15}Structural\newline Scaffolds}   & \textbf{\cellcolor{gray!15}Meaningful\newline Citations}   & \textbf{\cellcolor{gray!15}Multi-cite} & \textbf{\cellcolor{gray!15}Ours} \\     
        \cdashline{1-4}
            \colorbox{cyan!10}{\texttt{other}}\newline (background) & \colorbox{cyan!10}{\texttt{other}}\newline (non-meaningful) & \colorbox{orange!40}{\texttt{impact-revealing}} (motivation)  & \colorbox{orange!40}{\texttt{impact-revealing}}  (highlighting the gap in ideology detection methods)  \\   
         \hline 
                              \multicolumn{4}{|p{13cm}|}{Citation context: ``\textit{Recurrent networks are dedicated sequence models that maintain a vector of hidden activations that are propagated through time (Elman, 1990; Werbos, 1990; Graves, 2012).}''}\\
       \cdashline{1-4}
    \multicolumn{4}{|c|}{Intent predicted by:}\\
    \cdashline{1-4}
\textbf{\cellcolor{gray!15}Structural\newline Scaffolds}   & \textbf{\cellcolor{gray!15}Meaningful\newline Citations}   & \textbf{\cellcolor{gray!15}Multi-cite} & \textbf{\cellcolor{gray!15}Ours} \\     
        \cdashline{1-4}
            \colorbox{cyan!10}{\texttt{other}}\newline (background) & \colorbox{cyan!10}{\texttt{other}}\newline (non-meaningful) & \colorbox{cyan!10}{\texttt{other}}\newline (background)  & \colorbox{cyan!10}{\texttt{other}}  \newline (providing context on recurrent networks)  \\ 
            \hline
 
  \end{tabular}
  \caption{\label{external_methods_qualitative} Intent classes predicted by existing work on intent classification. The intents between parentheses are the actual labels predicted by the method, later mapped to either ``impact-revealing'' or ``other'' based on the mappings shown in Table~\ref{tab:schemamap}. For our method, the output is both the intent and its class.}
\end{table*}

\section{Details about the Ablation Study}
\label{app:detailsablation}

\subsection{Examples}
\label{app:ablationexamples}
Table~\ref{tab:faithfulness_analysis} shows examples of verified (faithful) impact descriptions. Table~\ref{tab:coverageexamples} shows examples for coverage evaluation. Note that we instruct the LLM to return the actual list of topics covered instead of only the number of themes covered, in order to determine its performance compared to a human evaluator (see Table~\ref{tab:humancorrelation}). Table~\ref{tab:citationsvalidityanalysis} shows examples from the citation year compliance evaluation.  In Table~\ref{tab:impactsummariesinformativness}, we show qualitative examples in informativeness.

\subsection{Long tail impact coverage}
\label{app:longtailcoverage}

We conduct a focused, quantitative analysis to evaluate the model’s ability to capture long-tail yet meaningful aspects of a paper’s impact. Specifically, we manually select 10 intent themes associated with 10 different papers that fall outside the top-3 most frequent themes (and are therefore not covered in the Coverage@3 metric) but nonetheless reflect important impact dimensions. We then analyze 50 generated summaries, 10 per variant, and measure the extent to which these summaries capture the selected long-tail themes. We refer to this metric as long tail impact coverage, or LTI for short. As shown in Table~\ref{table:longtailcoverage}, LTI ranges from 30\% for the ``all citation context, no intents'' variant to 50\% when citation intents are included. These findings suggest that the model is capable of recognizing and reflecting impact-revealing citations even when it appears less frequently in the input text, highlighting its sensitivity to semantically important but infrequent signals.

\begin{table}
\centering
\begin{tabular}{|l|c|c|}
\hline
\rowcolor{gray!15} \multicolumn{2}{|c|}{\textbf{Prompt variant}} & \textbf{LTI Coverage} \\
\rowcolor{gray!15} \textbf{Citations} & \textbf{Intents} & \\
\hline
None & \xmark & 0.1\\
All & \xmark & 0.3\\
All & \checkmark &0.5\\
Impact-rev. & \xmark & 0.4\\
\rowcolor{gray!15} Impact-rev. & \checkmark & 0.4\\
\hline
\end{tabular}
\caption{\textbf{LTI}: long tail impact coverage}
\label{table:longtailcoverage}
\end{table}

\subsection{Results per field}
\label{app:resultsperfield}
Table \ref{table:ablation_results_detailed} presents the results for papers of different research fields. We observe that faithfulness and citation year compliance exhibit similar patterns in specific fields as in the overall results, with variants with impact-revealing citations achieving better results. The only exception is the computer science, with a minor difference on year compliance of 1\% in favor of variants that receive all citations. We notice that providing citation intents has a negligible effect on faithfulness and decreases the year compliance. This is in line with the overall results. For informativeness, we notice that the same variant wins on all metrics with the exception of insightfulness in the psychology field, where the variant with all citation context and no intents wins by +1\%.

\subsection{Human-LLM correlation}
\label{app:humancorrelation}
In Table~\ref{tab:humancorrelation}, we report the correlation between LLM and human raters on our evaluation metrics and find that it ranges from moderate to strong, with statistical significance (reported too).

\subsection{G-Eval reasoning}
\label{app:gevalexamples}

We show a few examples of reasons for both high and low scores for each of our metrics in Table~\ref{tab:gevalexamples}. 

\subsection{Visualized impact statements}
\label{app:impactexamples}

\begin{figure*}
\centering
\includegraphics[width=\textwidth]{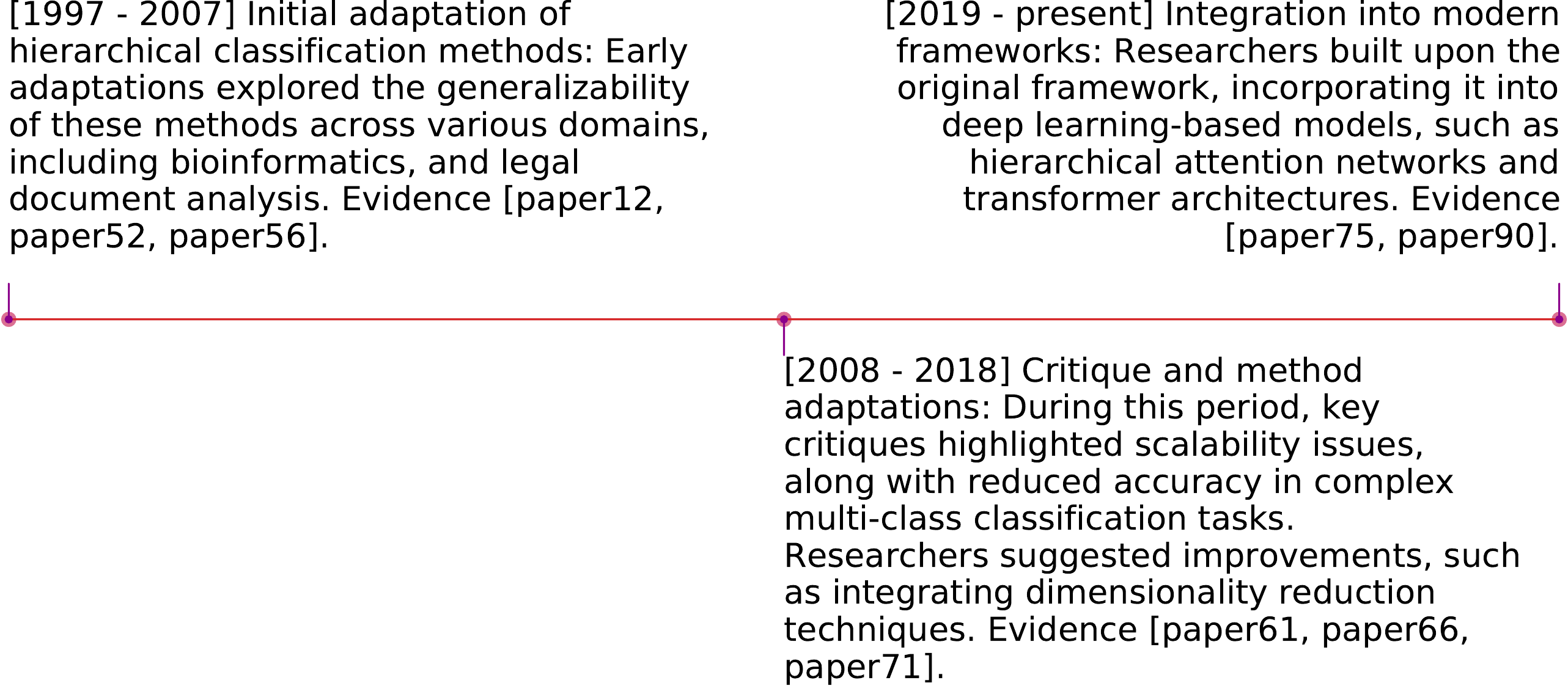}
  \caption{Impact of ``Hierarchically Classifying Documents Using Very Few Words'', published at ICML'97. (Full example from Figure~\ref{fig:motivation})}
  \label{fig:motivation2}
\end{figure*}

\begin{figure*}
\centering
\includegraphics[width=\textwidth]{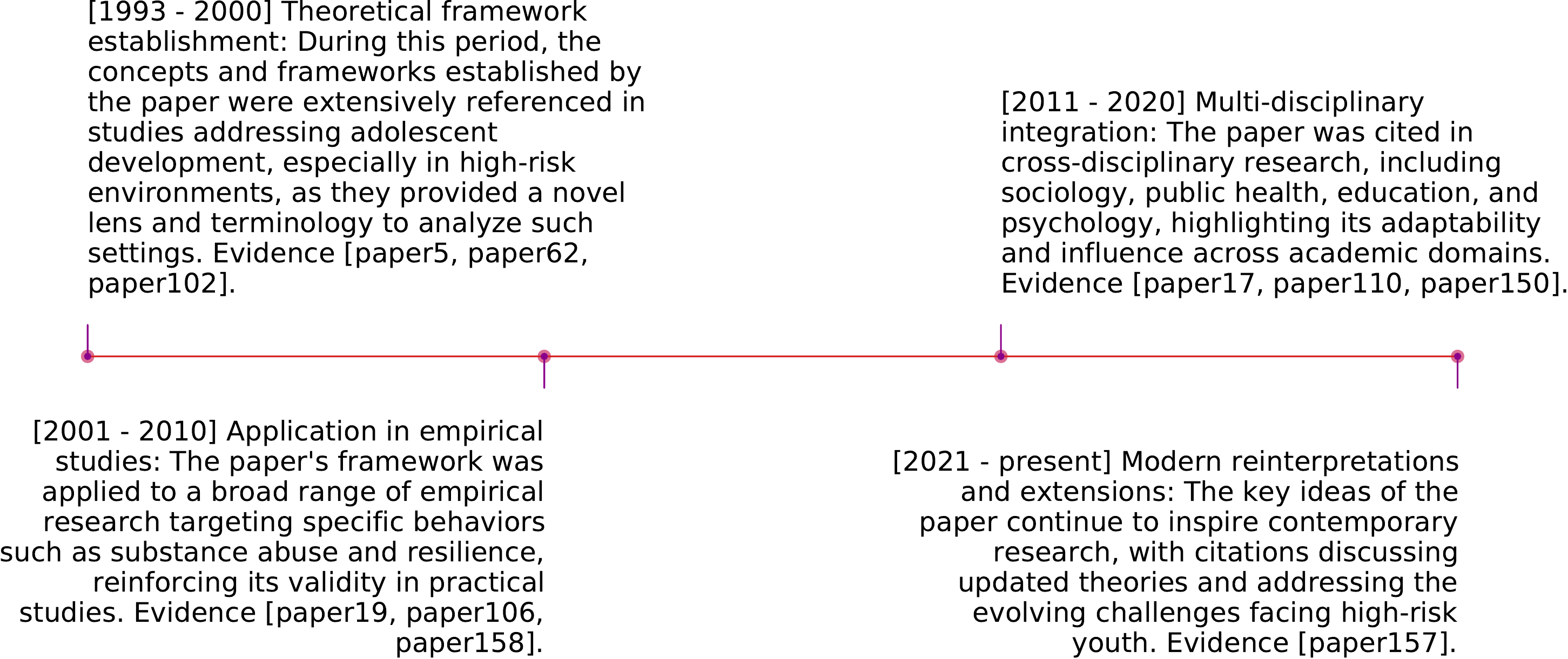}
  \caption{Impact of ``Successful Adolescent Development among Youth in High-Risk Settings'', published in American Psychologist'93.}
  \label{fig:psychologyexample1}
\end{figure*}

\begin{figure*}
\centering
\includegraphics[width=\textwidth]{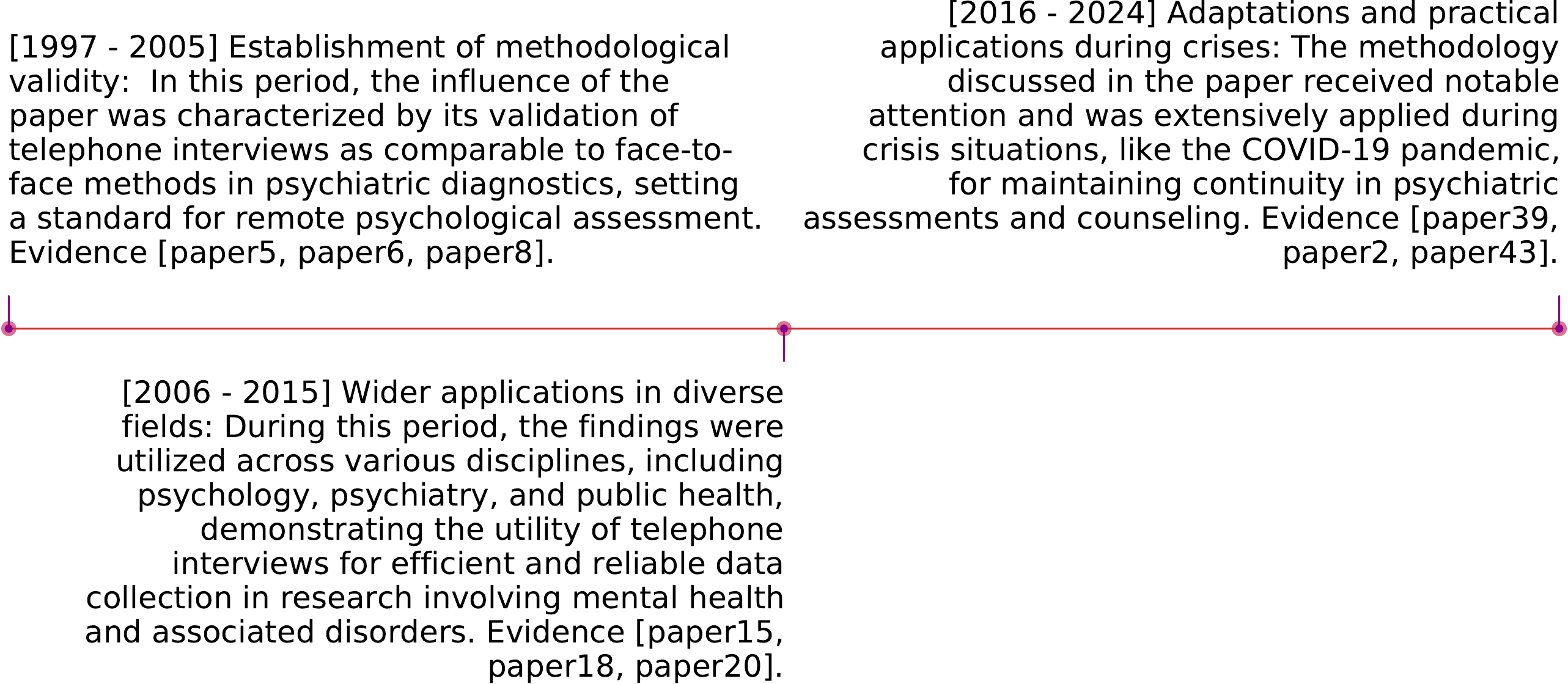}
  \caption{Impact of ``Comparability of telephone and face-to-face interviews in assessing axis I and II disorders'', published in The American Journal of Psychiatry'97.}
  \label{fig:psychologyexample2}
\end{figure*}

\begin{figure*}
\centering
\includegraphics[width=\textwidth]{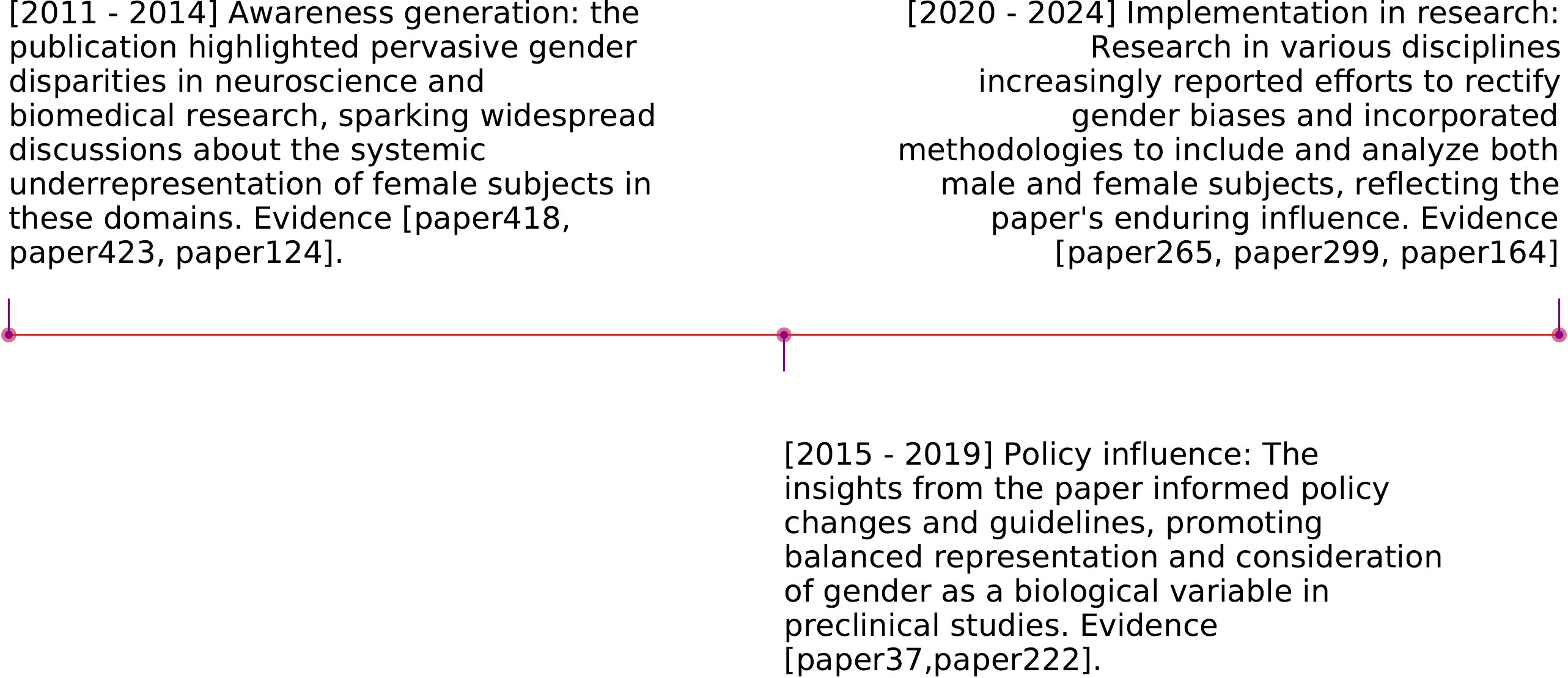}
  \caption{Impact of ``Sex Bias in Neuroscience and Biomedical Research'', published in Neuroscience \& Biobehavioral Reviews'11.}
  \label{fig:medicineexample}
\end{figure*}

Examples are shown in Figures~\ref{fig:motivation2} (computer science paper), ~\ref{fig:psychologyexample1} (psychology paper), ~\ref{fig:psychologyexample2} (psychology paper), and ~\ref{fig:medicineexample} (medicine paper). These impact summaries align closely with our definition. They capture the evolution of citations intents, from initial adoption and theoretical contributions to critiques, methodological refinements, policy influences, and integration into modern applications. For example, the impact summary of ``\textit{Hierarchically Classifying Documents Using Very Few Words}'' (ICML’97) illustrates how the paper initially shaped the research paradigm in document classification across fields like bio-informatics and legal document analysis. Later, it faced critiques related to scalability and multi-class accuracy, leading to refining the existing method. Another example is the impact summary of ``\textit{Sex Bias in Neuroscience and Biomedical Research}'' (Neuroscience \& Biobehavioral Reviews’11), which highlights the early discussions on gender disparities that this paper sparked. Those discussions later led to policy changes, promoting balanced gender representation in preclinical studies, influencing new research methods across various disciplines.

\subsection{Error analysis of our best variant}
\label{app:erroranalysis}

Although our intent classifier outperforms established baselines (Section~\ref{sec:intentexternalmethods}), misclassifications at the intent prediction stage can still compromise the quality of the generated impact summaries. For example, when an incidental citation context is mistakenly classified as impact-revealing, the resulting summary may include statements that do not reflect the true scholarly impact. For instance, in the following part of an impact summary, ``impact period: 2003-2010 \textbf{Initial Contextual Reference}: During this early period post-publication, the paper was often cited to provide comprehensive \textbf{background information} on the prevalence.. and risks associated with Alzheimer's and Parkinson's diseases...'', we can see that being mentioned for background information and context is listed under one of the impact periods. This is not in line with our definition of impact summaries, which only restrict impact to direct use of work. We inspect the reason for this mistake and track it back to a misclassification in the previous step, where intents such as ``providing context on the prevalence and significance of Parkinson's disease'' were classified as impact-revealing.

Conversely, when an impact-revealing citation is misclassified as incidental, the summary is deprived of key evidence of impact. For instance, we observed that a statement in the impact summary describing the use of a graph neural network methodology for drug discovery disappears when the related citation contexts, those referencing this methodological application, are removed from the input.

These cases illustrate the sensitivity of the generation step to errors in intent classification, underscoring the importance of continued research in this area. We hope our work lays the groundwork for future studies aimed at improving intent classification as a critical component of generating accurate and informative impact summaries.

\subsection{Generating impact summaries using other LLMs}
\label{app:otherllms}
We selected GPT-4o as our primary language model for this task due to its state-of-the-art performance in long-context reasoning and summarization. While we recognize the potential biases that may arise from relying solely on a single model family, the use of a consistent LLM and evaluation framework enables us to more clearly isolate the effects of input design. To evaluate the robustness of our method across model families, we also generate impact summaries using Qwen-2.5-72B~\cite{yang2025qwen3} and Gemini-2.5-flash~\cite{comanici2025gemini}, in addition to GPT-4o.

Quantitative results are presented in Table~\ref{tab:ablation_results_llms}. Interestingly, the baseline variant using Qwen achieves the highest insightfulness score. However, this comes at the cost of faithfulness, with the baseline producing 18\% fewer faithful claims than our best-performing variant. In contrast, Gemini demonstrates strong performance in citation year compliance, particularly with the impact-revealing with intents variant. Gemini tends to include numerous citations to substantiate its claims during impact periods and accurately aligns each citation with its corresponding time period. This variant also achieves the highest faithfulness (0.96, tied with all citations with intent variant), coverage (0.58), and specificity (0.83).

We present example summaries generated by both Qwen and Gemini in Tables~\ref{tab:otherllms_qualitative1} and ~\ref{tab:otherllms_qualitative2}.

\begin{table*}
\centering
\begin{tabular}{|l c|c|c|c|c|c|c|c|}
\hline
\rowcolor{gray!15}  \multicolumn{2}{|c|}{\textbf{Prompt variant}} &\multicolumn{3}{c|}{\textbf{Trustworthiness}} &\multicolumn{3}{c|}{\textbf{Informativeness}} \\
\hline
\rowcolor{gray!15} \textbf{Citations} & \textbf{Intents} & \textbf{Faith.} & \textbf{Cov.}  &\textbf{Cyc.}  & \textbf{Insi.} & \textbf{Trend.} & \textbf{Spec.}  \\ 
\hline
\rowcolor{gray!10} \multicolumn{8}{|c|}{\textit{Qwen}}\\
\hline
None & \xmark & 0.70 & 0.42 & n/a  & \textbf{0.73} & \textbf{0.84}& 0.76 \\
All & \xmark & 0.86 &  0.44&  \textbf{0.46}  &0.67 & 0.81& \textbf{0.78}\\
All & \checkmark & 0.86 & 0.44&  0.44 &0.66&0.80 &\textbf{0.78}\\
Impact-rev. & \xmark & 0.86 &  0.44& 0.43 &0.70 & 0.81& 0.77\\
\rowcolor{gray!15} Impact-rev. & \checkmark & \textbf{0.88}  & \textbf{0.45}& 0.42 &0.68 & 0.81& 0.76\\
\hline
\rowcolor{gray!10} \multicolumn{8}{|c|}{\textit{Gemini}}\\
\hline
None & \xmark & 0.91 & 0.44 &  n/a & 0.68& \textbf{0.84} & 0.82\\
All & \xmark & 0.95 & 0.52 &  0.85  & 0.70& 0.83&\textbf{0.83}\\
All & \checkmark & \textbf{0.96} & 0.57&  0.91 & \textbf{0.72}&\textbf{0.84}&0.79\\
Impact-rev. & \xmark & 0.93 &0.54  & 0.90 &0.70 &\textbf{0.84}&0.80\\
\rowcolor{gray!15} Impact-rev. & \checkmark & \textbf{0.96} &\textbf{0.58} & \textbf{0.93} &0.68 &0.83&\textbf{0.83}\\
\hline
\end{tabular}
\caption{Ablation results for Qwen and Gemini - Faithfulness: \textbf{Faith.}, Coverage: \textbf{Cov.}, Citation Year Compliance: \textbf{Cyc.}, Specificity: \textbf{Spec.}, Insightfulness: \textbf{Insi.}, Trend Awareness: \textbf{Trend.} These results show that adding impact-revealing contexts and their intents has an improvement on almost all metrics.}
\label{tab:ablation_results_llms}
\end{table*}

\begin{table*}
  \centering
  \begin{tabular}{|p{8cm}|p{8cm}|}
  \hline 
  \multicolumn{2}{|c|}{Paper: Hierarchically Classifying Documents Using Very Few Words, ICML, 1997.}\\
    \hline
  \textbf{\cellcolor{gray!15}Qwen}           & \textbf{\cellcolor{gray!15}Gemini}  \\
    \hline
\textbf{Impact period: 1998-2004} \textit{Initial Adoption and Methodological Foundations:} The paper laid the groundwork for hierarchical text classification by introducing a method that effectively classifies documents using very few words.. The initial adoption of this methodology was seen in several studies that built upon or compared their methods against the Koller and Sahami model. For example, the work by TaxE (2004) highlighted the challenge of comparing different hierarchical classification approaches due to the lack of a common dataset, underscoring the need for standardized evaluation metrics. Additionally, early works in 1998-2000 focused on extending and validating the use of hierarchical structures in text classification.\newline \textbf{Impact period: 2005-2010} \textit{Extension to Various Domains and Techniques:} During this period, the methodologies proposed in the paper were extended and applied to various domains, including medical vocabularies and web taxonomies. The work by Koller and Sahami was influential in demonstrating the effectiveness of hierarchical classification, which led to the development of new techniques such as the use of word embeddings and unsupervised learning. Research during this period also explored the integration of hierarchical structures with other machine learning algorithms like SVMs and neural networks. For instance, a 2007 study by Chakrabarti et al. drew parallels with Koller and Sahami's approach to attribute selection, while another paper in 2011 critiqued existing feature selection methodologies, suggesting improvements for hierarchical classification.\newline \textbf{Impact period: 2011-2020} \textit{Advanced Techniques and Real-World Applications:} In this period, the paper's influence continued to grow as researchers developed more advanced techniques for hierarchical text classification. The work by Koller and Sahami served as a foundational reference for numerous studies that sought to improve the robustness and efficiency of classification models. Specifically, research in 2018 explored the use of word embeddings for hierarchical text classification, building on the initial idea of leveraging hierarchical structures. Additionally, real-world applications of hierarchical classification were demonstrated in areas such as protein function prediction and image classification. The impact of the paper was also seen in the development of testbeds like TaxE, which aimed to standardize the evaluation of hierarchical classifiers.
& \textbf{Impact period: 1998-2005} \textit{Pioneering Hierarchical Text Classification and Feature Selection:} The paper rapidly established itself as a seminal work in text classification. The paper was widely recognized for introducing and validating the use of class hierarchies to improve classification efficiency and accuracy over traditional flat models (e.g., [15, 16]). Researchers often cited it as the first proper study on hierarchical text categorization (e.g., [12]) and a primary reference for feature selection strategies in high-dimensional text data. However, early discussions also began to identify potential limitations, such as susceptibility to overfitting, the inability to recover from errors made high in the hierarchy, and the constraints of existing datasets for comprehensive evaluation (e.g., [12, 14]).\newline  \textbf{Impact period: 2006-2015} \textit{Methodological Evolution \& Critical Analysis:} This period solidified the paper's status as a foundational reference for hierarchical classification, particularly its top-down approach, often referred to as the pachinko machine model. Researchers continued to build upon and compare against its principles, acknowledging its contributions to computational efficiency through techniques like subtree pruning (e.g., [33, 86]). However, there was a growing critical analysis of its inherent limitations, most notably the blocking issue or error propagation. The paper served as a baseline against which new algorithms were compared (e.g., [28, 92]), and its limitations, such as difficulties scaling to very large web taxonomies, were often highlighted to motivate new research (e.g., [8, 29]). The paper's influence also extended to other domains like protein function prediction and image classification (e.g., [37, 56]). \newline  \textbf{Impact period: 2016-Present} \textit{Enduring Relevance \& Benchmark Status:} The paper maintains its standing as a seminal and foundational benchmark for hierarchical text classification. Its principles are still acknowledged, even as the field evolves with advanced techniques like deep learning (e.g., [38, 41]). Crucially, the paper's identified limitations, such as the propagation of errors in deep hierarchies, are frequently highlighted as motivations for current research to develop more robust and scalable solutions, particularly in areas like hierarchical multi-label classification and for handling large numbers of classes (e.g., [0, 10]).\\
\hline
  \end{tabular}
  \caption{\label{tab:otherllms_qualitative1}
  Examples of impact summaries generated by Qwen and Gemini - Part 1.}
\end{table*}

\begin{table*}
  \centering
  \begin{tabular}{|p{8cm}|p{8cm}|}
  \hline 
  \multicolumn{2}{|c|}{Paper: Comparability of telephone and face-to-face interviews in assessing axis I and II disorders,}\\
  \multicolumn{2}{|c|}{The American Journal of Psychiatry, 1997.}\\
    \hline
  \textbf{\cellcolor{gray!15}Qwen}           & \textbf{\cellcolor{gray!15}Gemini}  \\
    \hline
\textbf{Impact period: 1997-2005} \textit{Validation of Telephone Assessments:} The study provided empirical evidence that telephone interviews could yield results comparable to face-to-face interviews, thereby expanding the methods available for clinical and epidemiological research. This validation was particularly important for large-scale studies and surveys where face-to-face interactions were logistically challenging or cost-prohibitive. The study's findings supported the reliability of tools such as the Personality Disorder Examination (PDE) and the Composite International Diagnostic Interview (CIDI) when administered via telephone, enhancing their utility in diverse settings.\newline \textbf{Impact period: 2006-2012} \textit{Equivalence and Reliability in Mental Health Research:} The influence of this paper extended into a broader validation of telephone assessments for various psychiatric disorders. Research during this period consistently cited the study's findings to support the equivalence and reliability of telephone interviews compared to face-to-face methods. This period saw the application of these validated methods in numerous mental health studies, including those on major depressive disorder, anxiety disorders, and substance abuse disorders. The equivalence of telephone assessments was particularly valuable in longitudinal studies and clinical trials, where maintaining consistent and high-quality data collection was essential. Tools such as the SCID (Structured Clinical Interview for DSM-IV) were commonly used in these applications.\newline \textbf{Impact period: 2013-2023} \textit{Practical Applications and Cost-Effectiveness:} The practical applications of telephone interviews in mental health research and clinical practice became more prominent. The validated equivalence of telephone assessments led to increased adoption in various settings, including telehealth services, crisis intervention, and routine mental health care. This period also emphasized the cost-effectiveness of telephone assessments, particularly in large-scale surveys and during global events like the COVID-19 pandemic. The practicality and efficiency of telephone interviews were highlighted, making them a preferred method for data collection, especially when face-to-face interactions were not feasible. 
& \textbf{Impact period: 1997-2007} \textit{Establishing Comparability and Validity of Telephone Interviews:} The paper provided crucial evidence demonstrating that telephone interviews yield comparable results to traditional face-to-face assessments for various disorders, including anxiety and major depressive disorders. Researchers frequently cited the paper to support the validity of using telephone methods for large-scale data collection in clinical and epidemiological studies. The paper also began to inform discussions on the practical advantages of telephone interviews, such as cost-efficiency, broader geographical coverage, and potential for higher response rates, while also highlighting the importance of interviewer-subject rapport and identifying early nuances like potential underreporting of sensitive information (e.g., substance abuse) and the comparability for specific diagnostic tools like the CIDI or SCID.\newline  \textbf{Impact period: 2008-2017} \textit{Broadened Acceptance and Practical Utility:} Over this period, the paper continued to be cited as authoritative evidence for the validity and reliability of telephone interviews, extending their application across a wider range of mental health conditions, including PTSD and eating disorders. Its influence shifted from merely establishing comparability to emphasizing the practical benefits of telephone methods, such as reduced participant burden, improved subject retention in longitudinal studies, and enhanced logistical efficiency for large-scale research. While broadly affirming the equivalence of telephone and face-to-face methods, citations during this time also reflected ongoing discussions about specific methodological considerations, such as the potential for differences in reporting stigmatizing behaviors or the role of nonverbal cues. \newline  \textbf{Impact period: 2018-2024} \textit{Enduring Relevance and Standard Practice:} In recent years, the paper's impact has solidified its position as a foundational reference for the routine use of telephone interviews in clinical assessment. It is cited as evidence for the established reliability and validity of telephone-based diagnostic methods, supporting their application as a standard practice for diagnosing various mental health conditions and even for monitoring adverse events. The citations highlight the practicality and effectiveness of telephone interviews, especially in contexts requiring reduced burden or remote data collection, such as during the COVID-19 pandemic.\\
\hline
  \end{tabular}
  \caption{{\label{tab:otherllms_qualitative2}
  Examples of impact summaries generated by Qwen and Gemini - Part 2.}}
\end{table*}

\section{Details on Model Usage}
\label{modelusage}

See Table~\ref{tab:modelusage}.

\begin{table*}
  \centering
  \begin{tabular}{|p{8cm}|c|p{5cm}|}
    \hline
    \textbf{\cellcolor{gray!15}Task/Component}             & \textbf{\cellcolor{gray!15}Model Name}   & \textbf{\cellcolor{gray!15}Details} \\
    \hline
intent classification and generation & gpt-4o-mini & temp=0, version=2024-07-18, Input: 128,000, Output: 16,384\\
\hline 
impact summary generation & gpt-4o & temp=0, version=2024-11-20, Input: 128,000, Output: 16,384\\
\hline 
impact summary evaluation  & gpt-4o & temp=0, version=2024-11-20, Input: 128,000, Output: 16,384\\
\hline 
textual pattern generation  (for groundtruth data extension)  & gpt-4o & temp=0, version=2024-11-20, Input: 128,000, Output: 16,384\\
 \hline
 practical applications (citation context analysis) & gpt-4o-mini & temp=0, version=2024-07-18, Input: 128,000, Output: 16,384\\
 \hline
 practical applications (author-level summary generation) & gpt-4o & temp=0, version=2024-11-20, Input: 128,000, Output: 16,384\\
 \hline
  \end{tabular}
  \caption{\label{tab:modelusage}
   Details about the LLMs used in our experiments.
  }
\end{table*}

\section{More on Practical Applications}
\label{app:practical}

\subsection{Equal citation count, different citing reasons}
\label{app:equalimpact}

We select 10 diverse research topics (e.g., LLMs for code generation, cultural sensitivity in clinical psychology), and analyze 3 papers within each. We ensure that the topic-specific papers have close citation counts. In this use case, we demonstrate how \textit{fine-grained} citation intent analysis provides a deeper understanding of research impact, moving beyond raw citation counts.   Results are in Tables~\ref{tab:unequal} and~\ref{tab:unequal2}. We observe that some papers' main impact is inspiring or motivating new work, some are used for their method or data, others are cited to point out remaining challenges and propose improvements, etc. For example, for papers on open-information extraction with citation counts of $\sim$ 200, ~\cite{cui2018neural} is frequently cited to motivate new work or highlight existing limitations, while ~\cite{gashteovski2017minie} and ~\cite{han2019opennre} are cited for their method. In commonsense knowledge mining, only one of the 3 papers is often cited for its dataset~\cite{hwang2021comet}, and even though the other 2 papers are cited to highlight limitations, the limitations are different, namely challenges with handling constraints and reasoning for one~\cite{davison2019commonsense} , but data noise and limited coverage for the other~\cite{speer2013conceptnet}. These examples show that even when citation counts are similar, citation intents reveal the specific citing reasons, shaping the perception of a paper’s true impact and assisting in the impact summary generation task. 

\subsection{Author-level impact summaries}
\label{app:authorlevel}
\begin{figure*}
\centering
\includegraphics[width=\textwidth]{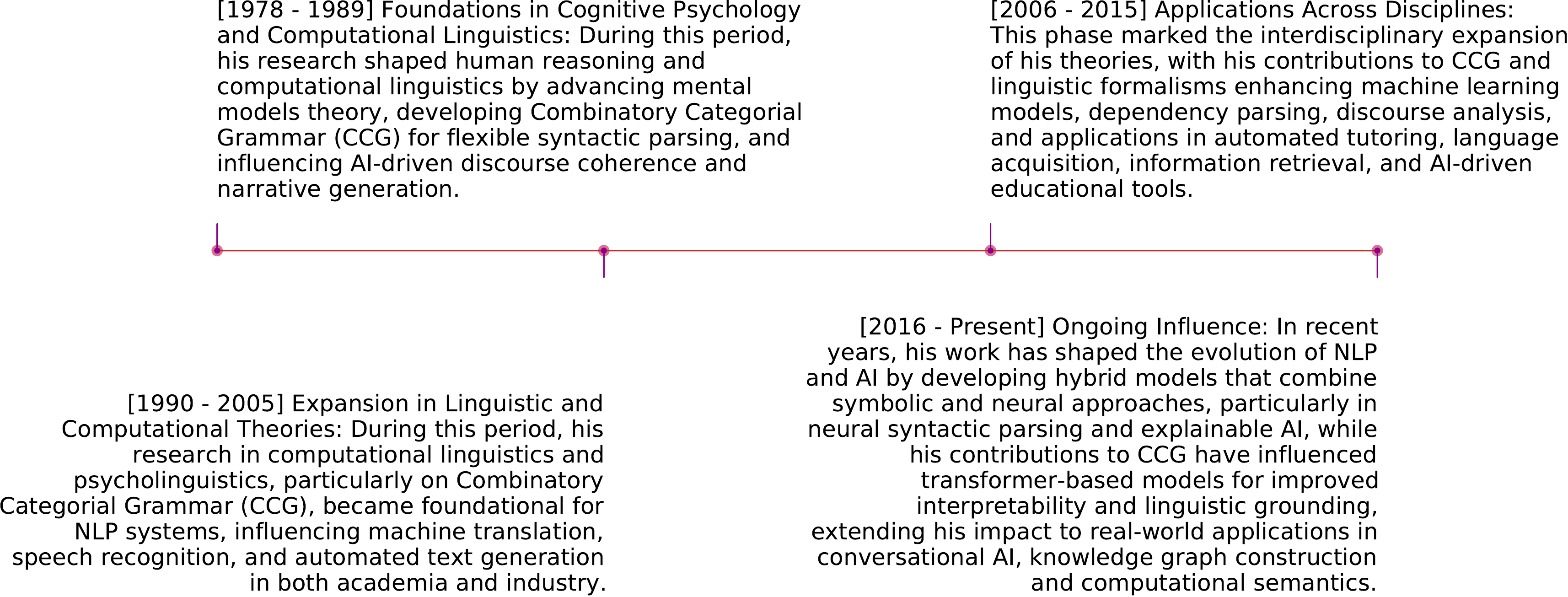}
  \caption{The impact summary of the work of a computer science researcher (focus on NLP, cognitive science), H-index=69.}
  \label{fig:mark}
\end{figure*}

\begin{figure*}
\centering
\includegraphics[width=0.9\textwidth]{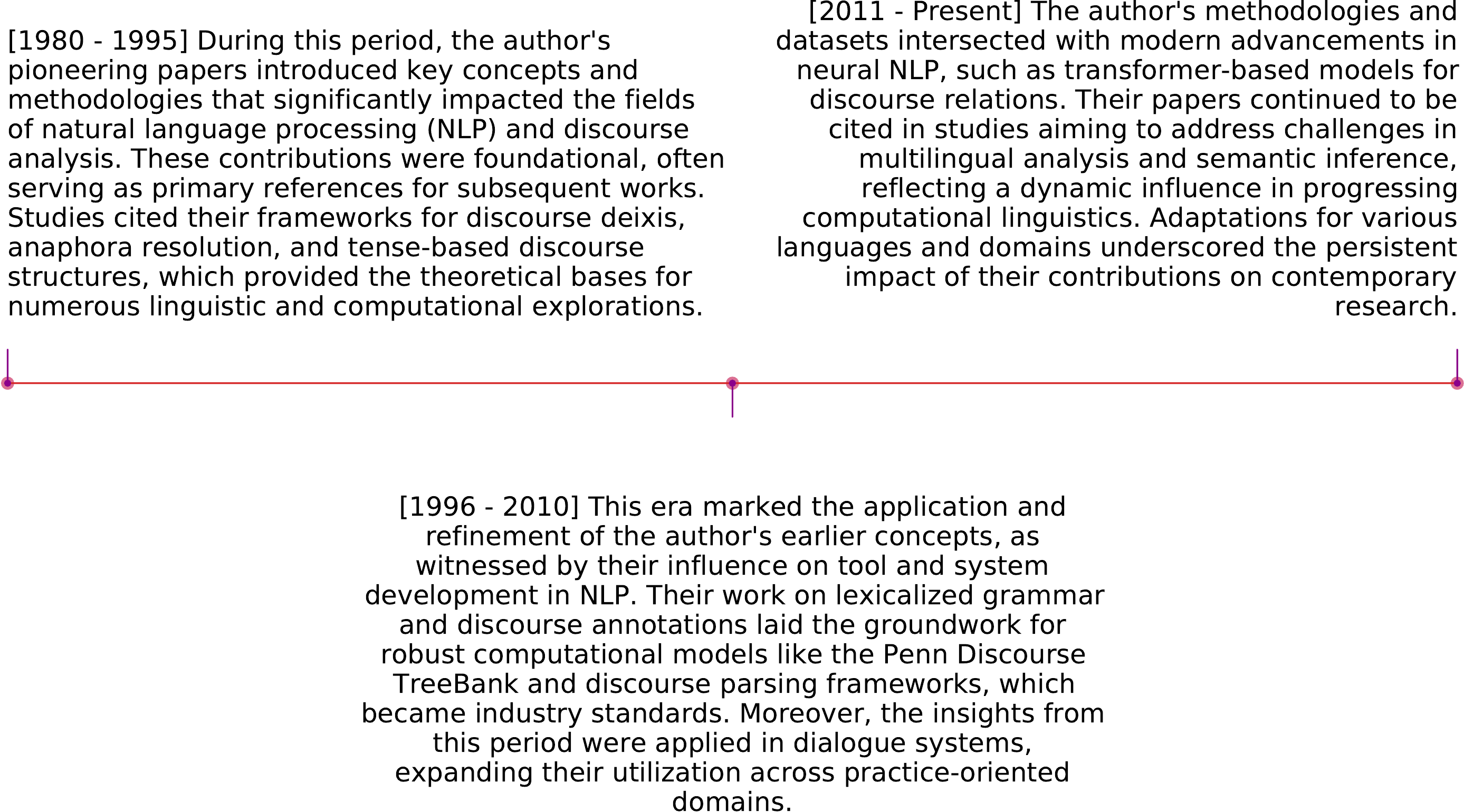}
  \caption{The impact summary of the work of a computer science researcher (focus on NLP, computational linguistics), H-index=61.}
  \label{fig:bonnie}
\end{figure*}

The focus of this paper is to generate impact summaries for individual papers. However, we would like to briefly show how these paper-level summaries can be used to create author-level summaries. Given the top-cited\footnote{Top-10 by citation counts.} papers of a given author, we generate impact summaries using our best variant, and then ask LLM to aggregate these summaries and infer the author’s overall impact.   The prompt used for generating author-level summaries in Figure~\ref{fig:prompt_author_impact}. We show two samples of such summaries for two senior NLP researchers in Figures~\ref{fig:mark} and ~\ref{fig:bonnie}. This application case hold great potential for future work, and could be extended to research labs, universities, venues, or countries. Methodologically, it would also be particularly interesting to explore different aggregating strategies to generate these summaries, more specifically whether these kind of impact summaries have better quality by aggregating existing paper-level summaries, or by directly use the citation context about an author's papers regardless of which paper they came from.

\begin{sidewaystable*}
  \centering
  \begin{tabular}{|p{7cm}|p{9cm}|p{9cm}|}
    \hline
  \textbf{\cellcolor{gray!15}Impact summary}           & \textbf{\cellcolor{gray!15} Analysis and model's verdict}  & \textbf{\cellcolor{gray!15}Proof (partial citations list for brevity)} \\ \hline

\textbf{Paper}: \textit{The importance of race and ethnic background in biomedical research and clinical practice}\newline

\textbf{[2011-2020] Expansion and debate}\newline
The discussion around the provided insights flourished, with broader interdisciplinary citations related to ethics, sociological interpretations, and medical trial designs highlighting its significance. &

The impact description discusses the flourishing discussion around insights related to race and ethnic background in biomedical research, emphasizing interdisciplinary citations that highlight its significance in ethics, sociological interpretations, and medical trial designs. Several citations in the provided list address the importance of race and ethnicity in biomedical research and clinical practice, indicating that these discussions are indeed significant and supported by various studies. \newline

\textbf{Verdict}: \textit{Verifiable} &

\textbf{<Associations Between Vitamin D Receptor Polymorphisms and Susceptibility to Ulcerative Colitis and Crohn's Disease: A Meta-analysis>}: "The importance of ethnic background has been raised in bio-medical research and clinical practice.(35) Genetic variation that predisposes to IBD appears to vary between different ethnic groups."\newline
\textbf{<Conceptual approaches to the study of health disparities.>}: "Some researchers have argued that this correlation between genetically identified classifications and self reported race justifies the continued use of self-reported race and ethnicity as a proxy for genetic differences in epidemiologic studies (11, 71)."

\\\hline
\textbf{Paper}: \textit{Myocardial infarction redefined--a consensus document of The Joint European Society of Cardiology/American College of Cardiology Committee for the redefinition of myocardial infarction.} \newline \newline

\textbf{[2006 - 2015] Integration into clinical practice guidelines} \newline
During this period, the criteria discussed in the input paper were incorporated into widely used clinical guidelines, affirming its essential role in healthcare practice and education.

&

The impact description states that the criteria discussed in the paper were incorporated into widely used clinical guidelines, affirming its essential role in healthcare practice and education. The citations provided indicate that the Joint European Society of Cardiology/American College of Cardiology criteria for diagnosing myocardial infarction have been referenced and utilized in various studies and guidelines, confirming their integration into clinical practice. This supports the claim that the criteria have been widely adopted in healthcare settings. \newline

\textbf{Verdict}: \textit{Verifiable} 
&
\textbf{<Trace element status in Saudi patients with established atherosclerosis.>}: "A diagnosis of a myocardial infarction was made in accordance with Joint European Society of Cardiology/American College of Cardiology Committee criteria [17]." \newline
\textbf{<Improved treatment and prognosis after acute myocardial infarction in Estonia: cross-sectional study from a high risk country>}: "The criteria applied for AMI diagnosis on 2001 and 2007 study populations were based on the consensus document published by the European Society of Cardiology in 2000 [12]."
\\ \hline
  \end{tabular}
  \caption{Examples of verified (faithful) impact summaries impact descriptions.}\label{tab:faithfulness_analysis}
\end{sidewaystable*}

\begin{table*}
  \centering
  \begin{tabular}{|p{9cm}|p{6cm}|}
    \hline
  \textbf{\cellcolor{gray!15}Impact summary}           & \textbf{\cellcolor{gray!15}Themes to be covered}  \\
    \hline
During this early period following the publication, the method proposed by the paper, such as the Spy technique, received attention by its developers who utilized it in the domain of semi-supervised learning. Researchers particularly explored its applications in areas like text document \hl{classification} and \hl{highlighting limitations} in existing learning paradigms. The study's methodology spurred advancements in \hl{classifications approaches} \hl{dealing with positive and unlabeled data}. This phase not only saw \hl{adaptation of the methodology into various applied fields} like web document classification but also \hl{critical analysis} and theoretical exploration particularly in \hl{precision improvement}. Researchers evolved the \hl{PU learning methods} by optimizing the constraints and extending it into more extensive datasets and alternative models leading towards scaling and efficiency enhancements \hl{in text and online information domains}. In this recent phase, the Spy technique from the discussed paper continues to be influential, \hl{inspiring modern machine learning techniques} like \hl{adaptive PU-learning} and text mining approaches. The bridging of neural methods with earlier methodologies showcases the adaptability and robustness over time. 
&Performance Comparison\newline
\hl{Novel Algorithm Development}\newline \hl{Limitations Acknowledgment}\newline \hl{Method Adaptation and Improvement}\newline \hl{Applications of PU Learning}\newline \hl{Classification Techniques}\newline \hl{Data Handling and Sampling}\\
\hline
During these years, the paper served as a cornerstone for advancing the implementation of \hl{individualized approaches in clinical decision-making}. Several studies cited the paper while exploring personalized medicine's integration. Focus on translational research and advances. This period saw the application of concepts introduced in the paper in \hl{fine-tuning medical treatments} and drug therapies \hl{for individual patients}. The references highlight \hl{adaptations in various domains from rheumatology to oncology}. The paper's concepts shaped the \hl{research direction towards the future application of precision medicine} utilizing technologies like AI and computational advances. &  \hl{Advancements in Precision Medicine}\newline Role of Genetic Variants and Pharmacogenomics\newline \hl{Technological Impact on Personalized Medicine}\newline Challenges and Gaps in Personalized Medicine\newline  \hl{Patient Variability and Individualized Treatment}\newline \hl{Applications of Personalized Medicine in Specific Fields}\newline Ethical and Regulatory Considerations\\
\hline
The input paper, published in 2020, provided a significant synthesis of knowledge about the \hl{public health importance of hookworm infections}. & 
Effectiveness of Treatment for Anemia\newline
\hl{Public Health Significance of Hookworm Disease}\newline
Limitations of Targeting School-Age Children\newline
Need for Studies on Military Infection Burdens\newline
..140 more..\\
\hline
  \end{tabular}
  \caption{\label{tab:coverageexamples}
  Examples of automated coverage results (\hl{themes actually covered}). }
\end{table*}

\begin{table*}
  \centering
  \begin{tabular}{|l|c|c|c|c|c|}
    \hhline{~|-|-|-|-|-|}
             \multicolumn{1}{c|}{} & \textbf{\cellcolor{gray!15}Faithfulness} & \textbf{\cellcolor{gray!15}Coverage} & \textbf{\cellcolor{gray!15}Insightfulness}  & \textbf{\cellcolor{gray!15}Trend Awareness} & \textbf{\cellcolor{gray!15}Specificity}  \\
    \hline
    Spearman & 0.668* & 0.481**  & 0.489* &	0.535* &	0.683*   \\
         \hline 
            Kendall-Tau & 0.668* & 0.443**	& 0.482* &	0.533*	& 0.676*   \\
\hline 
 
  \end{tabular}
  \caption{\label{tab:humancorrelation}
   Human and LLM correlation and agreement on evaluating results (Section~\ref{sec:ablation}). p-value: *$\le$0.001, **$\le$0.05.
  }
\end{table*}

\begin{sidewaystable*}
  \centering
  \begin{tabular}{|p{7cm}|p{9cm}|p{9cm}|}
    \hline
  \textbf{\cellcolor{gray!15}Impact summary}           & \textbf{\cellcolor{gray!15} Complying citations}  & \textbf{\cellcolor{gray!15}Noncomplying citations} \\ \hline
  \textbf{[2021 - 2022] supporting \hl{economic and societal discussions} of cardiovascular diseases} \newline\newline 
Studies during this period heavily relied on the paper for understanding the economic burden and healthcare strategies towards cardiovascular diseases, often referencing the comprehensive statistics presented. &

  \textbf{Title}: "Extrinsically Conductive Nanomaterials for Cardiac Tissue Engineering Applications",\newline
  \textbf{Year}: \textbf{2021},\newline
  \textbf{Intent}: "highlighting the significant economic impact of cardiovascular diseases"\newline &
  
  \textbf{Title}: "Optimising the treatment of chronic ischemic heart disease by training general practitioners to deliver very brief advice on physical activity (OptiCor): protocol of the systematic development and evaluation of a complex intervention", \newline
  \textbf{Year}: \textbf{2024}, \newline
  \textbf{Intent}: "highlighting \hl{the significant economic impact} of cardiovascular events on the German healthcare system" \\
  \hline

  \textbf{[2021 - 2023] guiding therapeutic practices and further clinical recommendation development}\newline \newline
  
  Multiple citations within this period reflect that the publication served as a cornerstone source for \hl{clinical trials and therapeutic evaluation studies aiming to manage PAD effectively.} &

  \textbf{Title}: "Clinical considerations after endovascular therapy of peripheral artery disease",\newline
  \textbf{Year}: \textbf{2021}, \newline
  \textbf{Intent}": "highlighting the significance of managing PAD as a CAD equivalent" &

    \textbf{Title}: "Ankle Brachial Index: An Easy and First-Choice Screening Marker of Peripheral Artery Disease and Physical Function",\newline
  \textbf{Year}: \textbf{2024}, \newline
  \textbf{Intent}: "\hl{highlighting the significance of ABI in identifying PAD and associated risks}"\\
  \hline

  \textbf{[2004 - 2010] Initial \hl{adoption} of stochastic gradient descent methods} \newline \newline
After its publication in 2004, the paper introduced stochastic gradient descent (SGD) as an efficient method for solving large-scale linear prediction problems, enabling significant computational advantages in machine learning applications. &

  \textbf{Title}: "Fast Stochastic Frank-Wolfe Algorithms for Nonlinear SVMs", \newline
  \textbf{Year}: \textbf{2010}, \newline
  \textbf{Intent}: "reporting efficiency of stochastic programming techniques" \newline
  &
 \textbf{Title}: "A Robbins-Monro Sequence That Can Exploit Prior Information For Faster Convergence", \newline
  \textbf{Year}: \textbf{2024}, \newline
  \textbf{Intent}: "highlighting the \hl{widespread application} and significance of a widely used algorithm" \\
  \hline

  \end{tabular}
  \caption{Examples of evidence citations from compliant and noncompliant years. Note the similarity between the summary impact description and the intent of noncompliant citations (\hl{highlighted}), which might be the source of the error.}\label{tab:citationsvalidityanalysis}
\end{sidewaystable*}

\begin{sidewaystable*}
  \centering
  \begin{tabular}{|p{5cm}|p{10cm}|p{10cm}|}
    \hline
  \textbf{\cellcolor{gray!15}Paper}           & \textbf{\cellcolor{gray!15}Baseline}  & \textbf{\cellcolor{gray!15}Ours} \\
    \hline
Liu, Bing, et al. Partially Supervised Classification of Text Documents.  2002.
& 
 ..\textbf{[2005-2012]: Development and application}\newline
 \textit{The approaches defined were adapted and extended in a number of practical applications to fine-tune frameworks for domain-specific text classification tasks} ..&  ..\textbf{[2008-2012]: Applications \& theoretical exploration}\newline
 \textit{The methods were applied to problems beyond text classification, such as bioinformatics, personalized recommendations, and data mining. This phase not only saw adaptation of the methodology but also critical analysis and theoretical exploration particularly in precision improvement} ..\\
         \hline 
         Martens, James. Deep learning via Hessian-free optimization. 2010.
         &
   ..\textbf{[2015-2020]: Adoption in new areas}\newline
 \textit{The techniques described in the paper facilitated advancements in optimizing models in diverse fields like natural language processing and reinforcement learning} ..&  ..\textbf{[2016-2020]: Integrating into broader optimization frameworks and comparisons}\newline
 \textit{As the field evolved, the approaches from this work became integrated into broader optimization discussions and comparison studies. Researchers began leveraging data-driven adaptations and contextualizing the paper's contributions relative to emerging advancements like Adam and other methods, enriching overall optimization frameworks} ..\\
         \hline 
          Lee, Honglak, et al. Convolutional deep belief networks for scalable unsupervised learning of hierarchical representations.  2009.
         &
   ..\textbf{[2013-2017]: Facilitating Advances in Image Recognition}\newline
 \textit{Techniques and methodologies from the paper influenced deep learning architectures, becoming foundational in computer vision applications} ..&  ..\textbf{[2012-2017]: Advancements in convolutional neural network methodologies}\newline
 \textit{This period marks significant enhancements in deep convolutional techniques, where the paper provided foundational principles that inspired advancements in methodologies across image processing, feature extraction, and specific application areas such as bioinformatics and health diagnostics} ..\\
  \hline 
          Myin-Germeys, Inez, et al. Experience sampling research in psychopathology: opening the black box of daily life. 2009.
         &
   ..\textbf{[2015-2023]: Application in diverse psychopathological studies}\newline
 \textit{The methodologies introduced by the paper were applied extensively in studies of mood disorders and other psychopathological conditions to dissect complex behavioral and cognitive interactions} ..&  ..\textbf{[2015-2019]: Expanding applications}\newline
 \textit{The method presented catalyzed its application across diverse practices such as genetic interaction studies, psychological intervention assessments, and computational behavioral analysis. Researchers recognized its role in enhancing ecological validity and precision of dynamic mental state evaluations}\newline\textbf{[2019-Present]: Technological Integration}\newline
 \textit{This period saw a surge in mobile and wearable technologies for real-time data collection, integrating ESM with machine learning and mobile health platforms to advance personalized medicine and therapeutic interventions.} ..\\
         \hline
  \end{tabular}
  \caption{\label{tab:impactsummariesinformativness}
  Examples of impact summaries from the informativeness evaluation, baseline = no-knowledge variant, our method = impact-revealing variants.}
\end{sidewaystable*}

\begin{table*}
\centering
\begin{tabular}{|l c|c|c|c|c|c|c|}
\hline
\rowcolor{gray!15}  \multicolumn{2}{|c|}{\textbf{Prompt variant}} &\multicolumn{3}{c|}{\textbf{Trustworthiness}} &\multicolumn{3}{c|}{\textbf{Informativeness}}  \\
\hline
\rowcolor{gray!15} \textbf{Citations} & \textbf{Intents} & \textbf{Faith.} & \textbf{Cov.}  &\textbf{Cyc.}  & \textbf{Insi.} & \textbf{Trend.} & \textbf{Spec.} \\ 
\hline
\rowcolor{gray!10} \multicolumn{8}{|c|}{\textit{Medicine}}\\
\hline
None & \xmark & 0.70 & 0.23& n/a  &0.75& 0.93 & 0.82\\
All & \xmark & 0.80 & 0.30& 0.52 &0.82&0.95&0.87\\
All & \checkmark & 0.82 &  0.30& 0.47 &0.83 &0.94&0.88\\
Impact-rev. & \xmark & \textbf{0.85}  & 0.31&  \textbf{0.57} & 0.82&0.95&0.86\\
\rowcolor{gray!15} Impact-rev. & \checkmark & \textbf{0.85} & \textbf{0.32}&  0.56&\textbf{0.87}&\textbf{0.96}&\textbf{0.89}\\
\hline
\rowcolor{gray!10} \multicolumn{8}{|c|}{\textit{Psychology}}\\
\hline
None & \xmark & 0.74 & 0.23& n/a  &0.75& 0.95 & 0.80\\
All & \xmark & 0.86 & 0.32& 0.54 &\textbf{0.84}&\textbf{0.98}&\textbf{0.89}\\
All & \checkmark & 0.86 &  0.32 & 0.52 &0.79 &0.94&0.86\\
Impact-rev. & \xmark & \textbf{0.90}  & \textbf{0.34} & \textbf{0.63} & 0.82&0.97&\textbf{0.89}\\
\rowcolor{gray!15} Impact-rev. & \checkmark & 0.89 & 0.32& 0.59 &0.83&\textbf{0.98}&\textbf{0.89}\\
\hline
\rowcolor{gray!10} \multicolumn{8}{|c|}{\textit{Computer science}}\\
\hline
None & \xmark & 0.87 & 0.30&n/a  &0.60& 0.94 & 0.63\\
All & \xmark & 0.83 & 0.35&\textbf{0.59} &0.75&0.96&0.83\\
All & \checkmark & 0.84 &  0.36&0.46 &0.77 &0.98&0.82\\
Impact-rev. & \xmark & 0.85  & 0.35& 0.58 & 0.74&0.95&0.86\\
\rowcolor{gray!15} Impact-rev. & \checkmark & \textbf{0.89} & \textbf{0.37}& 0.52 &\textbf{0.79}&\textbf{0.99}&\textbf{0.86}\\
\hline
\end{tabular}
\caption{Ablation results per field - Faithfulness: \textbf{Faith.}, Coverage: \textbf{Cov.}, Citation Year Compliance: \textbf{Cyc.}, Specificity: \textbf{Spec.}, Insightfulness: \textbf{Insi.}, Trend Awareness: \textbf{Trend.}}
\label{table:ablation_results_detailed}
\end{table*}

\begin{table*}
  \centering
  \begin{tabular}{|l|l|p{10cm}|}
    \hline
  \textbf{\cellcolor{gray!15}Criteria}           & \textbf{\cellcolor{gray!15}Score}  & \textbf{\cellcolor{gray!15}Reason} \\
    \hline
Insightfulness
& 0.9 & The output provides a clear description of the paper's influence, detailing its foundational, methodological, and expanded relevance. It highlights specific uses and integration into modern frameworks. Slightly more detailed examples could enhance insight further.\\
         \hline 
         Insightfulness
& 0.4 & While the output describes stages of adoption and application, it lacks detailed insight into specific ways the paper directly influenced or was used by other works.\\
         \hline 
         Trend Awareness & 1.0 & The output clearly identifies how the paper’s impact has changed over time with descriptive period titles. These titles are diverse and informative, fulfilling the evaluation criteria.\\
         \hline
         Trend Awareness & 0.4 & The output mentions distinct time periods with descriptions of the paper's impact, but the titles for the periods are not diverse, as both cover similar themes without clear differentiation.\\
         \hline 
         Specificity & 0.9 & The output provides detailed periods showcasing the paper's influence on foundational discussions, specific applications, and technological progress, meeting most evaluation criteria. A minor deduction is due to the lack of explicit mention of individual techniques influenced by the paper.\\
         \hline
                  Specificity & 0.3 & The output generically states the paper's influence and provides broad areas of impact, but it lacks concrete details, specific techniques, studies, or frameworks influenced by the paper.\\
\hline
  \end{tabular}
  \caption{\label{tab:gevalexamples}
  Examples of scores and reasons from the LLM evaluator. }
\end{table*}

\begin{table*}
  \centering
  \begin{tabular}{|p{4.5cm}|p{4.5cm}|p{4.5cm}|}
    \hline
\multicolumn{3}{|c|}{\cellcolor{gray!15}\textbf{\texttt{Topic:} Open information extraction; citation count: $\sim$200}}   \\
    \hline
            \textbf{\texttt{Paper:}} \cite{cui2018neural} & \cite{gashteovski2017minie} &  \cite{han2019opennre}    \\
         \textbf{\texttt{Top citation intents:}}  motivating new work in neural OIE, acknowledging limitations in generating facts and schema strength & comparing proposed methods with sota approach, method use for OIE&  method use and extension for relation extraction, highlighting performance issues in capturing various linguistic cues\\
 \hline
\multicolumn{3}{|c|}{\cellcolor{gray!15}\textbf{\texttt{Topic:} Commonsense knowledge mining; citation count: $\sim$350}}   \\
    \hline
            \cite{davison2019commonsense} & \cite{speer2013conceptnet} &  \cite{hwang2021comet}    \\
 acknowledging limitations in handling constraints and reasoning, promoting the perspective of LLMs as knowledge base &  highlighting limitations of data noise and limited relation coverage, emphasizing importance of commonsense knowledge graphs in AI applications   & reporting performance comparison and improvements,  use as data source for commonsense reasoning \\
\hline
\multicolumn{3}{|c|}{\cellcolor{gray!15}\textbf{\texttt{Topic:} Retrieval augmented generation; citation count: $\sim$[400-500]}}   \\
    \hline
            \cite{asai2023self} & \cite{ram2023context} &  \cite{mallen2023not}    \\
motivating the effectiveness of RAG in addressing hallucinations in LLMs, reporting variability in LLMs' performance with RAG methods &  motivating the need for well-chosen context in RAG, reporting limitations of retrievers in RAG & motivating research on LLMs for question answering, motivating the goal of improving LLMs' factuality\\
\hline
\multicolumn{3}{|c|}{\cellcolor{gray!15}\textbf{\texttt{Topic:} LLM as a judge; citation count:  $\sim$[370-470]}}   \\
    \hline
            \cite{fu2024gptscore} & \cite{chanchateval} &  \cite{wang2024large}    \\
motivating the growing interest in using LLMs for automatic evaluations, used on evaluating conversational agents and summary faithfulness &  highlighting the growing trend of using LLMs for evaluation, use of agent-based methods in evaluation processes & motivating the need for further investigation into biases in Judge LLMs, highlighting concerns regarding fairness and bias in using LLMs for evaluation\\
\hline
\multicolumn{3}{|c|}{\cellcolor{gray!15}\textbf{\texttt{Topic:} Medical domain LLMs; citation count:  $\sim$[1900-2100]}}   \\
    \hline
            \cite{gu2021domain} & \cite{singhal2023large} &  \cite{thirunavukarasu2023large}    \\
use of domain-specific pre-trained models, comparing performance of various transformer models in biomedical tasks &  highlighting a gap in existing research on interactive medical services, performance comparisons of LLMs in medical knowledge tasks & highlighting privacy concerns in data collection, performance comparison between LLMs and fine-tuned small models\\
\hline
\multicolumn{3}{|c|}{\cellcolor{gray!15}\textbf{\texttt{Topic:} Cultural sensitivity in clinical psychology; citation count:  $\sim$[1300-1500]}}   \\
    \hline
            \cite{bernal2009cultural} & \cite{sue2001multidimensional} &  \cite{sue1998search}    \\
highlighting the ongoing debate regarding cultural competency in treatment approaches, motivating the need for comparative analysis of culturally adapted evidence-based practices &  highlighting limitations in existing multicultural discourse research, motivating the application of cultural competency models in transgender care
 & motivating work on cultural sensitivity in therapy, highlighting the gap in research on treatment efficacy for diverse ethnic populations\\
\hline
  \end{tabular}
  \caption{\label{tab:unequal} Similar citation counts, different citation intents. Part I.}
\end{table*}

\begin{table*}
  \centering
  \begin{tabular}{|p{4.5cm}|p{4.5cm}|p{4.5cm}|}
    \hline
\multicolumn{3}{|c|}{\cellcolor{gray!15}\textbf{\texttt{Topic:} LMs for code generation; citation count: $\sim$[200-300]}}   \\
    \hline
            \textbf{\texttt{Paper:}} \cite{sun2020treegen} & \cite{le2022coderl} &  \cite{mastropaolo2021studying}    \\
         \textbf{\texttt{Top citation intents:}} drawing inspiration from prior work, comparing various state-of-the-art methods for code generation & suggesting potential integration of various search strategies for code generation, motivating the trend of using deep learning for code generation &  highlighting limitations in existing methods, building on previous ideas for code summarization\\
 \hline
\multicolumn{3}{|c|}{\cellcolor{gray!15}\textbf{\texttt{Topic:} Automated fact checking; citation count: $\sim$700}}   \\
    \hline
            \cite{ma2018rumor} & \cite{monti2019fake} &  \cite{bian2020rumor}    \\
motivating the growing interest in ML and NLP for rumor and fake news detection, comparing the proposed model with existing methods for effectiveness verification &  reporting shortcomings in fake news detection methods, comparing performance improvements of different methods   & highlighting the limitations of existing datasets for the task, drawing inspiration from existing strategies for early detection and localization \\
\hline
\multicolumn{3}{|c|}{\cellcolor{gray!15}\textbf{\texttt{Topic:} Multi-modal deep learning; citation count: $\sim$2000}}   \\
    \hline
            \cite{kim2021vilt} & \cite{liu2016coupled} &  \cite{li2021align}    \\
reporting performance comparisons among multimodal methods, highlighting advancements in multimodal pre-training and proposing new tasks &  reporting model performance compared to state-of-the-art, highlighting the success of GANs in image generation & highlighting the limitations of the current method in comparison to other visual LMs, drawing inspiration from existing multimodal models to propose a new method\\
\hline
\multicolumn{3}{|c|}{\cellcolor{gray!15}\textbf{\texttt{Topic:} Antibiotic resistance mechanisms; citation count:  $\sim$[2300-3000]}}   \\
    \hline
            \cite{aslam2018antibiotic} & \cite{pang2019antibiotic} &  \cite{wang2024large}    \\
motivating the urgent need for new antibiotics due to rising resistance, highlighting the potential therapeutic effects of the method &  highlighting the challenges and variations in antibiotic effectiveness against biofilms, motivating the urgent need for novel treatment strategies & motivating the need for ongoing antibiotic development and resistance study, highlighting the role of human behavior in antibiotic resistance\\
\hline
  \end{tabular}
  \caption{\label{tab:unequal2} Similar citation counts, different citation intents. Part II.}
\end{table*}

\begin{figure*}
\begin{tcolorbox}[colback=white!10!white,colframe=black!90!black,title=Generating a scientific impact summary about a researcher]
  The scientific impact of a researcher can be inferred from the impact of their publications and how they have been used by others.
Given a list of impact summaries about papers of a certain researcher, summarize their overall impact, and its evolution over time.\\

 This is the list of semi-structured paper summaries about the researcher:\\
  \texttt{\$paper\_impact\_summaries\$}

  Generate an impact summary that describes the overall impact of their papers. Focus more on how their papers have been used than what the content of their papers was.
\end{tcolorbox}
\caption{Prompt for generating author-level scientific impact summaries.}
 \label{fig:prompt_author_impact}
\end{figure*}
\newpage

\end{document}